\documentclass[12pt]{iopart}
\usepackage{epsf}
\usepackage{epsfig}

\begin{document}

\title[Fluctuations in the coarsening dynamics of the O(N) model]
{Fluctuations in the coarsening dynamics of the O(N) model: are they
  similar to those in glassy systems?}
\vskip 10pt
\author{
Claudio Chamon$^{1}$, Leticia F. Cugliandolo$^{2,3}$ and Hajime 
Yoshino$^2$\footnote{On leave from  Osaka University, 
Osaka, Japan}
}
\address{
$^1$ Physics Department, Boston University,
\\
590 Commonwealth Avenue, Boston, MA 02215, USA
\\
$^2$Laboratoire de Physique Th{\'e}orique  et Hautes {\'E}nergies, Jussieu, \\
5\`eme {\'e}tage,  Tour 25, 4 Place Jussieu, 75252 Paris Cedex 05, France
\\
$^3$Laboratoire de Physique Th{\'e}orique de l'{\'E}cole Normale
Sup{\'e}rieure,
\\
24 rue Lhomond, 75231 Paris Cedex 05, France
\\
{\tt chamon@bu.edu, leticia@lpt.ens.fr, yoshino@lpthe.jussieu.fr}
}
\date\today

\begin{abstract}

  We study spatio-temporal fluctuations in the non-equilibrium dynamics of
  the $d$ dimensional $O(N)$ in the large $N$ limit. We analyse the
  invariance of the dynamic equations for the global correlation and response
  in the slow ageing regime under transformations of time. We find that these
  equations are invariant under scale transformations.  We extend this study
  to the action in the dynamic generating functional finding similar results.
  This model therefore falls into a different category from glassy problems
  in which full time-reparametrisation invariance, a larger symmetry that
  emcompasses time scale invariance, is expected to be realised
  asymptotically.  Consequently, the spatio-temporal fluctuations of the
  large $N$ $O(N)$ model should follow a different pattern from that of
  glassy systems.  We compute the fluctuations of local, as well as spatially
  separated, two-field composite operators and responses, and we confront our
  results with the ones found numerically for the $3d$ Edwards-Anderson model
  and kinetically constrained lattice gases. We analyse the dependence of the
  fluctuations of the composite operators on the growing domain length and we
  compare to what has been found in super-cooled liquids and glasses.
  Finally, we show that the development of time-reparametrisation invariance
  in glassy systems is intimately related to a well-defined and finite
  effective temperature, specified from the modification of the
  fluctuation-dissipation theorem out of equilibrium. We then conjecture that
  the global asymptotic time-reparametrisation invariance is broken down to
  time scale invariance in {\it all} coarsening systems.

\end{abstract}

\newpage

\section{Introduction}
\label{sec:intro}

Many extended systems which consist in interacting microscopic degrees
of freedom exhibit non-trivial slow dynamics at low
temperatures. Macroscopic observables such as density-density or other
relevant correlations have extremely slow relaxations.  Magnetic,
dielectric or other susceptibilities slowly evolve in time. A large
amount of experimental and numerical data allow for a qualitative, and
sometimes also quantitative, description of these macroscopic
observables in a number of well-studied materials. A satisfying
understanding of the mechanism leading to such dramatic slowing down
is, however, still lacking. In order to get a better insight on the
relaxation of glassy systems it is important to investigate the
dynamics at length/times scales that range from the microscopic to the
macroscopic, through proper
experimental~\cite{exp-heter}-\cite{exp-aging-heter},
numerical~\cite{num-heter}-\cite{dynamical-legnth-sg},
and theoretical tools \cite{Fisher-Huse0}-\cite{Chamon-etal2}.

Systems with a clear mechanism for slow relaxations are the ones that
evolve through coarsening of domains. They may thus provide a useful
guideline to understand the dynamics of, in principle, more
complicated systems. After a transient, systems undergoing
phase-ordering kinetics enter a scaling regime in which the
order-parameter morphology and its correlation functions depend on
time only through a time-dependent length $L(t)$, that characterises
the mean size of the domains~\cite{Bray}. Interestingly enough, 
all microscopic details are absorbed in $L(t)$.
It is tempting to speculate that such space-time scaling 
also exists asymptotically in glassy systems.
This is the starting point, for example, in the dynamic droplet theory
of  spin-glasses~\cite{Fisher-Huse0} (see \cite{droplet-scaling} for
a detailed numerical examination).

Independently, analytical studies of dynamical mean-field theories of
glassy systems demonstrated that the relaxation of global two-time
correlation functions follows a self-similar structure, with a
long-times scaling given by a ratio between a function of time
evaluated at the two times involved, $C(t,t') \approx
f_C[h(t')/h(t)]$~\cite{Cuku}.  In these models, there is no
interpretation of the function $h(t)$ as a length-scale. Even more
generally, one can argue that any monotonic two-time correlation,
independently of the origin of the slow dynamics, should depend on
times only through a ratio $h(t')/h(t)$ within a given 
correlation scale~\cite{Cuku2}.

It has been noticed by several
authors~\cite{Somp1}-\cite{Kennett-etal2}
that the dynamic equations for the slow decay of the global
correlations and responses of mean-field disordered models with glassy
features acquire time-reparametrisation invariance once the
time-derivatives (and other irrelevant terms) are dropped in the long
times limit, in which the scaling in $h(t')/h(t)$ actually holds. 
This symmetry is not exactly realised since one function $h(t)$ is 
selected by the dynamic evolution; in other words, the time-derivative
and other irrelevant terms, act as (asymptotically vanishing) pinning 
fields that select the time-scaling $h(t)$. 
The development, at long times, of an {\it approximate} 
invariance under generic
reparameterisation of time has hindered the complete solution of the
dynamic problem, for fixing the choice of reparametrisation involves a
proper matching of the short-time and long-time dynamics that should be done 
by taking into account the effect of the time-derivative -- and other terms.

More recently it has been
suggested that a global time-reparametrisation invariance may also
exist in finite dimensional glassy systems and that it may be
responsible for the main spatio-temporal
fluctuations~\cite{Chamon-etal}-\cite{Chamon-etal2}. In this way, 
the inconvenience generated by the time-reparametrisation invariance 
was transformed into a tool with predictive power. Some consequences
of this proposal were listed in these articles
together with  their numerical checks in finite dimensional
spin-glasses~\cite{Castillo-etal1,Castillo-etal2} and kinetically
facilitated models~\cite{Chamon-etal2}. Interestingly enough, a
kind of `universality' emerged in the sense that the time 
evolution and form of the distributions of local correlations and responses 
followed a similar pattern for these rather different systems.

The global time-reparametrisation  $t \to h(t)$ we are referring to 
acts on all
spatial positions in identical way and it does not involve
transforming space simultaneously. It is then simpler in form than the
usual space-time rescaling that holds in coarsening systems at long
times and large scales. In several stages of this paper we compare the
time reparametrization invariance to the usual space-time rescaling.  We also
stress that time-reparametrisation invariance is a different
transformation from Henkel's local scale invariance
hypothesis~\cite{Henkel-etal} (see
\cite{Calabrese-Gambassi,Zannetti3} for a discussion on the validity
of the latter).

The aim of this paper is to investigate the similarities and
differences between fluctuations in simple coarsening and glassy
systems.  Specifically, we study analytically the coarsening dynamics
of the $d$ dimensional $O(N)$ model in the large $N$ limit.  This model
has been studied in a large number of papers, see
e.g.~\cite{Zannetti3}-\cite{Corberi-review} and references therein.  In
Sect.~\ref{sec:model} we review its static and dynamic behaviour.  We
explain in special detail the separation of the field in two
components, as presented by Corberi, Lippiello and
Zannetti~\cite{Corberi-etal}, and how this helps understanding the
condensation phenomenon and thermal fluctuations.  Next, we analyse
the fluctuating dynamics. In Sect.~\ref{sec:action} we derive the
dynamic generating functional and write it in terms of the slow and fast
fields introduced in Sect.~\ref{sec:model}. We also derive closed
dynamic equations for the global correlation and linear response of
{\it any}  $O(N)$ model in the large $N$ limit or spherical model. Then, in
Sect.~\ref{sec:time-transf} we examine the symmetries of the dynamic
equations for the global correlation and response, and the dynamical
generating functional, under global transformations of time. We compare with
the time-reparametrisation invariance suggested for glassy
systems. Section~\ref{sec:local-corr} is devoted to the study of the
probability distributions of the fluctuations at various mesoscopic
length/time scales through several dynamical observables. We confront
the latter to the results obtained for disordered
spin~\cite{Chamon-etal}-\cite{Castillo-etal2} and kinetically constrained
models~\cite{Chamon-etal2} and with the usual space-time scaling invariance
of pure ferromagnetic coarsening.
 In Sect.~\ref{sec:fourpoint} we compute a
four-point correlation function similar to the one that is usually
used in the context of super-cooled 
liquids~\cite{num-heter}-\cite{num-heter-aging}, 
\cite{Franz-Parisi}-\cite{Mayer-etal}, \cite{Ludovic-liquids} to
extract a dynamic growing length. We study its behaviour as a function
of the two times involved and discuss its relation to a response.
Finally, in Sect.~\ref{sec:conclusions} we present our conclusions
together with some speculations.

\section{The $O(N)$ model}
\label{sec:model}

The $d$-dimensional $O(N)$ non-linear sigma model is a coarse-grained 
approximation to a
lattice spin model with nearest-neighbour ferromagnetic interactions.  
Its Hamiltonian reads
\begin{displaymath}
H = \int_V d^dx \; \left[ \frac12 (\nabla \vec \phi(\vec x))^2 + 
\frac{g}{4N} (\phi^2(\vec x))^2 + \frac{r}{2} \phi^2(\vec x) 
- \vec h(\vec x,t) \vec \phi(\vec x) \right]
\; .
\end{displaymath}
The spatial dependence
is given by the continuous $d$-dimensional 
vector $\vec x=(x_1,\dots, x_d)$ and $V$ is the volume of the system.
The field $\vec \phi$ is an $N$-dimensional vector, $\vec \phi =
(\phi_1,\dots,\phi_N)$ with $-\infty < \phi_\alpha < \infty$. 
A subindex $\alpha$ labels 
its $N$ components, $\alpha=1,\dots,N$. 
The interplay between the quadratic and quartic terms (with couplings $r$
and $g>0$, respectively) favours the $\phi^2(\vec x,t)\equiv \sum_{\alpha=1}^N 
\phi_\alpha^2(\vec x,t)
=-Nr/g$ configurations for $r<0$.
$h_\alpha$ is a magnetic field coupled linearly to the field. 
In the infinitesimal limit $\vec h$ serves 
to compute the linear response, see eq.~(\ref{eq:linear-resp-def}).
A soft Ising, XY or Heisenberg model correspond to $N=1,2$ and $3$,
respectively.
In principle, the large $N$ limit is the starting point 
for a systematic $1/N$ expansion, although this may be difficult to 
control~\cite{Newman-Bray}. 

In the absence of the magnetic field $\vec h$, the Hamiltonian $H$ 
is invariant under uniform rotations of $\vec \phi$:
\begin{displaymath}
\phi_\alpha(\vec x)  \to \tilde \phi_\alpha(\vec x) = {\cal R}_{\alpha\beta} 
\phi_\beta(\vec x) \; , \;\;\;\; 
\forall \, \vec x
\; ,
\end{displaymath}
${\cal R}\in O(N)$.
The summation convention over repeated indeces is used here 
and in what follows.

Dynamics is attributed to the field via the 
Langevin equations of motion:
\begin{eqnarray}
\gamma \dot \phi_\alpha(\vec x,t) &=& \nabla^2 \phi_\alpha(\vec x,t) -
\left(\frac{g}{N}\, \phi^2(\vec x,t)+r \right)
\phi_\alpha(\vec x,t)+h_\alpha(\vec x,t) 
+\eta_\alpha(\vec x,t) 
\; . \nonumber
\end{eqnarray}
Henceforth 
we measure time in units of the inverse of the friction coefficient $\gamma$.
$\eta_\alpha(\vec x,t)$ is a spatially uncorrelated
Gaussian white noise with zero mean, 
$\langle \, \eta_\alpha(\vec x,t) \, \rangle = 0$ for all 
$\vec x$ and $t$,  and variance
\begin{eqnarray*}
\langle \, \eta_\alpha(\vec x,t) \eta_\beta(\vec x',t') \, \rangle 
&=& 2 k_B T \, 
\delta_{\alpha\beta} \, \delta^d(\vec x-\vec x') \, \delta(t-t') 
\; ,
\end{eqnarray*}
where $T$ is the temperature of the bath and $k_B$ is the Boltzmann
constant. It is convenient to regularise the spatial correlations of the
noise including a finite short-distance cut-off
\begin{eqnarray*}
\langle \, \eta_\alpha(\vec x,t) \eta_\beta(\vec x',t') \, \rangle 
&=& 2 k_B T \, 
\delta_{\alpha\beta} \, \,
\frac{e^{-\frac{1}{4}(\vec x-\vec x')^2
\Lambda^2}}{(4\pi\Lambda^{-2})^{d/2}}\;
\, \delta(t-t') 
\; ,
\end{eqnarray*}
that introduces correlations over a typical length $1/\Lambda$
simulating the lattice spacing and cures 
some short distance divergences. 
$1/(2\Lambda^2)$ will define a microscopic 
time scale $t_0$ that regularises divergent equal-time correlations. 
Hereafter the angular brackets indicate an average over the thermal noise
and we set $k_B=1$.

The stochastic evolution has to be supplemented with 
the initial condition $\vec\phi(\vec x,0)$. Since we are interested 
in phase-ordering dynamics, we typically 
choose initial conditions that are uncorrelated in the $N$ dimensional
space, 
$[\phi_\alpha(\vec x,0)\phi_\beta(\vec x,0)]_{ic}\propto \delta_{\alpha\beta}$,
and in real space, 
and have a Gaussian distribution 
\begin{equation}
P[\vec\phi(\vec x,0)] = 
(2\pi\Delta^2)^{-NV/2}
\; e^{-\frac{1}{2\Delta^2}
\sum_\alpha \int d^d x \; \phi_\alpha^2(\vec x,0)}
\; .
\label{eq:Gaussi-initial-pdf-x}
\end{equation}
Hereafter we use square brackets, $[\dots ]_{ic}$,  
to represent an average over 
initial conditions.

In the large $N$ limit one expects that the sum over components in
$\phi^2(\vec x,t)$ averages away the $\vec x$ dependence.
One then looks for a solution such that
\begin{equation}
z(\vec x, t) 
\equiv 
\frac{g}{N} \, \phi^2(\vec x,t)+r 
\approx z(t) \equiv
\frac{g}{N} \; [\, \langle \, \phi^2(\vec x,t) \, \rangle \, ]_{ic} + r
\; ,
\label{eq:z-def}
\end{equation} 
where the average in the last term is taken over thermal histories and
initial conditions. The functional form of $z(t)$
has to be determined self-consistently. As we shall see below the
time-dependence of $z(t)$ determines the scaling in time of most of the
interesting dynamic quantities. 
Note that we are implicitly assuming 
that $N\to\infty$ in that we are not letting $z$ fluctuate.
All results in this paper have been derived in this limit. 
As discussed by Newman and Bray \cite{Newman-Bray},
fluctuations of $z(t)$ appear at order $1/N$.

Under the assumption~(\ref{eq:z-def}), that has to be verified {\it a posteriori},
one can Fourier transform the Langevin equation
 and the noise-noise correlation.
We use the following conventions:
\begin{eqnarray*}
f(\vec k) = \int d^dx \; e^{-i\vec k \vec x} \, f(\vec x)
\; ,
\;\;\;\;\;\;\;\;\;\;\;\;
f(\vec x) = \int \frac{d^dk}{(2\pi)^d} \; e^{i\vec k \vec x} \, f(\vec k)
\; ,
\end{eqnarray*}
and we obtain
\begin{eqnarray}
&& \dot \phi_\alpha(\vec k,t) = -k^2 \phi_\alpha(\vec k,t) - z(t) 
\phi_\alpha(\vec k,t)
+ \eta_\alpha(\vec k,t) 
\; ,
\label{eq:Lang-ON-k}
\\
&& 
\langle \, \eta_\alpha(\vec k,t) \eta_\beta(\vec k',t') \, \rangle = 2 T \, 
\delta_{\alpha\beta} \;e^{-\frac{k^2}{\Lambda^2}} \, (2\pi)^d \;  
\delta^d(\vec k+\vec k') \, \delta(t-t') 
\nonumber
\; .
\label{eq:noise-noise-k}
\end{eqnarray}
In terms of the Fourier components $\vec \phi(\vec k,0)$, the initial
conditions are distributed according to
\begin{equation}
P[\vec\phi(\vec k,0)] = 
(2\pi\Delta^2)^{-NV/2}
\; e^{-\frac{1}{2\Delta^2} 
\int \frac{d^d k}{(2\pi)^d} \, \vec\phi(\vec k,0) \vec\phi(-\vec k,0)}
\; .
\label{eq:Gaussi-initial-pdf-k}
\end{equation}
Thus, the coupled dynamics in $x$ space transforms 
into a set of $N$ {\it independent} first-order differential 
equations for the $k$-components of the field.
The label $\alpha$ is now superfluous and we omit it 
unless otherwise stated.

The $O(N)$ model is intimately related to the spherical ferromagnet on a
lattice and the fully-connected spherical spin-glass with two-body
interactions.  The main difference between these models is the form of the
density of states of the quadratic interaction matrix and how it decays to
zero at its edge. In the case of the $O(N)$ model the density of states
is given by
\begin{equation}
g(\epsilon) \sim \epsilon^{\nu} \qquad \nu=d/2-1.
\label{eq:density-of-state}
\end{equation}
at low energies $\epsilon$.  Many papers have been devoted to the study of
the relaxation dynamics and global properties of the $O(N)$
model~\cite{Zannetti3}-\cite{Corberi-review}, the spherical
ferromagnet~\cite{Godreche-Luck}-\cite{Horner}, and the fully-connected
spin-glass with two-body interactions~\cite{Ciuchi},
\cite{Horner}-\cite{Secu}.  In the rest of this section we recall the main
features of the statics and dynamics of the $O(N)$ model while in the rest of
the paper we focus on the study of fluctuations and of symmetries under time
transformations.

\subsection{Statics}

Let us briefly review the static behaviour of the $O(N)$ model (see
Refs.~\cite{Zannetti3,Corberi-etal} for more details). 
If the volume $V$ is finite, the system
equilibrates in finite time and the probability distribution function ({\sc
  pdf}) of the order parameter approaches the Gibbs-Boltzmann form
\begin{eqnarray}
&& P_{eq}(\vec \phi) = 
Z^{-1} e^{-\frac{\beta}{2V} \sum_{\vec k}\; (k^2+ \xi^{-2})\;
\vec \phi(\vec k) \cdot \vec \phi(-\vec k)} 
\; ,
\label{eq:equilibrium-ON}
\\
&& Z \equiv \int {\mathcal D}\vec\phi \; 
e^{-\frac{\beta}{2V} \sum_{\vec k} \;(k^2+ \xi^{-2})\;
\vec \phi(\vec k)\cdot \vec \phi(-\vec k)} 
\; ,
\label{eq:equilibrium-ON-1}
\end{eqnarray}
meaning that the 
Fourier components are independent Gaussian random variables.
The path-integral measure is ${\cal D}\vec \phi
\equiv \prod_\alpha \prod_k d\phi_\alpha(\vec k)$.
$\xi$ is the static correlation length
\begin{equation}
\xi^{-2} = \frac{g}{N} \langle \, \phi^2(\vec x) \, \rangle_{eq} +r
\label{eq:corr-length}
\end{equation}
where the subindex `eq' indicates that the average 
has to be computed using the measure 
(\ref{eq:equilibrium-ON})-(\ref{eq:equilibrium-ON-1}). 
(We shall see below that $z(t)$ plays a similar role to $\xi^{-2}$.)
All modes have vanishing thermal average,
$\langle \, \vec \phi(\vec k)\,\rangle_{eq} =0$ for all 
$\vec k$. 
The static structure factor
\begin{eqnarray}
C_{\sc eq}(\vec k) \equiv \frac1{N} \;
\langle \, \vec\phi(\vec k) \cdot \vec\phi(-\vec k) \, \rangle_{eq} 
= \frac{TV}{k^2 + \xi^{-2}}
\label{eq:structure factor}
\end{eqnarray}
shows the ordering in the low temperature phase. 
The correlation length, $\xi$, is determined by eq.~(\ref{eq:corr-length})
with $\langle \, \phi^2(\vec x) \, \rangle_{eq}=
\langle \, \phi^2(\vec 0) \, \rangle_{eq}$ replaced by $V^{-1}$ times 
the sum over $\vec k$ of (\ref{eq:structure factor}). The detailed 
analysis of this equation has been presented elsewhere~(see 
{\it e.g.}~\cite{Corberi-etal}). 
One finds that in $2 < d  $ there is a finite 
critical temperature $T_c$ defined by 
\begin{equation}
r + g T_c \int \frac{d^dk}{(2\pi)^d} \; 
\frac{e^{-\frac{k^2}{\Lambda^2}}}{k^2} =0
\; ,
\label{eq:Tc}
\end{equation}
where the correlation length changes from a volume independent value
at $T>T_c$ to a volume-dependent one at $T\leq T_c$.  In $d=2$ the
integral over $k$ has a logarithmic divergence and the critical
temperature is pushed down to zero.  Above but near criticality $\xi$
behaves as
\begin{eqnarray*}
\xi &\sim \left(\frac{T-T_c}{T_c}\right)^{-\nu}
\; ,
\;\;\;\;\;\;\;\;\;\;
\nu=
\left\{
\begin{array}{ll}
1/2 \;\;\; &d>4
\; ,
\\
(d-2)^{-1} \;\;\; &d<4
\; ,
\end{array}
\right.
\end{eqnarray*} 
with logarithmic corrections in $d=4$. At $T_c$, 
$\xi\sim V^\zeta$ with $\zeta=1/4$ for $d>4$ and $\zeta=d^{-1}$
for $d<4$, again with logarithmic corrections in $d=4$. Below $T_c$,
the order parameter $m_{eq}$,
\begin{equation}
V^{2} m^{2}_{eq} \equiv N^{-1}\langle \, \phi^2(\vec k=\vec{0}) 
\, 
\rangle_{eq}.
\end{equation}
becomes non-zero and one finds,
\begin{eqnarray}
m^{2}_{eq} & = & -\frac{r}{g} \; \frac{T_c-T}{T_c}
\;\;\;\; \mbox{and} \;\;\;\;
\xi^2 \sim m_{eq}^2 \; \frac{V}{T} 
\; .
\label{eq:meq}
\end{eqnarray}

The temperature and volume dependence of $\xi$ dictates that of the 
structure factor. When $T>T_c$ the variance of all modes grows
linearly with the volume. Instead, when $T\leq T_c$, $\xi^{-2}$ is
negligible with respect to $k^2$ except at $\vec k=0$ yielding
\begin{eqnarray*}
C_{eq}(\vec k) =
\left\{
\begin{array}{ll}
V \,T_c \,k^{-2} \; (1-\delta_{\vec k, \vec 0}) + \overline c \,T_c\,
V^{2\zeta+1} \,\delta_{\vec k, \vec 0}
\;\;\;\;\;\;\; & T=T_c
\\
V \, T \, k^{-2} \; (1-\delta_{\vec k, \vec 0}) + m_{\sc eq}^2 \,V^2 \,
\delta_{\vec k, \vec 0}
\;\;\;\;\;\;\; & T<T_c
\end{array}
\right.
\end{eqnarray*}
where $\overline c$ is a constant and $m_{eq}$ is given in eq.~(\ref{eq:meq}). 
The transition is characterised by a zero wave-vector mode that 
condenses and has a variance, 
$\langle \, \phi^2(\vec k=\vec 0)\, \rangle_{eq}$,
that grows as $V^2$.
Below $T_c$ the equilibrium susceptibility $\chi_{eq}$ per unit volume is 
\begin{equation}
\chi_{eq}= \frac{m^{2}_{0}-m^{2}_{eq}}{T}=-\frac{r}{g}T^{-1}_{c}
=(4 \pi)^{-d/2} \frac{2}{d-2} \Lambda^{d-2}
\label{eq:chieq}
\end{equation}
where 
\begin{equation}
m_0^{2}=-r/g.
\label{eq:m0}
\end{equation}

In conclusion, the $O(N)$ model has a 
phase transition from a paramagnetic to a 
ferromagnetic phase. The low temperature phase is characterised by a 
condensation phenomenon that signals ordering. 
The upper critical dimension is $d=4$ and 
the lower critical dimension is $d=2$. 

\subsubsection{Separation of the field}{$\;$}

The nature of the phase transition and low-temperature phase
can be well understood by splitting the real-space
order parameter in a constant contribution and a space-varying 
one~\cite{Corberi-etal}:
\begin{eqnarray*}
&&
\vec \phi(\vec x) = \vec \sigma + \vec \psi(\vec x) 
\;\;\;\;\;\;\;\;
\mbox{with}
\nonumber\\
&&
\vec \sigma \equiv V^{-1} \; \vec \phi(\vec k=0)
\;\;\;\;\;\;
\mbox{and}
\;\;\;\;\;\;
\vec \psi(\vec x)  \equiv V^{-1} \; \sum_{\vec k\neq 0}
\vec \phi(\vec k) \, e^{i\vec k \cdot \vec x}
\; .
\end{eqnarray*}
It is clear that the fields $\vec \sigma$ and $\vec \psi$ are
independent.
The Gibbs-Boltzmann  measure factorises:
\begin{eqnarray*}
&& P[\vec \phi(\vec x)] = P(\vec \sigma) \;P[\vec \psi(\vec x)] 
\\
&& P(\vec \sigma) = (2\pi m_{\sc eq}^2)^{-N/2} \; 
e^{-\sigma^2/(2m_{\sc eq}^2)}
\; , 
\;\;\;\;\;\;
P[\vec \psi(\vec x)] =
Z_\psi^{-1} e^{-\beta/2 \int_V d^d x \; [\vec \nabla \vec \psi(\vec x)]^2}
\; ,
\end{eqnarray*}
with 
$Z_\psi = \int {\cal D} \vec \psi
\; e^{-\beta/2 \int_V d^d x \; [\vec \nabla \vec \psi(\vec x)]^2}$.
The first factor describes the condensate with macroscopic variance
$\langle \, \sigma_\alpha^2 \, \rangle_{\sc eq} = m_{\sc eq}^2$. 
The second factor describes thermal fluctuations about the condensate.
Consequently, the static correlation function separates in two terms:
\begin{equation}
C_{\sc eq}(\vec r) \equiv N^{-1} \langle \, \vec \phi(\vec x)
\cdot \vec \phi(\vec x+\vec r) \,\rangle_{eq}
= m^2_{\sc eq} + N^{-1} 
\langle \, \vec \psi(\vec x) \cdot \vec \psi(\vec x+\vec r) \,\rangle_{eq}
\label{eq:Ceq}
\end{equation}
where the first term represents the macroscopic variance of the 
condensate and the second one is the correlation of 
thermal fluctuations.

\subsection{Dynamics}

The set of linear differential equation (\ref{eq:Lang-ON-k})
can be easily solved:
\begin{eqnarray}
\phi({\vec k},t) &=& 
e^{-k^2 t -\int_0^t dt' z(t')} 
\;\phi({\vec k},0) 
\nonumber\\
&&
+ 
\int_0^t dt' \; e^{-k^2 (t-t') -\int_{t'}^t dt' \, 
z(t')} \; [\eta(\vec k, t')+h(\vec k, t')]
\; ,
\label{eq:sol-k}
\end{eqnarray}
where we dropped the component index $\alpha$, since all components satisfy
the same equations due to rotational symmetry.

The function $z(t)$ is self-consistently determined.
Indeed, 
\begin{displaymath}
Y^2(t) \equiv e^{\;2\int_0^t dt' \; z(t')}
\end{displaymath} 
satisfies the differential equation
\begin{equation}
\frac{dY^2(t)}{dt} = 
2 \left(\frac{g}{N} \; 
[ \langle \, \phi^2(\vec x,t) \, \rangle ]_{ic}+r \right)
Y^2(t)
\label{eq:Y2}
\end{equation}
that, using the equation of motion to represent 
$[\langle \, \phi^2(\vec x,t) \, \rangle]_{ic}$, transforms into a 
closed first-order differential equation for $Y^2(t)$ complemented 
by the initial condition $Y(0)=1$. One finds 
that $Y^2$  grows exponentially at high temperatures, as 
a power law at criticality, and it decays as a power law, 
$Y^2(t) \sim t^{-d/2}$, below the critical temperature
(see~\cite{Zannetti3} for a careful study of the preasymptotic
behaviour of $Y(t)$ at low temperature).

The solution (\ref{eq:sol-k}) 
takes a specially appealing form when written in terms
of 
\begin{eqnarray}
\left. 
\frac{\delta \phi_\alpha(\vec k,t)}
{\delta h_\beta(-{\vec k}',t')} \right|_{h=0}
= \delta_{\alpha\beta} \, 
\delta^d(\vec k+\vec k') \; \frac{Y(t')}{Y(t)} \; 
e^{-k^2 (t-t')}\;\theta(t-t')
\; ,
\label{eq:linear-resp-def}
\end{eqnarray}
which can be Fourier transformed to give
\begin{eqnarray*}
&&
\left. 
\frac{\delta \phi_\alpha(\vec x,t) }
{\delta h_\beta({\vec x}',t')} 
\right|_{h=0}
=
\int {\frac{d^dk}{(2\pi)^d}}\;{\frac{d^dk'}{(2\pi)^d}}\;
e^{i{\vec k}\cdot{\vec x}}\;e^{i{\vec k'}\cdot{\vec x'}}\;
(2\pi)^d\;\left. 
\frac{\delta \phi_\alpha(\vec k,t)}
{\delta h_\beta(-{\vec k}',t')} \right|_{h=0}
\end{eqnarray*}
Note that these quantities depend on the 
noise realisation and the initial condition only through the value of 
$Y(t)$. They are also identical to the linear response
function,
\begin{eqnarray}
R_{\alpha\beta}(\vec x,{\vec x}'; t, t') 
\equiv 
\left\langle \, \left. 
\frac{\delta  \phi_\alpha(\vec x,t)   }
{\delta h_\beta({\vec x}',t')} 
\right|_{h=0}\, \right\rangle
\; .
\label{eq:resp-def}
\end{eqnarray}
This property is special of (quasi) quadratic models.
Calling now 
\begin{displaymath}
r(k;t,t') \equiv \frac{Y(t')}{Y(t)} \; e^{-k^2 (t-t')}
\end{displaymath}
the solution (\ref{eq:sol-k}) can be rewritten as
\begin{equation}
\phi({\vec k},t) = 
r(k;  t,0)\; \phi({\vec k},0) + 
\int_0^t dt' \; r(k; t,t') \, 
\left[ \eta(\vec k, t') + h(\vec k, t') \right]
\; .
\label{eq:sol-k-resp}
\end{equation}

\subsection{Evolution of the distribution of Fourier components}

Let us consider the evolution of initial configurations
distributed according to the Gaussian law (\ref{eq:Gaussi-initial-pdf-x}) 
[and (\ref{eq:Gaussi-initial-pdf-k})] in the absence of the 
perturbing field $\vec{h}$. 
Expression (\ref{eq:sol-k-resp}) indicates that the field 
configuration at time $t$ is in linear relation with the initial 
condition and the thermal noise. Since these fields are independent 
and Gaussian distributed, $\vec \phi(\vec k,t)$ is also Gaussian 
distributed with zero mean and time-dependent variance 
\begin{eqnarray*} 
&& [\, \langle \, \phi(\vec k,t) \;\phi(\vec k', t) \, \rangle \, ]_{ic}
=
r(k,t,0) \,r(k',t,0)\; [\,  \phi(\vec k,0)\;\phi(\vec k',0) \, ]_{ic}
\nonumber\\
&& 
\;\;\;\;\;\;\;\;\;\;\;\;\;\;\;\;\;\;\;\;\;\;\;\;\;\;
+ \int_0^t dt' \int_0^t dt'' \; 
r(k,t,t')\, r(k',t,t'') \; 
\langle \, \eta(\vec k,t') \eta(\vec k',t'')\, \rangle 
\nonumber\\
&& \;\;\;\;\;\;\;\;\;\;\;
=
(2\pi)^d \, \delta^d(\vec k + \vec k')
\left[
\frac{\Delta^2}{Y^2(t)} \, e^{-2k^2 t}  + 2T \; 
\int_0^{t} dt' \; 
\frac{Y^2(t')}{Y^2(t)} \; e^{-2k^2 (t-t'+t_0)}
\right]
\; ,
\end{eqnarray*}
where we set $t_0\equiv(2\Lambda^2)^{-1}$.
Note that this result holds for each component in the $N$-dimensional 
space while $[\, \langle \, \phi_\alpha(\vec k,t) \;\phi_\beta(\vec k',t') \; 
\rangle \, ]_{ic}=0$ for all $t$ and $t'$ if $\alpha\neq \beta$.

\subsection{Correlations and responses}
\label{subsec:sigma}

Let us discuss the relaxation of the correlations
\begin{displaymath}
C_{\alpha\beta}(\vec x,\vec{x'};t,t') \equiv 
[\, \langle \, \phi_\alpha(\vec x,t) 
\phi_\beta(\vec{x'},t') \, \rangle \, ]_{ic}
\; ,
\end{displaymath} 
and the linear response defined in eq.~(\ref{eq:resp-def}).  Due to the
decorrelation of the initial conditions and noise in the $N$
dimensional space, these quantities are proportional to
$\delta_{\alpha\beta}$ and we henceforth omit the internal indeces
assuming that we take $\alpha=\beta$ in all quantities studied.

\subsubsection{Asymptotic behaviour.}{$\;$}

In $2 < d$ one finds a dynamic phase transition at the static critical
temperature, $T_c$ given in eq.~(\ref{eq:Tc}), 
where the asymptotic behaviour of $Y(t)$ changes.
At high temperature each mode and, hence, the global correlation decay
exponentially
\begin{eqnarray*}
&& 
C(t,t') \equiv V^{-1} \int d^d x \; C(x,x;t,t') 
\sim C_{\sc eq} \; e^{-(t-t')/t_{\sc eq}} 
\;\;\;\;
\mbox{with}
\nonumber\\
&&
C_{\sc eq} \equiv \langle \, \phi^2(\vec x) \, \rangle_{eq} 
\;\;\;\;
\mbox{and}
\;\;\;\;
t_{\sc eq} = 2 \xi^{-2}
\; ,
\end{eqnarray*}
where $\xi$ is the static correlation length given in
eq.~(\ref{eq:corr-length}). The linear response is related to the correlation
by the fluctuation dissipation theorem, $R(t,t') = T^{-1} \partial_{t'}
C(t,t') \theta(t-t')$.  At the transition one finds interrupted ageing. Since
we shall not discuss the critical dynamics, we do not give a detailed
description of the scaling laws at criticality.  Below the transition, after
a {\it transient} ({\it i.e.} $t'\gg \tau_t \gg 1$),
\begin{eqnarray*}
R(\vec k,\vec k';t,t') &\sim& 
(2\pi)^d \; \delta^d(\vec k+\vec k') \; \left(\frac{t}{t'}\right)^{d/4} \; 
e^{-k^2 (t-t')}
\; \theta(t-t')
\; ,
\\
C(\vec k,\vec k';t,t') &\sim& 
(2\pi)^d \; \delta^d(\vec k+\vec k') \; 
(tt')^{d/4} \; e^{-k^2 t-{k'}^2 t'} 
\\
&&
\, \times \,  \left[ 
\Delta^2
+
2 T  \, 
\int_0^{\min(t,t')} dt'' \; 
Y^2(t'') \; e^{(k^2+{k'}^2) (t''-2t_0)}
\right]
\; .
\end{eqnarray*}
Note that the asymptotic linear response does not depend on
temperature.  The first term in the correlation represents the decay
of the initial conditions while the second one has its origin in the
thermal noise.  Each Fourier component with $k >0$ decays
exponentially in time (with power law corrections).  The $k=0$
component behaves differently since the exponential factor disappears.
The slow decay of the low wave-vector components generates the
non-trivial dynamics of the global correlation and response.

From the expressions above one easily recovers the real-space
behaviour of the response and correlation. Using the large wave-vector
cut-off $\Lambda$, [see eq.~(\ref{eq:noise-noise-k})] -- 
that will be important in $d\geq 4$,
\begin{eqnarray}
R(\vec{x},\vec{x'}, t,t') \equiv \int \frac{d^dk}{(2\pi)^d} \, 
\int \frac{d^dk'}{(2\pi)^d} \; R(\vec k, \vec k'; t,t') 
e^{-i\vec k \vec x-i\vec k' \vec x'} \, e^{-(k^2+{k'}^2)/\Lambda^2}
\; , 
\end{eqnarray}
and similarly for the correlation, one finds
\begin{eqnarray}
&& R(\vec x,\vec x';t,t') \sim 
\left(\frac{t}{t'}\right)^{d/4} \; (t-t'+t_0)^{-d/2}
\; e^{-(\vec x-\vec x')^2/[4(t-t'+t_0)]}
\; ,
\label{eq:Rxttp}
\\
&& C(\vec x,\vec x';t,t') 
\sim 
(tt')^{d/4} 
\left[ \frac{\Delta^2}{(t+t')^{d/2}} \; e^{-(x-x')^2/[4(t+t')]}
\right.
\nonumber\\
&&
\;\;\;
\left.
+ 2T 
\int_0^{\min(t,t')} dt'' \; \frac{Y^2(t'')}{[t+t'+2(t_0-t'')]^{d/2}} \;
e^{-(x-x')^2/\{4[t+t'+2(t_0-t'')]\}}
\right]
\; ,
\label{eq:Cxttp}
\end{eqnarray}

The local response and correlation 
on the same spatial point, $\vec x=\vec x'$,
are independent of $\vec x$; thus they are also equal to the global values,
$R(\vec x,\vec x; t,t') = R(t,t')$ and $C(\vec x,\vec x; t,t') = C(t,t')$
with  
\begin{eqnarray}
R(t,t') \sim
\left(\frac{t}{t'}\right)^{d/4} \; (t-t'+t_0)^{-d/2}
\; ,
\\
C(t,t')
\sim 
\left(\frac{tt'}{(t+t')^2}\right)^{d/4} 
\left[ \Delta^2
+ 2 T \int_0^{\min(t,t')} dt'' \; 
\frac{Y^2(t'')}{\left[1-\frac{2t''}{(t+t')}\right]^{d/2}}
\right]
\; ,
\label{eq:Cglobal}
\end{eqnarray}
where we assumed that $t+t' $ is larger than $t_0=1/(2\Lambda^2)$ and we 
neglected the dependence of the correlation on this time-scale.
The contribution from the small wave vectors lead to a non-trivial dynamics 
of the global correlation and response, with no exponential
decay, and a separation of time scales shown below.

The case $d=2$, the lower critical dimension, may seem  slightly
different~\cite{Corberi-review}.  Interesting dynamics occurs only at
zero temperature, {\it i.e.} at the critical point.  However, the
dynamics is not typically critical but it corresponds to the zero
temperature limit of the coarsening phenomena observed in higher
dimensions. More precisely, there is still an additive separation of
time-scales, as opposed to what occurs in critical relaxations where
the separation is multiplicative and the ageing contribution to the 
correlation progressively disappears as time elapses.

One can check that these results are valid for all initial conditions
with short-range correlations.  Initial configurations with long-range
correlations (as for an ordered configuration) lead to different
scaling forms~\cite{Newman-Bray,p2spherical}.

\subsubsection{Separation of time-scales.}{$\;$}

At low temperature, $T< T_c$, 
and for very long waiting-time, $t'\gg \tau_t$, 
the global linear response and correlation have 
two distinct two-time regimes depending 
on the relation between the times $t$ and $t'$. 
These are defined by 
\begin{eqnarray*}
t-t' \ll t' \;\; & \;\;\;\;\;\; \mbox{stationary regime} \; ,
\\
\lambda \equiv \frac{t'}{t} \in [0,1) \;\; &  \;\;\;\;\;\;
\mbox{ageing regime} 
\; ,
\end{eqnarray*}
In the limit $t'\gg \tau_t$ 
the global (and local) correlation and linear response are 
well described by an additive separation
\begin{eqnarray}
&& C(t,t') = C_{\sc st}(t-t') + C_{\sc ag}(t,t') 
\; ,
\label{eq:Cglobal-sep}
\\
&& C_{\sc st}(t-t') \equiv 
(m_0^2-m_{eq}^2) \; (\Lambda^{2}(t-t')+1)^{1-d/2} 
\; ,
\label{eq:Cfast-global}
\\
&&C_{\sc ag}(t,t') \equiv 
m_{eq}^2 \left[ \frac{4\lambda}{(1+\lambda)^2} \right]^{d/4}
\; ,
\label{eq:Cag-global}
\end{eqnarray}
and
\begin{eqnarray}
&& R(t,t') = 
R_{\sc st}(t-t') + R_{\sc ag}(t,t') 
\; ,
\label{eq:Rglobal-sep}
\\
&& R_{\sc st}(t-t') = (4\pi)^{-d/2} \, (t-t'+t_0)^{-d/2}
= T^{-1} \partial_{t'} C_{\sc st} (t-t')
\label{eq:Rfast-global}
\\
&& R_{ag}(t,t') \equiv (4\pi)^{-d/2} \,
t^{-d/2} \, 
\left[\left(\frac{t'}{t}\right)^{-d/4} -1\right] \,  
\left(1-\frac{t'}{t}+\frac{t_0}{t} \right)^{-d/2} 
\; .
\label{eq:Rag-global}
\end{eqnarray}
$m_{\sc eq}$ is the equilibrium magnetisation
given in (\ref{eq:meq}) and $m_0$ is its value at $T=0$
given in (\ref{eq:m0}).  The relation (\ref{eq:chieq}) between 
$m^2_{eq}-m^2_0$ and $t_0$ ensures the validity of the fluctuation-dissipation 
theorem in the stationary regime.
For long-time differences, such that the
ratio between the two times $t$ and $t'$ is held fixed, $t'/t=\lambda
\in [0,1)$ the correlation and response ``age'', {\it i.e.}
they depend on the waiting-time $t'$. 

The detailed scaling of the correlation of the field evaluated at
different times and spatial points for the (simpler) Gaussian scalar
model was presented in \cite{Cukupa}. Here we just recall that
eqs.~(\ref{eq:Rxttp})-(\ref{eq:Cxttp}) can be put in a scaling
form~\cite{Bray}:
\begin{eqnarray}
&& C(\vec x,\vec x'=\vec x+\vec r;t,t') 
\sim 
f_{C_r}\left( \frac{r}{L(t')},\frac{L(t')}{L(t)}  
\right)
\; ,
\label{eq:scaling-corr}
\\
&& R(\vec x,\vec x'=\vec x+\vec r;t,t') 
\sim 
t^{-d/2} f_{R_r}\left( \frac{r}{L(t')},\frac{L(t')}{L(t)},\frac{t_0}{t}  
\right)
\; ,
\nonumber
\end{eqnarray}
with $r=|\vec x-\vec x'|$. 
$L(t)$ is the `domain length' at time $t$, which in the relaxation
$O(N)$ model with non-conserved order parameter
is given by $L(t)\sim \sqrt{t}$. In particular, the ageing contribution to
the global correlation (\ref{eq:Cag-global}) scales as
\begin{equation}
C_{\sc ag}(t,t') = 
f_C\left( \frac{L(t')}{L(t)} \right)
\label{eq:scaling-corr-global}
\;\;\;\;\;\;\;
\mbox{with} 
\;\;\;\;\;\;\;
f_C(x) = m_{eq}^2 \left( \frac{2 x}{1 + x^2} \right)^{d/2} 
\; ,
\end{equation}
$f_C(x) \sim x^{\frac{d}{2}}$ when $x\sim 0$ and $f_C(x) \sim
m^{2}_{eq} (1 - d \epsilon^2/4)$ when $x\sim 1-\epsilon$ and $\epsilon
\ll 1$. Note that the regime of very separated times, $t \gg t'$ or
$x\sim 0$, is characterised by a power law decay with an exponent
$\overline\lambda=d/2$~\cite{BK92,Huse-Jenssen}.  The global linear
response (\ref{eq:Rag-global}) also takes a scaling form~\cite{Zannetti3}:
\begin{eqnarray}
R_{ag}(t,t') 
\sim 
t^{-d/2} f_R\left(x,y\right)
\; ,
\;\;\;\;\;
\mbox{with} 
\nonumber\\
f_R(x,y) = (4\pi)^{-d/2} \, (1-x)^{-d/4}\, (1-x+y)^{-d/2}
\; ,
\;\;\;\;\;
\mbox{and} 
\nonumber\\
x=t'/t =L^2(t')/L^2(t) \; , \;\;\; y=t_0/t 
\label{eq:scaling-resp-global}
\; .
\end{eqnarray}

In the coarsening regime
the global correlation and response are not related by the fluctuation
dissipation theorem. One defines the ratio~\cite{Leto}
\begin{equation}
X(t,t') \equiv 
\frac{T R_{ag}(t,t')}{\partial_{t'} C_{ag}(t,t')} 
\sim t^{1-d/2} f_X(\lambda)
\;.
\label{eq:Xdef}
\end{equation}
Note that this is a decreasing function of time that tends to zero 
in all $d>2$ and to a function of the times ratio, $f_X(\lambda)$, 
taking finite values when $d\to 2^+$.

The dc susceptibility or zero field cooled magnetisation, 
defined as the integral of the linear response over a time period:
\begin{equation}
\chi(t,t') = \int_{t'}^t dt'' \, R(t,t'') 
\end{equation}
can be expressed as a sum of two terms, a stationary and an 
ageing contribution,
\begin{equation}
\chi(t,t') \approx
\chi_{st}(t,t') + \chi_{ag}(t,t') 
=
\int_{t'}^t dt'' \, R_{st}(t-t'') + \int_{t'}^t dt'' \, R_{ag}(t,t'')
\end{equation}
given by~\cite{Zannetti3}
\begin{eqnarray}
\chi_{st}(t-t') &=&
\chi_{\rm eq}\; 
\{ 1- [\, (t-t')/t_0+1 \, ]^{1-d/2} \}
\; , 
\label{eq:chi-st}
\\
  \chi_{ag}(t,t') &\sim&
\label{eq:chi-ag}
\left\{
\begin{array}{ll}
{t}^{1-d/2} \;\;\; & d<4
\; , 
\\
{t}^{1-d/2} \ln(t/t_0) \;\;\; & d=4
\; , 
\\
{t}^{-1} t_0^{1-d/2} \;\;\; & d>4
\; .
\end{array}
\right.
\end{eqnarray}
where $\chi_{\rm eq}=(4\pi)^{-d/2} \,t_0^{1-d/2} /(d/2-1)$ is the
equilbrium susceptibility given in eq.~(\ref{eq:chieq}). There are several
features to be noticed in these expressions.  The first one is that the
stationary integrated response approaches a value proportional to
$t_0^{1-d/2}$ in the long $t-t'$ limit. If one takes the cut-off $\Lambda$ to
infinity this value {\it diverges} as a power law, $t_0^{1-d/2}$ in all
$d>2$. Instead, in $d=2$ $\chi_{st}$ {\it diverges} as a logarithm of the
time difference, $\chi_{st}(t,t') \sim \ln[(t-t')/t_0]$, for $t-t'\gg t_0$.
The approach to this asymptotic value is given by a power law,
$[(t-t')/t_0]^{1-d/2}/(1-d/2)$, that will play an important role in the
analysis of the invariances of the slow dynamics.  Above $d=4$ the decay of
the ageing part of the integrated linear response does not depend on
dimensionality any longer. As discussed by Corberi {\it et al} a similar
upper dimension $d_\chi$ is expected to exist in other coarsening
systems~\cite{Zannetti3}. In all $d>2$ the ageing contribution to the total
susceptibility vanishes at long times, {\it i.e.} when $t\to\infty$.

$d=2$ is the lower critical dimension. 
  But the dynamic behaviour at zero
  temperature can be reached as the zero temperature limit of the
  finite temperature coarsening dynamics just described~\cite{Corberi-review}. 
  The additive
  separation of the correlation and response also holds in this
  case. In particular, the Edwards-Anderson parameter, $q_{ea} =
  m_{eq}^2$, that separates the stationary from the ageing regime in
  the correlation, remains finite (and equal to $m_0^2$ at zero
  temperature) in the limit of long waiting-time, $t'\to\infty$. Note
  that the stationary response in (\ref{eq:Rfast-global}) does not
  depend on temperature and thus these soft `spins' respond even at
  zero temperature. In particular, one has
\begin{equation}
R_{st}(t-t') = - \lim_{T\to 0} T^{-1}\; \frac{d}{dt} C_{st}(t-t')
\;, \;\;\;\; \mbox{for} \;\; t-t' >0\; .
\end{equation}

\subsubsection{Separation of the field}{$\;$}
\label{subsec:sigma-psi}

Interestingly enough, Corberi, Lippiello and Zannetti showed
that in $d>2$ the above results can also be found by 
using a splitting of the space and time dependent
field $\vec \phi(\vec x,t)$ in two
components~\cite{Corberi-etal}:
\begin{eqnarray*}
\vec \phi(\vec x,t) &=& \vec \sigma(\vec x,t) +\vec \psi(\vec x,t)
\; .
\end{eqnarray*}
Indeed, the solution (\ref{eq:sol-k-resp}) 
can be rewritten as
\begin{eqnarray*}
\vec \phi(\vec k,t) &=& \vec \sigma_h(\vec k,t) +\vec \psi(\vec k,t)
\; ,
\\
\vec\sigma_h(\vec k,t)
&\equiv& r(k;t,t_1) \vec \phi(\vec k,t_1)
+ \int_{t_w}^t dt' \; r(k; t,t') \, \vec h(-\vec k,t')
\; ,
\\
\vec \psi(\vec k,t) 
&\equiv& \int_{t_1}^t dt' \; r(k;t,t') \, \eta(-\vec k,t')
\; .
\end{eqnarray*}
$t_1$ is an arbitrary time satisfying $t_1 \ll t' \leq t$ and sufficiently long
so that the scaling limit has been established between the initial
quench and $t_1$, {\it i.e.} $t_1\gg \tau_t$. 
The second term in $\vec\sigma$ represents the effect 
of an external field applied from $t_w$ (another long time $t_w\gg t_1$)
on.
Note that, in the absence of the field $\vec{h}$, the ``slow'' component 
$\vec \sigma$ can also be written as the 
evolution of the initial condition since 
\begin{eqnarray*}
\vec \sigma(\vec k,t)
&& 
\equiv r(k;t,t_1) \,\vec \phi(\vec k,t_1)
\nonumber\\
&&
=
r(k;t,t_1)\,r(k;t_1,0) \,\vec \phi(\vec k,0)
+ \int_{0}^{t_1} dt' \; r(k;t,t_1)\,r(k;t_1,t') \,\vec \eta(\vec k,t')
\nonumber\\
&& 
=
r(k;t,0) \,\vec \phi(\vec k,0)
+ \int_{0}^{t_1} dt' \; r(k;t,t') \,\vec \eta(\vec k,t')
\; .
\end{eqnarray*}
Any typical initial condition that is the result of a quench 
from high temperatures can be thought of as a random noise
of Gaussian type. Thus, the first term is statistically 
`identical' to the contribution of the lower limit of the integral
in the second term.

The field $\vec \sigma$ is associated to local condensation of the
order parameter while $\vec \psi$ describes thermal fluctuations within
the domains.  These fields are statistically independent ($\langle \, \vec
\sigma(\vec x,t) \vec \psi(\vec x', t) \, \rangle =0$) and have
zero average.  The explicit calculations in \cite{Corberi-etal}
demonstrate that, in the long times limit, $t \geq t' \gg t_1$ with $t_1$
itself diverging, the global correlation of $\vec \sigma$,
$N C_\sigma(t,t',t_1) = \langle \, \vec\sigma(t)  
\vec\sigma(t') \, \rangle$,  yields the
ageing component of the global correlation of the field $\vec \phi$,
while the global correlation of $\vec \psi$,
$N C_\psi(t,t') = \langle \, \vec\psi(t)  
\vec\psi(t') \, \rangle$, yields the stationary
components of the global correlation of the field $\vec \phi$. More
precisely, for $t, t' \gg t_1$ one finds
\begin{eqnarray}
C_{\psi}(t,t';t_1) &=&
(m_0^2-m_{eq}^2) \; 
\left[ \, (4t_1/t_0)^{1-d/2} 
\left(\frac{4 tt'}{(t+t')^2} \right)^{d/4}
\right.
\nonumber \\
&& \left. \hspace*{2cm}+ 
(2(t-t')/t_0+1)^{1-d/2} \, \right] 
\; , 
\label{eq:Cpsi}
\\
C_{\sigma}(t,t';t_1) &=&
\left[ \, m_{eq}^2 - (m_0^2-m_{eq}^2) (4t_1/t_0)^{1-d/2} \, \right]
\left(\frac{4 tt'}{(t+t')^2} \right)^{d/4}
\; .
\label{eq:Csigma}
\end{eqnarray}
In the limit $t_1/t_0 \gg 1$ and for $d>2$ 
the first term in (\ref{eq:Cpsi}) vanishes and $C_{\psi}$
describes
the time-difference variation of the correlation in the 
stationary approach to the plateau at $m_0^2-m_{eq}^2$.
Similarly, $C_\sigma(t,t',t_1)$ becomes $C_{\sc ag}(t,t')$.
Indeed, in the stationary regime, $t-t' \ll t'$, $C_\psi$ 
varies from $m_0^2-m_{eq}^2$ to zero while $C_\sigma$ takes the constant
value $m_{eq}^2$. In the ageing regime $t'/t=\lambda$, $C_\psi$ has already
decayed to zero while $C_\sigma$ varies from $m_{eq}^2$ to zero.
 
The linear response is simply obtained as 
\begin{displaymath}
\left.
\left\langle\;
\frac{\delta  \phi_\alpha(\vec k, t)}{\delta h_\beta (-\vec k', t')}
\;  \right\rangle 
\right|_{h=0}
=
\left.
\left \langle \; 
\frac{\delta {\sigma_h}_\alpha(\vec k, t)}
{\delta h_\beta (-\vec k', t')}
\right\rangle 
\right|_{h=0}
\; .
\end{displaymath}

A similar separation has been used by Franz and Virasoro in a more general 
context~\cite{Franz-Virasoro}.

\section{The action}
\label{sec:action}

Let us now write the dynamic generating functional in terms of a path
integral over the fields $\vec\sigma$ and $\vec\psi$. This will be useful to
identify the symmetries of the ``slow'' action 
under transformations of time.

The dynamic generating functional is
\begin{eqnarray}
&& 
Z = \int {\mathcal D}\vec\phi(\vec k,t) 
{\mathcal D}i\vec{\hat\phi}(\vec k,t) 
{\mathcal D}\vec\eta(\vec k,t) 
\; 
e^{-\int \frac{d^d k}{(2\pi)^d} \int_0^\infty  dt \; S_{kt}^{(1)}}
\; ,
\nonumber\\
&& S_{kt}^{(1)}
\equiv 
i \vec{\hat\phi}(\vec k,t)
\left[
\left(\partial_t + k^2 + z(t) \right) \vec \phi(-\vec k,t)
- \vec \eta(-\vec k,t) 
\right.
\nonumber\\
&& 
\;\;\;\;\;\;\;\;\;\;\;\;\;\;\;\;\;\;\;\;\;\;
\left.
-
\vec h(-\vec k,t) \theta(t-t_w)  
\right]
- (4T)^{-1}
\; \vec \eta(\vec k,t)  \vec \eta(-\vec k,t)
\label{action1}
\end{eqnarray}
where, for simplicity, we took the cut-off $\Lambda$ in the noise-noise 
correlation to infinity. The external field $\vec h$ is applied from 
$t_w$ onwards. 
We call $R^{-1}(k;t,t')$ the differential operator
$\delta(t-t') \left[\partial_{t'} + k^2 + z(t')\right]$
whose inverse is the {\it retarded} 
linear response function, see eq.~(\ref{eq:resp-def}), 
\begin{displaymath}
\int dt' \; \delta(t-t') \left(\partial_{t'} + k^2 + z(t') \right) 
R(k; t',t'') =\delta(t-t'')
\; ,
\end{displaymath}
$R(k;t,t') = r(k;t,t') \theta(t-t')$,
for each $k$. The action in eq.~(\ref{action1})  can be rewritten as
\begin{eqnarray}
&& 
\int \frac{d^d k}{(2\pi)^d} \int_0^\infty  dt \; S_{kt}^{(1)}
=
\int \frac{d^d k}{(2\pi)^d} \int_0^\infty  dt\int_0^\infty  dt' \; 
S^{(2)}_{ktt'}
\label{eq:action2}
\\
&& 
S^{(2)}_{ktt'}
=
i \vec{\hat\phi}(\vec k,t) R^{-1}(k;t,t')
\left[\vec \phi(-\vec k,t') - 
\int_0^{t_1} dt'' \; R(k;t',t'') \vec \eta(-\vec k,t'')
\right.
\nonumber\\
&&
\;\;\; 
\left.
- \int_{t_1}^{\infty} dt'' \; R(k;t',t'') \vec\eta(-\vec k,t'') 
- \int_0^\infty dt'' \; R(k;t',t'') \vec h(-\vec k,t'') \theta(t''-t_w)
\right]
\nonumber\\
&&
\;\;\;\;\;
- (4T)^{-1} \; \vec \eta(\vec k,t) \delta(t-t') \vec\eta(-\vec k,t')
\; .
\nonumber
\end{eqnarray}
Defining 
\begin{eqnarray*}
\vec \sigma(\vec k,t) 
&\equiv& 
\int_0^{t_1} dt'' \; R(k;t,t'') \vec\eta(-\vec k,t'')
+
\int_{t_w}^\infty dt'' \; R(k;t,t'') \vec h(-\vec k,t'') 
\\
&=& 
\int_0^{t_1} dt'' \; r(k;t,t'') \vec\eta(-\vec k,t'')
+
\int_{t_w}^t dt'' \; r(k;t,t'') \vec h(-\vec k,t'') \theta(t-t_w) 
\; , 
\\
\vec \psi(\vec k,t) 
&\equiv& 
\int_{t_1}^{\infty} dt'' \; R(k;t,t'') \vec\eta(-\vec k,t'')
=
\int_{t_1}^{t} dt'' \; r(k;t,t'') \vec\eta(-\vec k,t'')
\; , 
\end{eqnarray*}
along the lines of what has been reviewed in Sect.~\ref{subsec:sigma-psi}, 
and introducing these definitions with delta functions
in the generating functional one has 
\begin{eqnarray*}
&& Z=\int {\mathcal D}\vec\phi  
{\mathcal D}i\vec{\hat\phi} {\mathcal D}\vec\eta  
{\mathcal D}\vec\sigma  {\mathcal D}\vec{\hat\sigma}  
{\mathcal D}\vec\psi  {\mathcal D}\vec{\hat\psi}  
\; 
e^{\int \frac{d^d k}{(2\pi)^d} \int_0^\infty  dt\int_0^\infty  dt'
\; S_{ktt'}^{(3)}}
\\
&& 
S_{ktt'}^{(3)}=
i \vec{\hat\phi}(\vec k,t) R^{-1}(k;t,t')
\left[\vec\phi(-\vec k,t') - \vec\sigma(-\vec k,t') -
\vec\psi(-\vec k,t')\right]
\nonumber
\\
&&
\;\;\;\;\;\;\;\;
+
i\vec {\hat\sigma}(\vec k,t) \delta(t-t') 
 \left[ \vec \sigma(-\vec k,t') -
\int_{0}^{t_1} dt'' \; r(k;t',t'') \vec\eta(-\vec k,t'')+
\right.
\nonumber\\
&& 
\left.
\;\;\;\;\;\;\;\;\;\;\;\;\;\;\;\;\;\;\;\;\;\;\;\;\;\;\;\;\;\;\;
\;\;\;\;\;\;\;\;\;
-
\int_{t_w}^\infty dt'' \; R(k;t',t'') \vec h(-\vec k,t'')
\right]
\nonumber\\
&& 
\;\;\;\;\;\;\;\;
+
i\vec {\hat\psi}(\vec k,t) \delta(t-t') 
 \left[ \vec \psi(-\vec k,t') -
\int_{t_1}^\infty dt'' \; R(k;t',t'') \vec\eta(-\vec k,t'')
\right]
\nonumber\\
&& 
\;\;\;\;\;\;\;\;
- (4T)^{-1} \; \vec \eta(\vec k,t) \delta(t-t') \vec\eta(-\vec k,t')
\; .
\end{eqnarray*}
In the end we shall focus on the behaviour of the two-time
action for long times; more explicitly we shall take $t$ and $t'$ to 
be longer than a long but otherwise arbitrary time $t_1$
but, for the moment, $t_1$ is just an arbitrary time scale.

The integration over the $\vec{\hat \phi}$ field yields a
functional delta-function.  
Integrating next over the thermal noise (note that the 
noises multiplying $\vec\sigma$ and $\vec\psi$ are independent
since they are evaluated at different times),
one finds
\begin{eqnarray*}
&& Z=\int {\mathcal D}\vec\phi  
{\mathcal D}\vec\sigma  {\mathcal D}\vec{\hat\sigma}  
{\mathcal D}\vec\psi  {\mathcal D}\vec{\hat\psi}  
\; 
e^{\int \frac{d^d k}{(2\pi)^d} \int_0^\infty  dt\int_0^\infty  dt'
S_{ktt'}^{(4)}}
\; 
\nonumber\\
&&
\;\;\;\;\;\;\;\;
\times \; 
\delta\left[\int_0^t  dt'\; R^{-1}(k;t,t') 
\left( \vec\phi(\vec k, t')-\vec\sigma(\vec k,t')-
\vec\psi(\vec k,t')\right)\right] 
\\
&& 
S_{ktt'}^{(4)}=
\delta(t-t') \left[ 
i\vec {\hat\sigma}(\vec k,t)   \vec \sigma(-\vec k,t') 
+
i\vec {\hat\psi}(\vec k,t)  \vec \psi(-\vec k,t')
\right.
\nonumber\\
&& 
\left.
\;\;\;\;\;\;\;\;\;\;\;\;\;\;\;\;\;\;\;\;\;\;\;\;\;\;\;\;\;\;\;\;
- i \vec {\hat \sigma}(\vec k, t)  
\int_{t_w}^\infty dt'' R(k; t',t'') \vec h(-\vec k, t'') 
\right]
\nonumber\\
&&
\;\;\;\;\;\;\;\;
+ T \int_0^{t_1} dt'' \; 
i\vec {\hat \sigma}(\vec k,t) r(k;t,t'')  
i\vec {\hat \sigma}(-\vec k,t') r(k;t',t'') 
\nonumber\\
&&
\;\;\;\;\;\;\;\;
+ T \int_{t_1}^\infty dt'' 
\; i\vec {\hat \psi}(\vec k,t) R(k;t,t'') 
 i\vec {\hat \psi}(-\vec k,t') R(k;t',t'') 
\; .
\end{eqnarray*}
From this expression one can relate the linear response 
of the field $\vec \sigma$ to the correlation 
between the fields $\vec {\hat \sigma}$ and $\vec {\sigma}$:
\begin{eqnarray*}
&& \left. \left\langle 
\frac{\delta \sigma_\alpha(\vec k,t)}{\delta h_\beta(-\vec k',t')} 
\right\rangle\right|_{h=0}
=
- \int_{t_w}^\infty dt'' \, R(k',t'',t')
\left\langle \sigma_\alpha(\vec k, t) 
i \hat \sigma_\beta(-\vec k',t'')  \right\rangle
\theta(t'-t_w)
\; .  
\end{eqnarray*}
The average has to be computed with the action $S_{ktt'}^{(4)}$ evaluated 
at zero field ($\vec h=\vec 0$). Since this action is diagonal 
in $\vec k$ and time the cross-correlation
between the fields is just 
$\langle i\hat\sigma_\beta(-\vec k',t'') \sigma_\alpha(\vec k,t)\rangle=
- \delta^d(\vec k+\vec k') \delta_{\alpha\beta} \delta(t-t'')$. Thus, 
\begin{eqnarray}
&& \left. \left\langle 
\frac{\delta \sigma_\alpha(\vec k,t)}{\delta h_\beta(-\vec k',t')} 
\right\rangle \right|_{h=0}
=
r(k,t,t') \, \theta(t-t') \, \delta^d(\vec k+\vec k') \, \theta(t'-t_w)  
\label{eq:result-response-sigmahat}
\end{eqnarray}
as it should. 

The action $S_{ktt'}^{(4)}$ is quadratic in the 
fields $\vec {\hat \sigma}$ and 
$\vec {\hat \psi}$. Integrating them out one finds
\begin{eqnarray}
&& Z \propto \int {\mathcal D}\vec\phi  
{\mathcal D}\vec{\sigma}  
{\mathcal D}\vec{\psi}  
\; 
e^{\int \frac{d^d k}{(2\pi)^d} \int_0^\infty  dt\int_0^\infty  dt'
S_{ktt'}^{(5)}}
\;
\nonumber\\
&&
\;\;\;\;\;\;\;\;
\times \; 
\delta\left[\int_0^t  dt'\; R^{-1}(k;t,t') 
\left( \vec\phi(\vec k, t')-\vec\sigma(\vec k,t')-
\vec\psi(\vec k,t')\right)\right] 
\; , 
\nonumber\\
&& 
S_{ktt'}^{(5)}=
-(4T)^{-1} 
\vec \sigma(\vec k,t)  
\left[ \int_0^{t_1} dt'' \; r(k;t,t'') r(k;t',t'') \right]^{-1} 
\vec \sigma(-\vec k,t') 
\nonumber\\
&&
\;\;\;\;\;\;\;\;
- 
(4T)^{-1} 
\vec \psi(\vec k,t)  
\left[ \int_{t_1}^\infty dt'' \; R(k;t,t'') R(k;t',t'') \right]^{-1} 
\vec \psi(-\vec k,t') 
\; .
\label{eq:S5}
\end{eqnarray}
(The proportionality sign is due to the fact that the integration over
the fields $\vec {\hat \sigma}$ and $\vec {\hat \psi}$ also yields a
determinant that is just a `numerical constant' that we ignore.)  From
this action one easily derives the correlations of the Fourier
components of the $\vec \sigma$ and $\vec \psi$ fields:
\begin{eqnarray}
N^{-1} \langle \, \vec \sigma(\vec k,t)
  \vec \sigma(-\vec k,t') \, \rangle
&=&  
 2T \int_0^{t_1} dt'' \; r(k;t,t'') r(k;t',t'')
\nonumber\\
&=& 
r(k;t,t_1) r(k;t',t_1) \; 
\langle \, \vec \sigma(\vec k,t_1)   
\vec \sigma(-\vec k,t_1) \, \rangle
\; ,
\label{eq:Csigmak}
\\
N^{-1} 
\langle \, \vec \psi(\vec k,t)
 \vec \psi(-\vec k,t') \, \rangle
&=&  
 2T \int_{t_1}^{\min(t,t')} dt'' \; r(k;t,t'') r(k;t',t'')
\; .
\label{eq:Cpsik}
\end{eqnarray}

In Sect.~\ref{subsec:sigma} we showed that the separation in fast and slow 
time-scales is clear in the spatial domain. 
Going back to real space one has 
\begin{eqnarray}
S_{xytt'}^{(5)} &=&
-\frac{1}{2} \, 
\vec \sigma(\vec x,t)  
K_\sigma(\vec x, \vec y;t,t')
\; \vec \sigma(\vec y,t') 
\nonumber\\
&&
-\frac{1}{2} \,
\vec \psi(\vec x,t)  
K_\psi(\vec x, \vec y;t,t')
\; \vec \psi(\vec y,t') 
\; ,
\label{eq:S6}
\end{eqnarray}
with  
$K^{-1}_\sigma(\vec x, \vec y;t,t')=
N^{-1} \langle \, \vec \sigma(\vec x, t)  
\vec \sigma(\vec y,t') \, \rangle $
and
$K^{-1}_\psi(\vec x, \vec y;t,t')=N^{-1}\langle \, \vec \psi(\vec x, t)  
\vec \psi(\vec y,t') \, \rangle$
the Fourier transforms of 
(\ref{eq:Csigmak}) and (\ref{eq:Cpsik}), respectively.
 
Focusing on equal space points, $\vec x=\vec y$, and 
taking the limit $t\geq t' \gg t_1 \gg \tau_t$ as in~\cite{Corberi-etal},
eq.~(\ref{eq:S6}) is the action in the generating functional for the 
slow and fast components of the global correlation
given in eqs.~(\ref{eq:Cfast-global}) and (\ref{eq:Cag-global}), 
respectively.

\section{Time transformations}
\label{sec:time-transf}

Long ago it was realised that the dynamic equations of motion of
mean-field disordered models acquire, in the long waiting time limit
and for large separations of times, an invariance under generic
reparameterisation of time~\cite{Somp1}-\cite{Kennett-etal2}. 
This symmetry initially
appeared as a nuisance since it was related to the impossibility of
determining the equivalent of the scaling function $L(t)$
analytically. More recently, we tried to use this symmetry as a
guideline to predict the main fluctuations in finite dimensional
systems undergoing glassy dynamics~\cite{Chamon-etal}-\cite{Chamon-etal2}. 
With this aim we first analysed
the symmetry properties of the action of the $d$-dimensional
Edwards-Anderson spin-glass~\cite{Chamon-etal}.  Let us here recall
the definition of the time-reparametrisation, how it acts on the
fields, and check whether this invariance exists in the $O(N)$ model.

\subsection{Global time-reparametrisation}

Global monotonic time-reparametrisation is defined 
as~\cite{Chamon-etal}
\begin{equation}
t \to \tilde t \equiv h(t)
\label{eq:trep}
\end{equation}
with $h(t)$ any monotonic function of time. 
A particular subset of transformations are 
re-scalings of time
\begin{equation}
t \to \zeta t \;\;\;\; \mbox{that correspond to}
\;\;\;\; h(t) = \zeta t 
\; .
\label{eq:rescaling}
\end{equation}
The transformation (\ref{eq:trep}) acts on the
fields $\vec\phi(\vec x,t)$ and $\vec{\hat\phi}(\vec x,t)$
as
\begin{eqnarray}
\vec\phi(\vec x,t) &\to& 
\tilde{\vec\phi}(\vec x, t) \equiv \vec\phi(\vec x, h(t))
\; ,  
\label{eq:RpG-phix}
\\
\vec{\hat\phi}(\vec x,t) &\to& 
\tilde{\vec{\hat\phi}}(\vec x,t) \equiv 
\frac{dh(t')}{dt'} \;
\vec{\hat\phi}(\vec x,h(t))
\; .
\label{eq:RpG-hatphix}
\end{eqnarray}
Consequently, the
space and time dependent two-point functions transform as
\begin{eqnarray}
C(\vec x, \vec x'; t,t') &\to& 
\tilde C(\vec x, \vec x'; t,t') \equiv C(\vec x, \vec x'; h(t),h(t'))
\; ,  
\label{eq:RpG-Cx}
\\
R(\vec x, \vec x';t,t') &\to& 
\tilde R(\vec x, \vec x';t,t') \equiv \frac{dh(t')}{dt'} \;
R(\vec x, \vec x';h(t),h(t'))
\; .
\label{eq:RpG-Rx}
\end{eqnarray}
All spatial positions transform in the same way
under the simultaneous transformation of the two-times.
The Fourier components transform in an identical way.

The choice of the transformation of the fields is such that the integrated
linear response transforms as the correlation under these reparametrisations
of time:
\begin{eqnarray}
&& 
\chi(\vec x, \vec x'; t,t') \to
\tilde \chi(\vec x, \vec x'; t,t') \; = \;
\int_{t'}^t dt'' \; \tilde R(\vec x, \vec x';t,t')
\nonumber\\
&& 
\;\;\;\;\;\;\;\;
=  \int_{t'}^t dt'' \; \left(\frac{dh(t'')}{dt''}\right)  \;
R(\vec x, \vec x';h(t),h(t''))
= 
\;\;
\int_{h'}^h dh'' \; 
R(\vec x, \vec x';h,h'')
\nonumber\\
&&
\;\;\;\;\;\;\;\;
=
\;\;
\chi(\vec x, \vec x'; h,h')
\; .
\nonumber
\end{eqnarray}

It is interesting to notice that the transformation in (\ref{eq:RpG-phix}) 
and (\ref{eq:RpG-hatphix}) does not leave all terms in the Martin-Siggia-Rose
action invariant. If we write this action in its most general form
\begin{displaymath}
S \equiv \int d^d x \int dt \; 
\left[ T (i\hat \phi(\vec x, t))^2 + i \hat \phi(\vec x, t) 
\frac{\partial \phi(\vec x,t)}{\partial t} 
+ i \hat \phi(\vec x, t) 
\frac{\delta V[\phi(\vec x, t)]}{\delta \phi(\vec x,t)} \right] 
\end{displaymath}
we note that the first and second terms are not invariant while the
last one is. This is not surprising since a particular evolution, {\it
i.e.} a particular $h(t)$, has to be chosen by the dynamic action. It
is only the {\it slow} dynamics, which is generated in some models,
that may acquire full time-reparametrisation (or a reduced)
invariance. We shall come back to this important point below.

\subsection{Symmetries in the dynamic equations}

Following the same route as in the study of the dynamics of 
disordered spin models, let us first examine whether the 
dynamic equations for the {\it global} correlation and 
response of the $O(N)$ model become invariant under 
generic reparametrisation of time in the scaling regime of 
long waiting-time ($t'\gg \tau_t$)
and for very separated times ($t-t'\gg t'$). 

With this aim, we first derive closed-form dynamic equations for the
global correlation and response of the $O(N)$ model. We then show that
these are not invariant under the most generic
time-reparametrisation defined in eqs.~(\ref{eq:RpG-Cx}) and
(\ref{eq:RpG-Rx}) but only under the subgroup of time-re-scalings given
in eq.~(\ref{eq:rescaling}).

\subsubsection{Dynamic equations for the global correlation and response}{$\;$}

In~\ref{app:eq-of-motion} we show that the dynamic 
equation for the global linear response takes the form
\begin{eqnarray}
&& \frac{\partial R(t,t')}{\partial t} 
= - z(t) R(t,t') 
\nonumber
\\
& & 
\;\;\;\;\;
+
\sum_{n=0}^{\infty} A_n \int dt_n\int dt_{n-1} \dots 
\int dt_1 \; R(t,t_1) R(t_1,t_2) \dots R(t_{n},t')
\label{eq:series-with-A}
\end{eqnarray}
for {\it all} spherical models with arbitrary two-body interactions and {\it
  all} $O(N)$ models in the limit $N\to\infty$ with arbitrary two-body
elastic energy. The only requirement for this result to hold is that the
energy band must have a {\it finite edge}. The coefficients $A_n$ are
determined by the density of states 
of the interaction matrix or elastic
`coefficients' and thus depend on dimensionality. In particular, for the
$p=2$ spherical spin-glass with interactions chosen from a Gaussian
distribution with zero mean and variance of order $1/N$, $A_0=0$, $A_1\ne 0$,
$A_{n\ge2}=0$ and the series truncates at $n_{max}=1$. For a general density of
states $n_{max}\to\infty$. Dilute models are excluded from this
family since their densities of states have
long-tails~\cite{Guilhem,Guilhem2}. It is interesting to 
note that the dynamics of the response is decoupled from that of the
correlation for all these models.

Putting this equation in the Schwinger-Dyson form
\begin{eqnarray}
&& \frac{\partial R(t,t')}{\partial t} 
= - z(t) R(t,t') + 
\int dt_n \;  \Sigma(t,t_n) R(t_n,t')
\label{eq:SD-R}
\end{eqnarray}
allows us to identify the self-energy:
\begin{equation}
\Sigma(t,t') = \sum_{n=0}^{\infty} A_n \int dt_{n-1}\int dt_{n-2} \dots 
\int dt_{1} \; R(t,t_1) \dots R(t_{n-1},t')
\; .
\label{eq:Sigma}
\end{equation}
Since we are not considering the possibility of applying non-potential
forces, the Schwinger-Dyson equation for the global correlation should
read
\begin{eqnarray}
&& \frac{\partial C(t,t')}{\partial t} 
= - z(t) C(t,t') + \int dt_n \; 
[\Sigma(t,t_n) C(t_n,t') + D(t,t_n) R(t',t_n)]
\label{eq:SD-C}
\end{eqnarray}
with $D(t,t')$ the vertex kernel. 
If the model {\it has} an equilibrium high
temperature phase the vertex should be related to the self-energy in such a
way that the solution verifies the fluctuation-dissipation theorem. This is
achieved by
\begin{eqnarray*}
\Sigma(t-t') = \frac{1}{T} \, \frac{\partial D(t-t')}{\partial t'}
\;\theta(t-t') 
\end{eqnarray*}
in the high $T$ phase. One can then guess that
\begin{equation}
D(t,t_n) = \sum_{n=0}^{\infty} A_n \int dt_1\dots 
\int dt_{n-1} R(t,t_1) \dots R(t_{n-2},t_{n-1}) C(t_{n-1},t_{n})
\; .
\label{eq:D}
\end{equation}
Note that 
\begin{displaymath}
\Sigma(t,t_n) = \int dt_a dt_b \; 
\frac{\delta D(t,t_n)}{\delta C(t_a,t_b)} \, R(t_a,t_b)
\; .
\end{displaymath}
The equal-time global correlation $C(t,t) \equiv \int d^d x \; [
\langle \, \phi^2(\vec x,t)\, \rangle ]_{ic}$ may not, in general, be
fixed to a constant value. The dynamic equation determining its
time-evolution is obtained by writing $d_t C(t,t)\equiv \lim_{t'\to t}
[\partial_t C(t,t') + \partial_{t'} C(t,t') ] $ using
eq.~(\ref{eq:SD-C}).  The Lagrange multiplier $z(t)$ is in general
determined by eq.~(\ref{eq:Y2}) while one should use the equation for
$C(t,t)$ to compute the average $[\langle \, \phi^2(\vec x,t)\,
\rangle]_{ic}$.

\subsubsection{Solution in the ordered phase}{$\;$}
\label{sec:solution-dyn-eqs}

Let us assume that the global correlation and response on the one 
side, and the self-energy and the vertex on the other, separate in a 
fast and a slow component 
as in eqs.~(\ref{eq:Cglobal-sep})-(\ref{eq:Rag-global}). 
We then introduce this {\it Ansatz} in eqs.~(\ref{eq:SD-R}) 
and (\ref{eq:SD-C}) to derive dynamic equations for the fast and slow 
parts~\cite{Cuku}. The equations of motion for the 
slow parts have the form
\begin{eqnarray*}
&& 
\frac{\partial C_{ag}(t,t')}{\partial t}
=
-z(t) C_{ag}(t,t') + int_C
\nonumber\\
&& 
\frac{\partial R_{ag}(t,t')}{\partial t}
=
-z(t) R_{ag}(t,t') + int_R
\end{eqnarray*}
with $int_C$ and $int_R$ being two series of rather complicated terms
involving $n$-order convolutions of the response and the correlation
over the times. Clearly, the time-derivatives on the left-hand-side
are not invariant under a generic reparametrisation of time. A necessary 
step in trying to prove time-reparametrisation invariance is to
assume that asymptotically they are much smaller than {\it each term}
on the right-hand-side, drop them, and check the invariance of the
remaining terms. This is an assumption that should be checked {\it a
posteriori} once the solution for $C_{ag}$ and $R_{ag}$ is derived
from the remaining equations. In the case of the $O(N)$ model we already
know the exact solution for all times, from which we can derive the
approximate form that holds in the scaling limit of very long times
and separations among them, and check whether this form allows 
for the time-reparametrisation invariance of the equations.

Let us first focus on the equation for the global response which is 
easier to deal with. 
The slow ageing part of the linear response behaves asymptotically as
$t^{-d/2} f_R(\lambda)$, see eq.~(\ref{eq:scaling-resp-global}). 
Its time-derivative is
\begin{displaymath}
\frac{\partial R_{ag}(t,t')}{\partial t} \sim 
- t^{-d/2-1} \, 
\left[ \frac{d}{2} \, f_R(\lambda) + \lambda f^{'}_R(\lambda) \right]
\; .
\end{displaymath}
The `mass' $z(t)$ decays as 
\begin{equation}
z(t) = \frac{d}{dt} \left( \frac12 \ln Y^2(t) \right) \sim
\frac12 \frac{d}{dt} \ln t^{-d/2} = -\frac{d}{4} t^{-1} 
\; ,  
\label{eq:z-decay}
\end{equation}
consequently, the first term in the right-hand-side goes as 
\begin{eqnarray*}
z(t) R_{ag}(t,t') \sim 
-\frac{d}{4} t^{-d/2-1} f_R(\lambda)
\end{eqnarray*}
and it is of the same order as the time-derivative in the left-hand-side.

The terms in the series can also be analysed by separating the stationary and
ageing contributions to the integrals in eq.~(\ref{eq:series-with-A})
(see~\cite{Leto} for a detailed explanation). Such
separation, as carried out in \ref{app:eq-of-motion-aging}, leads to
\begin{eqnarray}
&& \frac{\partial R_{ag}(t,t')}{\partial t} 
= - z(t) R_{ag}(t,t') 
\label{eq:series-with-A-tilde}
\\
& & 
\;\;\;\;\;\;\;\;\;\;
+
\sum_{n=0}^\infty {\tilde A}_n \int dt_n\int dt_{n-1} \dots 
\int dt_1 \; R_{ag}(t,t_1) R_{ag}(t_1,t_2) \dots R_{ag}(t_{n},t')
\;,
\nonumber
\end{eqnarray}
where the coefficients ${\tilde A}_n$ are given by [see
eq.~(\ref{eq:tilde-A-appendix})]
\begin{equation}
\tilde A_n \equiv -\frac{1}{(n+1)!} \left( \frac{d}{d\chi_{st}}\right)^n 
\epsilon(\chi_{st})
\;,
\end{equation}
with $\chi_{st}$ the integrated stationary response evaluated at time scales
of order $\Delta t=t-t'$. The function $\epsilon(h)$ is the inverse of the
function
\begin{displaymath}
h(\epsilon)=
\int_0^\infty \!\!d\epsilon'\;\frac{g(\epsilon')}{\epsilon'-\epsilon}
\end{displaymath}
which is obtained from the density of states $g(\epsilon)$ of the model
(see \ref{app:eq-of-motion} and \ref{app:eq-of-motion-aging}).

In the $O(N)$ model the coefficients $\tilde A_n$ scale as a power of
$\Delta t$, and the precise power is controlled by 
the form of the density of states 
at low energies, $g(\epsilon)\propto \epsilon^\nu$, 
with 
$\nu=d/2-1$ [see eq.~(\ref{eq:density-of-state})].
The scaling of $\tilde A_n$ at long time differences can be
obtained as follows.
First, notice that 
$h(0)-h(\epsilon)\sim\epsilon^\nu$ by taking into account the power law
dependence of the density of states at low energies. 
Then, the inverse function $\epsilon(h)$ is 
$\epsilon=h^{-1}(h(\epsilon)) \sim (h(0)-h(\epsilon))^{1/\nu}$. 
On the other hand, the 
stationary susceptibility $\chi_{st}(\Delta t)$ is intimately related 
to the density of states. Using eq.~(\ref{eq:G-tilde-G}) one finds,
$$
\chi_{st}(\Delta t)=
\int_{-\infty}^{\Delta t}d\Delta t' \; G(\Delta t')
=
\int_{0}^{\infty} d \epsilon \; 
g(\epsilon)\;\frac{1-e^{-\epsilon \Delta t}}{\epsilon}
=
\chi_{st}(\infty)-\int_{0}^{\infty} d\epsilon \; 
\frac{g(\epsilon)}{\epsilon}\; e^{-\epsilon \Delta t}\;\;
\; . 
$$
Using now the power law decay  
of the density of states at low energies one finds
\begin{equation}
  \label{eq:bar_h-h}
\chi_{st}(\infty)-\chi_{st}\sim \Delta t^{-\nu}
\qquad
{\rm or}
\qquad
1/\Delta t \sim [\chi_{st}(\infty)-\chi_{st}]^{1/\nu}
\;,
\label{eq:chist-tau}
\end{equation}
with $\chi_{st} \equiv \chi_{st}(\Delta t)$. 
From  eq.~(\ref{eq:tilde-G}) one notices that
$\chi_{st}(\infty)=h(0)$. Thus, 
\begin{equation}
\epsilon(\chi_{st}) \sim [\chi_{st}(\infty)-\chi_{st}]^{1/\nu}
\; ,
\end{equation}
so that
$d^n\epsilon/d\chi_{st}^n\sim 
[\chi_{st}(\infty)-\chi_{st}]^{1/\nu-n} \sim \Delta t^{-1+n\nu}$ which 
yields 
\begin{equation}
  \label{eq:tilde-A-scaling}
\tilde A_n\sim \Delta t^{-1+n\nu}
\;.
\end{equation}

Notice that the coefficients $\tilde A_n$ keep a power law dependence on time
differences. This anomalous dependence is in {\it remarkable} contrast to
glassy systems, such as the $p$-spin disordered 
models with $p\geq 3$, in which the coefficients 
$\tilde A_n$ reach
finite constants as $\Delta t\to \infty$ and $\chi_{st}\to\chi_{st}(\infty)$.
We discuss the consequences of this difference below, when we consider the
scaling dimensions of the global correlation and response under time
re-scalings.

\subsubsection{Scaling dimensions}{$\;$}

Consider the situation in which under, say, a scale transformation,
the global correlation and response in the ageing regime transform
according to
\begin{eqnarray}
t\to\zeta t
\\
C_{ag}(t,t')\to \tilde C_{ag}(t,t') = C_{ag}(\zeta t,\zeta t')
\\
R_{ag}(t,t')\to \tilde R_{ag}(t,t') = \zeta^{\Delta_R}\;R_{ag}(\zeta t,\zeta
t')
\;,
\end{eqnarray}
where $\Delta_R$ is the retarded dimension~\cite{Kennett-etal1} of the
response (the advanced dimensions for both response and correlation,
as well as the retarded dimension for the correlation, are zero in the
above). In systems in which correlation and response are related by an
off-equilibrium fluctuation-dissipation relation with a finite
effective temperature~\cite{Cukupe}, the retarded dimension takes the
value $\Delta_R=1$. In this case, one can show that if the equations
of motion are invariant under these scale transformations, they are
{\it also} invariant under time-reparametrisation $t\to h(t)$. In
other words, for the special case $\Delta_R=1$, scale invariance {\it
implies} reparametrisation invariance~\cite{Chamon-etal}. The
situation is similar in character to what happens with scale invariant
field theories, where in the special case of two dimensional systems
local scale invariance implies conformal invariance, a much larger
symmetry.

Let us discuss concisely why the symmetry is larger when $\Delta_R=1$.
Consider the ageing contribution to a generic term $I_n$
with $n$ integrals and
$n+1$ responses as an example:
\begin{eqnarray}
  \label{eq:example1}
\tilde I_n(t,t') &\equiv&
\int dt_n\int dt_{n-1} \dots 
\int dt_1 \; \tilde R_{ag}(t,t_1) \tilde R_{ag}(t_1,t_2) 
\dots \tilde R_{ag}(t_{n},t')
\nonumber\\
&=&
\int dt_n \dots 
\int dt_1 \; \zeta^{(n+1)\Delta_R}\;R_{ag}(\zeta t,\zeta t_1) 
\dots R_{ag}(\zeta t_{n},\zeta t')
\nonumber\\
&=&
\zeta^{(n+1)\Delta_R-n}
\int d(\zeta t_n) \dots 
\int d(\zeta t_1) \;R_{ag}(\zeta t,\zeta t_1) 
\dots R_{ag}(\zeta t_{n},\zeta t')
\nonumber\\
&=&
\zeta^{n(\Delta_R-1)}\;\zeta^{\Delta_R} I_n(\zeta t,\zeta t')
\;.
\label{eq:example-conv-int}
\end{eqnarray}
Thus, under the rescaling transformation, $I_n(t,t')$ has dimension
\begin{equation*}
\Delta_{I_n}=n(\Delta_R-1)+\Delta_R.
\end{equation*}
There are two special features that arise when $\Delta_R=1$. The first is
that all the $I_n(t,t')$ have the same dimension, $\Delta_{I_n}=1$ for all
$n$. The second is that the change of variables inside the integrals can be
carried out for a more general change of variables $t\to h(t)$ with an
arbitrary monotonic function $h(t)$, because
\begin{equation*}
\int dt_i \;\left(\frac{dh(t_i)}{dt_i}\right)^{\Delta_R=1} \dots =
\int dt_i \;\frac{dh(t_i)}{dt_i} \dots =
\int dh_i\dots
\end{equation*}
holds for each of the times that are being integrated over. Therefore, for
$\Delta_R=1$ the rescaling of the correlations and responses can be absorbed
in the Jacobian for the changes of integration variables.

Hence {\it for
$\Delta_R=1$ scale invariance implies reparametrisation invariance.}

\subsubsection{Fixing the retarded dimension $\Delta_R$}{$\;$}

Let us discuss now how the asymptotic behaviour of $z(t)$ fixes the retarded
dimension $\Delta_R$.

\label{sec:DeltaR}

\vspace{0.3cm}
\noindent{\it The case of glassy dynamics}{$\;$}
\vspace{0.3cm}

In mean-field spherical models displaying glassy dynamics, such as for
example the spherical $p$-spin model for $p\ge 3$, the right-hand-side
of the dynamic equation for the response is much simpler that
eq.~(\ref{eq:series-with-A}) in that the series actually 
has only one term of the form $I_1$. A supplementary difficulty arises 
from the fact that $C$ enters the integral but this is not 
very difficult to deal with if we assume that the scaling 
dimensions of $C$ vanish. 

The function $z(t)\to z_\infty\ne 0$ as $t\to\infty$. In this case, the term
\begin{equation*}
z(t)\;R_{ag}(t,t')\to z_\infty\;R_{ag}(t,t')
\end{equation*}
has the same scaling dimension as the response itself, {\it i.e.},
$\Delta_R$. In these glassy systems, the coefficients ${\tilde A}_n$ in front
of the integral terms are finite constants in the limit of $\Delta
t=t-t'\to\infty$. (In the specific case of the $p$-spin models, only ${\tilde
  A}_0$ and ${\tilde A}_1$ are non-vanishing.)

The time-derivative term
\begin{equation*}
\frac{\partial}{\partial t} R_{ag}(t,t')
\end{equation*}
has scaling dimension $\Delta_{\rm deriv.}=\Delta_R+1$, because the
$\frac{\partial}{\partial t}$ has dimension 1. So in this case the
time-derivative term is irrelevant, and it can be dropped in the long-time
limit, as long as one finds a non-trivial solution to the remaining
equations. Indeed, there can be a non-trivial solution of the long-time
dynamical equations if at least one of the integral terms, $I_n(t,t')$ for
some $n\ge 0$, can balance the $z_\infty\;R_{ag}(t,t')$ term.  As we
discussed previously, the contribution to a generic integral $I_n$ with
$n\geq 0$ has scaling dimension $\Delta_{I_n}=n(\Delta_R-1)+\Delta_R$. The
cancellation can be achieved if and only if
\begin{equation*}
\Delta_{I_n}=n(\Delta_R-1)+\Delta_R=\Delta_R \quad \mbox{for some}\quad n\ge
0
\;.
\end{equation*}
There are two ways of achieving this scope. The contribution with $n=0$
trivially satisfies this identity for any value of $\Delta_R$.  Besides,
terms of the same order arise from $n\geq 1$ only if $\Delta_R=1$, in which
case the $n$ dependence disappears and the condition is actually satisfied
for all $n\ge 1$.
The second possibility is realized by the $p>2$
spherical Gaussian spin-glass, a model with $$\Delta_R=1$$
and for which
reparametrisation invariance develops. The $p=2$ spherical
spin-glass with Gaussian interactions is discussed in detail in
\ref{app:p=2}.

One can argue that the scaling dimensions zero for the correlation (both
retarded and advanced dimensions) and retarded dimension $\Delta_R=1$ for the
response are consistent with a factor $X(t,t')=T/T_{\sc eff}$ that remains
finite for fixed $C_{ag}$ in the long time limit. Consider the
out-of-equilibrium fluctuation-dissipation relation:
\begin{equation}
R_{ag}(t,t')=\frac{X(t,t')}{T}\;
\frac{\partial}{\partial t'}C_{ag}(t,t')\;\theta(t-t')
  \;.
\label{eq:oefdr-dim}
\end{equation}
If the factor $X(t,t')\to X(C_{ag})$ for fixed $C_{ag}$ in the large $t,t'$
limit, without vanishing with some anomalous extra powers of $t$, then it
follows that the retarded dimension of the response is one more than that of
the correlation, because of the $\partial/\partial t'$. So if the correlation
has retarded dimension zero, the response will have retarded dimension
$\Delta_R=1$ as long as $X$ remains finite and has no anomalous power law
dependence on $t$ for fixed $C_{ag}$. This is the situation in glassy
systems, where finite factors $X=T/T_{\sc eff}$ have been observed in
experiments and simulations (see~\cite{Crisanti} for a review).

\vspace{0.3cm}
\noindent{\it The case of the $O(N)$ model}
\vspace{0.3cm}

Consider a case in which the function $z(t)\sim t^{-\Delta_z} \to 0$ as
$t\to\infty$. Particularly, $\Delta_z=1$ for the $O(N)$ model [see
eq.~(\ref{eq:z-decay})]. Now, the term
\begin{equation*}
  z(t)\;R_{ag}(t,t')
\end{equation*}
has scaling dimension $\Delta_R+\Delta_z$. The time-derivative term,
just as in the case of glassy systems above, has scaling dimension
$\Delta_{\rm deriv}=\Delta_R+1$. Therefore, as opposed to the cases
discussed above, one {\it cannot} naively neglect the time-derivative
term, because it has the {\it same} scaling dimension as the
$z(t)\;R_{ag}(t,t')$ term.

In the $O(N)$ model the series in the r.h.s. of 
eq.~(\ref{eq:series-with-A}) does not truncate.
The prefactor of the integral term $I_n(t,t')$ 
depends on $\Delta t=t-t'$, $\tilde A_n\sim \Delta t^{-1+n\nu}$
with $\nu=1-d/2$ [see eq.~(\ref{eq:tilde-A-scaling})], 
and there is an additional scaling dimension arising 
from the anomalous scaling of the prefactors $\tilde A_n$:
\begin{equation*}
\Delta_{{\tilde A}_n}=1-n\nu
\;.
\end{equation*}

In order to determine the dimension $\Delta_R$, one must balance at
least one of the integral terms, $\tilde A_n(\Delta t)\;I_n(t,t')$ for
some $n\ge 0$, against the $z(t)\;R_{ag}(t,t')$ term and the
time-derivative $\partial R_{ag}(t,t')/\partial t$. This can be
achieved if and only if
\begin{equation}
\Delta_{{\tilde A}_n}+\Delta_{I_n}
=n(\Delta_R-1-\nu)+\Delta_R +1=\Delta_R +1 \quad \mbox{for some}\quad n\ge
0
\;.
\label{eq:conditionDeltaR}
\end{equation}
Notice that this condition is satisfied in particular by $n=0$, but it
can also be satisfied {\it for any} $n$ if $$\Delta_R=\nu+1=d/2\;,$$
which is indeed consistent with the exact result given in
eq.~(\ref{eq:Rag-global}).

Notice that {\it all} terms in the equation of motion of $R_{\rm ag}(t,t')$
have the same scaling dimension $\Delta_{R}+1$ as the time derivative term,
which thus cannot be dropped in any dimension $d$, in contrast to the case in
glassy systems. Notice also that $\Delta_R\ne 1$ for $d>2$, so
reparametrisation invariance {\it does not} develop; only scale invariance is
a symmetry of the long-time dynamical equations of motion. A retarded scaling
dimension $\Delta_R>1$ implies, using eq.~(\ref{eq:oefdr-dim}), that the
factor $X(t,t')\to 0$ for long times and fixed $C_{ag}$, if the correlation
has retarded and advanced dimensions zero. This result is in agreement with
the direct calculation of the factor $X$ in eq.~(\ref{eq:Xdef}).

In $d=2$, one obtains that $\Delta_R=1$, but in contrast to the case of
glassy dynamics where the prefactors $\tilde A_n$ were constant,
$\Delta_{{\tilde A}_n}=1$. For reparametrisation invariance to develop, it is
necessary that $\Delta_R=1$ {\it and} that $\Delta_{{\tilde A}_n}=0$. Hence,
reparametrisation invariance does not develop even in the $d=2$ case. Notice,
however, that $\Delta_R=1$ in $d=2$ implies a non-trivial $X(t,t')$ which is
actually found in the exact solution \cite{Zannetti3}; there is still an
additive separation of correlation and linear response in a stationary and an
ageing part at $T=0$ (as opposed to the multiplicative scaling found in
critical relaxations~\cite{Corberi-review}). Nevertheless, it is important to
remark that this $X(t,t')$ depends continuously on the ratio $t/t'$ [see
eq.~(\ref{eq:Xdef})], which implies a $C_{ag}$ dependent effective temperature
instead of a constant effective temperature; the latter is expected for a
problem with a single correlation scale.

\subsection{Conjecture}

We argued that $\Delta_R=1$ is a necessary condition for having an asymptotic
time-reparametrization invariance - though this condition is not sufficient,
as shown by the $d=2$ $O(N)$ case. In addition, $\Delta_R=1$ implies a finite
integrated linear response and a finite effective temperature, as can be
derived from eq.~(\ref{eq:oefdr-dim}).

The $O(N)$ model has a weaker response than that of glassy models, as for
example the $p$-spin spherical disordered system. Indeed, in the $O(N)$ model
the ageing contribution to the integrated response vanishes asymptotically in
all $d>d_L=2$ -- and this can be related to the development of an infinite
effective temperature~\cite{Cukupe} at long times; while it approaches a
finite $C$-dependent value in $d=d_L=2$ -- and this cannot be interpreted in
terms of an effective temperature since one would have a $C$ dependent value
within a single correlation scale~\cite{Cuku2}.  Other solvable coarsening
problems have a similar integrated response (see {\it
  e.g.}~\cite{Corberi-review}).

On the basis of the discussion above, we {\it conjecture} that models with a
finite and well-defined effective temperature, such as the $p$ spin spherical
disordered system with $p\geq 3$ or the more complex Sherrington-Kirkpatrick
spin-glass, develop time-reparametrisation invariance asymptotically, while
this does not occur in systems with a diverging or ill-defined effective
temperature, such as the $O(N)$ model.

\subsubsection{Space-time rescaling}{$\;$}
\label{sec:subsubsec-fourier}

For the sake of
comparison, in the following we consider the standard dynamical
scaling~\cite{Bray} which consists of {\it simultaneous} rescaling of time
and space.

So far we discussed time rescaling and time-reparametrisation
invariance properties in the real space representation.  This is
because we have been interested in making contact with glassy systems
for which composite fields, that are related to the two-time
correlation and response functions at equal space points, might be the
natural order parameters~\cite{Chamon-etal}.  In the case of the
$O(N)$ model, however, one knows that the original field
$\phi_{\alpha}(\vec x,t)$ is already the natural order
parameter. Especially, one expects its Fourier space representation
$\phi_{\alpha}(\vec{k},t)$ to be easier to handle.  

Let us take the response function $\delta
\phi_{\alpha}(\vec{k},t)/\delta h_{\beta}(-\vec{k}',t')|_{h=0}=
r(k;t,t')\delta_{\alpha\beta}\delta(\vec{k}+\vec{k}')$ and the
two-time composite field
$\phi_{\alpha}(\vec{k},t)\phi_{\beta}(-\vec{k}',t')=
c(k;t,t')\delta_{\alpha\beta}\delta(\vec{k}+\vec{k}')$ where
$k=|\vec{k}|$. 
In the $T\to 0$ limit \footnote{This is general since the domain growth 
scaling is controlled by a ``zero temperature fixed point''~\cite{Bray}.} 
these quantities satisfy the same 
evolution equation,
\begin{eqnarray}
&& \frac{\partial}{\partial t} r(k;t,t') = -[k^{2}+z(t)] \; r(k;t,t')\;,
\label{eq:rkttp}
\\
&& \frac{\partial}{\partial t} c(k;t,t') = -[k^{2}+z(t)] \; c(k;t,t')
\;.
\label{eq:ckttp}
\end{eqnarray}
Now let us suppose that these equations admit a set of asymptotic
solutions with the following scaling forms\footnote{The simple analysis that 
we present here yields the correct behaviour for non-conserved
but it is does not in the case of dynamics with conserved order-parameter.} 
\begin{eqnarray}
r(k;t,t')=\zeta^{\Delta^R_{r}+\Delta^A_{r}-d \Delta_s}
\; r(k\zeta^{-\Delta_{s}},\zeta t,\zeta t')
\; ,
\label{eq:r-scaling-ansatz}\\
c(k;t,t')=\zeta^{\Delta^R_{c}+\Delta^A_{c}-d \Delta_s}
\; c(k\zeta^{-\Delta_{s}},\zeta t,\zeta t')
\; ,
\label{eq:c-scaling-ansatz}
\end{eqnarray}
and
\begin{displaymath}
z(t)=\zeta^{\Delta_{z}} z(\zeta t)
\; .
\label{eq:z-scaling-ansatz}
\end{displaymath}
$\Delta_r^R$ and $\Delta_r^A$ are the retarded and advanced dimensions
of the response $r$; similarly, $\Delta_c^R$ and $\Delta_c^A$ are the
retarded and advanced dimensions of the correlation $c$.  The scaling
dimensions are then fixed by inserting this {\it Ansatz} in
eqs.~(\ref{eq:rkttp}) and (\ref{eq:ckttp}).  First, focusing on the
$k=0$ component, one finds $\Delta_{z}=1$.  Then considering
the $k>0$ components, one finds two possibilities.
The first one is that the $k^{2}$ term has the same scaling dimension
as the other two terms. In this case 
one finds the exponent for spatial scaling
$\Delta_{s}=1/2$~\footnote{Its inverse corresponds to the dynamical
exponent $z$ in critical dynamics.}.  
The other possibility is $\Delta_{s} > 1/2$
which means that the $k^{2}$ term becomes irrelevant. Which of the two
cases appear in the large times regime depends on the initial
conditions. For usual random initial condition of the form given in
eq.~(\ref{eq:Gaussi-initial-pdf-k}) with the same statistical weight
on all $k$ components, the exact solution summarised 
in Sect.~\ref{subsec:sigma} tells us that $\Delta_{s}=1/2$ is actually 
selected. In the following we only consider this case.

In the asymptotic
regime all the terms in eqs.~(\ref{eq:rkttp}) and (\ref{eq:ckttp})
have the {\it same} scaling dimensions.  Thus none of them can be
dropped irrespective of the scaling dimensions of the response and
correlation functions which will be determined below. 

We still need to determine the retarded and advanced
scaling dimensions of the response and
correlation functions. Since the solution of the Langevin equation at
$T=0$ can be written as [see eq.~(\ref{eq:sol-k-resp})]
\begin{displaymath}
\phi({\vec k},t) = r(k;  t,t')\; \phi({\vec k},t'),
\end{displaymath}
the overall scaling factor in $r$ must be identical to one, and 
one has that retarded and advanced scaling 
dimensions of the response functions must satisfy
$\Delta_{r}^R+ \Delta_{r}^A=d \Delta_{s}=d/2$.
This is achieved by 
$\Delta_{r}^R=d \Delta_{s}=d/2$
and  $\Delta_{r}^A=0$.
The analysis of the self-consistent equation for $z(t)$ 
fixes the scaling dimensions $\Delta_c^R$ and $\Delta_c^A$. Indeed, 
eq.~(\ref{eq:z-def}) reads
\begin{displaymath}
z(t) = g \int \frac{d^{d}k}{(2\pi)^{d}} \; c(k,t,t) +r
\; .
\end{displaymath}
In the large time limit $z(t) \to 0$ and the integral
converges to $-r/g$. Therefore
$\Delta_{c}^R+\Delta_{c}^A =0$ that implies the natural choice
$\Delta_{c}^R=\Delta_{c}^A =0$.

It is instructive to consider the inverse Fourier transform of
the scaling {\it Ansatz} in 
eqs.~(\ref{eq:r-scaling-ansatz}) and (\ref{eq:c-scaling-ansatz}) which reads
\begin{eqnarray*}
R(|\vec{x}-\vec{y}|;t,t')
&=&
\zeta^{\Delta^A_{r}+\Delta^R_{r}} \; 
R(|\vec{x}-\vec{y}| \zeta^{\Delta s}, \zeta t, \zeta t')\; , 
\\
C(|\vec{x}-\vec{y}|;t,t')&=&
\zeta^{\Delta^A_{c}+\Delta^R_{c}} \; 
C(|\vec{x}-\vec{y}| \zeta^{\Delta s}, \zeta t, \zeta t')
\; .
\label{eq:r-c-scaling-ansatz-real-space}
\end{eqnarray*}
Thus the solution can be written in the scaling form [see
(\ref{eq:scaling-corr})],
\begin{eqnarray*}
R(|\vec{x}-\vec{y}|;t,t')
&\approx&
\frac{1}{L^{d}(t)} f_{R_r}\left(\frac{|\vec{x}-\vec{y}|}{L(t)}
,\frac{L(t)}{L(t')}\right) 
\; ,
\\
C(|\vec{x}-\vec{y}|;t,t')&\approx&
f_{C_r}\left(\frac{|\vec{x}-\vec{y}|}{L(t)} ,\frac{L(t)}{L(t')}\right)
\; ,
\label{eq:r-c-scaling-ansatz-real-space2}
\end{eqnarray*}
with the domain growth law
\begin{displaymath}
L(t) \propto t^{\Delta_{s}}=\sqrt{t}
\; .
\end{displaymath}

Thus, the analysis of the invariance of the dynamic
equations for the {\it global} $C$ and $R$ under rescaling of time 
presented in the previous sections, that serves to fix the scaling
dimension of the global response $\Delta_R$, can be extended 
to study the scaling dimensions of the 
space-dependent correlation and response $C(r;t,t')$ and
$R(r;t,t')$ under simultaneously rescaling of space and time. 
We find $\Delta^R_{R_r}=d/2$ and $\Delta^A_{R_r}=0$, 
and $\Delta^A_{C_r}=0$ and $\Delta^R_{C_r}=0$, for any $d$.

\subsection{Symmetries of the (long-times) action}
\label{sec:sym-action}

We have already mentioned that a generic Martin-Siggia-Rose action is
not invariant under reparametrisations of the times and fields defined
in eqs.~(\ref{eq:RpG-phix})-(\ref{eq:RpG-hatphix}).  Let us now
analyse the symmetries of the action in the long times limit in which
there is a separation of time-scales in the global correlation and
response.

Let us first focus on the case $d>2$.  Using the simple transformations
described in Sect.~\ref{sec:action}~\cite{Corberi-etal} 
the dynamic generating function
of the $O(N)$ model can be expressed as a path integral over the fields
$\vec\sigma(\vec x,t)$ and $\vec \psi(\vec x,t)$ 
only. On the one hand one
can argue that the `fast' $\vec \psi$-part `renormalises' to zero
under generic time-reparametrisation and becomes asymptotically
irrelevant. On the other hand, while the $\vec \sigma$ field transforms in
such a way that it ensures the correct transformation of the
integration measure, the kernel in its action is just the 
inverse of the global correlation itself that is time dependent and
transforms in a non-trivial manner under generic reparametrisations of
time. The quadratic action for $\vec \sigma(\vec x,t)$ at equal space points
is not invariant under generic reparametrisations of time.

The local action of the $\vec \sigma$ field is, however, 
invariant under a reduced subset of transformations, namely 
time re-scalings. 
Since the kernel is a function of $t'/t$ one finds that the 
slow action written in terms of the fields only, 
see eq.~(\ref{eq:S5}),  
is invariant under
\begin{eqnarray*}
&& t \to \zeta \, t 
\; ,
\\
&&
\vec \sigma(\vec x,t) \to \vec \sigma(\vec x, \zeta \, t)
\; .
\end{eqnarray*}

But one can also go one step back and check whether the 
action for the field, $\vec \sigma(\vec x,t)$, and response
field, $i \vec {\hat \sigma}(\vec x,t)$, is invariant under 
time re-scalings that change the response field as
\begin{equation}
i \vec {\hat \sigma}(\vec x,t) 
\to 
\zeta  i \vec {\hat \sigma}(\vec x, \zeta t)
\; ,  
\end{equation}
{\it i.e.} the reduction of (\ref{eq:RpG-hatphix})
to time re-scalings. Transforming the `slow' part of 
$S_{ktt'}^{(4)}$ into spatial coordinates and 
writing explicitly all integrals one has
\begin{eqnarray}
S^{(4)} &=& \int d^d x \int d^d y \int dt \int dt'
\left[ 
i \vec {\hat \sigma}(\vec x,t) K_\sigma(\vec x-\vec y; t,t') 
 i \vec {\hat \sigma}(\vec y,t')
\right.
\nonumber\\
&& 
\left.
\;\;\;\;\;\;\;\;\;\;\;\;\;\;
+ 
i \vec {\hat \sigma}(\vec x,t) \delta(t-t')\delta^d(\vec x-\vec y)
 \vec \sigma(\vec y,t') 
\right]
\end{eqnarray}
with 
\begin{equation}
K_\sigma(\vec x-\vec y; t,t')
\equiv T
\int \frac{d^dk}{(2\pi)^d} 
\; e^{i\vec k  (\vec x-\vec y)} \, e^{-k^2/\Lambda^2}
\int_0^{t_1} dt'' \, r(k,t,t'') r(k,t',t'') 
\; .
\label{eq:Ksigma}
\end{equation}
One can then easily check that the action
at equal space points, $\vec x=\vec y$, 
remains invariant under the time re-scalings proposed above.
Indeed, in terms of the transformed fields the local 
action reads
\begin{eqnarray}
\tilde S_x^{(4)} &=& 
\int dt \int dt'
\left[ 
i \tilde{\vec {\hat \sigma}}(\vec x,t) K_\sigma(\vec 0; t,t') 
 i \tilde{\vec {\hat \sigma}}(\vec x,t')
+ 
i \tilde{\vec {\hat \sigma}}(\vec x,t) \delta(t-t')
 \tilde{\vec \sigma}(\vec x,t') 
\right]
\nonumber\\
&=&
\int dt \int dt'
\left[ 
\zeta \, i  \vec {\hat \sigma}(\vec x, \zeta \, t) 
K_\sigma(\vec 0;\zeta \, t,\zeta \, t') 
 \zeta \, i  \vec {\hat \sigma}(\vec x, \zeta \, t')
\right.
\nonumber\\
&&
\;\;\;\;\;\;\;\;\;\;\;\;\;\;\;\;\;\;
\left.
+ 
\zeta \, i \vec {\hat \sigma}(\vec x, \zeta \, t) 
\zeta \, \delta(\zeta \, t-\zeta \, t')
 \vec \sigma(\vec x,\zeta \,  t') 
\right]
\nonumber\\
&=& S_x^{(4)}
\end{eqnarray}
where we used the fact that $ K_\sigma(\vec 0; t,t') $ is just 
identical to the global correlation function, $C(t,t')$. In the 
limit $t,t'\gg t_0$ this is a function 
of $t'/t$ and thus invariant under time-rescaling.
The last identity follows simply from changing the integration 
variables form $t$ to $\zeta \, t$. 

A similar treatment in $d=2$ is much more delicate. The explicit
calculations in \cite{Corberi-etal} show that the separation of the 
field is achieved by taking advantage of the fact that a factor 
proportional to $(\Lambda^2 t_1)^{1-d/2}$ vanishes 
[see eqs.~(\ref{eq:Cpsi}) and (\ref{eq:Csigma})]. 
This, however, is no longer true in 
$d=2$. Besides, the interesting dynamics in this case arises only at
$T=0$, another non-trivial limit to be taken in the asymptotic expressions.
For these reasons, we cannot simply carry through the arguments above
to $d=2$. Another way to attack the same problem would be to write 
an action in terms of $R(\vec{x},\vec{x};t,t')$ and $C(\vec{x},\vec{x};t,t')$
and use a similar reasoning to the one we used for the analysis of the 
equations of motion for $R(t,t')$ and $C(t,t')$. We shall not 
pursue this study here.

Let us note that the invariance under time rescaling discussed above can also
be understood as a part of the usual space-time scaling invariance discussed
in Sect.~\ref{sec:subsubsec-fourier}.  To this end, we perform a 
renormalisation group (RG) analysis on the Fourier space
representation of the slow part of the action,
\begin{eqnarray}
S^{(4)}_{ k < \Lambda} [i\hat{\sigma},\sigma] &=&
\int_{0}^{\Lambda} d^{d}k \int^{\infty}_{0} dt \int^{\infty}_{0} dt'
\left[ 
i \hat{\sigma}(k,t) K_\sigma(k,t,t') i \hat{\sigma}(k,t')
\right.
\nonumber\\
&& 
\qquad
\left.
+ \delta(t-t') i \hat{\sigma}(k,t){\sigma}(-k,t') 
\right]
\end{eqnarray}
$K_\sigma$ the Fourier transform of (\ref{eq:Ksigma}).
First, by integrating out the ``fast modes'' in $\Lambda/b < k < \Lambda$ 
we obtain $S^{(4)}_{ k < \Lambda/b}$. Next, we choose a set of rescaled
variables
\begin{eqnarray}
\begin{array}{ll}
\tilde{k}=kb
\; ,  
\;\;\; & \qquad 
\tilde{t}=t/b^{2}
\; , 
\nonumber \\
i \tilde{\hat{\sigma}}(\tilde{k},\tilde{t})=b^{-d/2+2} i\hat{\sigma}(k,t)
\; , 
\;\;\; & \qquad
\tilde{\sigma}(\tilde{k},\tilde{t})=b^{-d/2} \sigma(k,t)
\; , 
\nonumber \\ 
\tilde{z}(\tilde{t})=b^{2} z(t) \; . 
\;\;\;
& 
\end{array}
\label{eq:slow-rg-rescaling}
\end{eqnarray}
In terms of the new variables the cut-off is put back to $\Lambda$ 
and the action of the original form is recovered. Converting the above results
to the real space representation and equating the scaling parameter
of space $b$ and time $\zeta$ as $b^{2}=\zeta$.
The space dependent slow action is invariant under simultaneous rescaling
of space and time. 

\subsection{How the $O(N)$ escapes reparametrisation invariance 
but displays scale invariance}

It was shown in~\cite{Chamon-etal} that under rather mild assumptions
(namely, causality, a separation of time-scales as the one discussed
in Sect.~\ref{sec:model}, the fact that the remaining free field
action does not lead itself to slow dynamics and the use of the naive
scaling dimensions of the fields) the slow part of the action of the
$3d$ Edwards-Anderson spin-glass, when written in terms of the
two-time dependent dynamic order parameters, remains invariant under
global time-reparametrisation. We have shown above that the action
for the slow evolution of the $O(N)$ model {\it is not}
invariant under these transformations. One would like to identify
which of the assumptions used in~\cite{Chamon-etal} is (are) violated in the
$O(N)$ case. Indeed, 
in~\cite{Chamon-etal} we used the naive dimensions for $Q_R$, that is
to say $\Delta_{Q_R}^A=0$ and $\Delta_{Q_R}^R=1$. This 
assumption should be correct for systems
that develop a finite and well-defined effective temperature
in the ageing regime. The O(N) falls out of this class and this 
assumption does not apply to it.

\section{The distribution of local two-time observables}
\label{sec:local-corr}

During the ageing relaxation of glassy systems one expects important
temporal and spatial fluctuations.  The distribution of
local coarse-grained correlations and linear responses in
spin-glasses~\cite{Castillo-etal1,Castillo-etal2} and kinetically
facilitated models~\cite{Chamon-etal2} were computed numerically. The
comparison of these probability distribution functions ({\sc pdf}s) 
with the theoretical framework developed
in~\cite{Chamon-etal}-\cite{Chamon-etal2} was
also discussed. In short, the main features of these distributions
are:

{\it i.} The {\sc pdf}
of  coarse-grained local two-time 
correlations is a function that depends on the two times
and, when these are chosen to lie in the ageing regime, 
the {\sc pdf} scales in time just as the global correlation itself.

{\it ii.} The functional form of the {\sc pdf} of coarse-grained local
two-time correlations changes with the two times.  It can be
approximately described with a Gumbel-like function with a two-time
dependent parameter, which in the ageing regime is simply a function of
the global correlation~\cite{Chamon-etal2}. The parameter $a$
characterising the Gumbel-like form is positive for values of $C$ that
are relatively large and close to the maximum given by $q_{ea}$: the
distribution is negatively skewed. The parameter $a$ increases when
decreasing $C$ and diverges at some value of $C$ signalling and
approximately symmetric and Gaussian-like distribution. For still
lower values of $C$ the {\sc pdf} becomes positively skewed and this can be
described with a negative value of the parameter $a$.

{\it iii.} The joint {\sc pdf} of local two-time correlations and linear 
responses follows the global $\chi(C)$ curve in the ageing regime.
This means that the longitudinal fluctuations that take the 
points out of this ``master'' curve become rare when the coarse-graining size
increases while the transverse fluctuations along the 
master curve become more and more important when the waiting-time
increases.

In this Section we compute these and similar distributions and we check
whether the same features are observed in the $O(N)$ model.  For the sake of
simplicity we work at $T=0$ and we analyse the fluctuations induced by a
Gaussian distribution of initial conditions keeping in mind that all
calculations can be generalised to the finite temperature case. We take the
$N\to\infty$ limit strictly and we do not let the ``constraint'' $N^{-1}
\sum_{\alpha}\phi_\alpha^2(\vec x, t)$ fluctuate
(see~\cite{Yoshino-etal,Sollich-etal} for more details).

\subsection{Coarse-graining the field}

The $O(N)$ model yields  a mean-field description of ferromagnetic 
ordering. It is then worth starting by studying the distribution 
of the local magnetisations coarse-grained 
within a region of volume
$V_{x_0}\equiv \ell^{d}$ around $\vec{x}_{0}$. This quantity is 
defined as,
\begin{displaymath}
\vec{m}_{\ell}(\vec{x}_{0},t)
\equiv \int \frac{d^{d}x}{(2\pi\ell^{2})^{d/2}} 
\; e^{-\frac{|\vec{x}-\vec{x}_{0}|^2}{2\ell^2}} \; \vec{\phi}(\vec{x},t)
\; .
\end{displaymath}
In terms of the Fourier transform $\vec{\phi}(\vec{k},t)$
of the original field
we find
\begin{displaymath}
\vec{m}_{\ell}(\vec{x}_{0},t)=
\int \frac{d^{d}k}{{2\pi}^{d/2}}
\; 
e^{i \vec{k} \vec{x_{0}}} 
\; 
\vec{\phi}(\vec{k},t) \; e^{-k^{2}\ell^{2}/2}
\; .
\end{displaymath}
Using the solution of the equation
of motion at $T=0$
we find
\begin{eqnarray*}
&& \vec{m}_{\ell}(\vec{x}_{0},t) = m_{\ell}(t)
\vec{\phi}(\vec{x}_{0},t+\ell^{2}/2)
\;\;\;
\mbox{with} 
\;\;\;
m_{\ell}(t) = \frac{Y(t+\ell^{2}/2)}{Y(t)}
\; .
\end{eqnarray*}
The coarse-grained local magnetisation has the same statistical
properties as the original field $\vec{\phi}(\vec{x},t)$ but with time
increased from $t$ to $t+\ell^2/2$ and amplitude reduced from $1$ to
$m_{\ell}(t)$.  Namely, the fluctuations of each of its components
obey a Gaussian distribution but the amplitude of the vector in $N$
space does not fluctuate at all due to the limit $N \to \infty$.
\footnote{We do not find a 
non-trivial Gumbel-like
distribution $P(m)$ of the amplitude of the magnetisation as found  
for the finite-volume  two dimensional XY model 
in equilibrium in the KT phase~\cite{Peter}. This is again due to 
the $N\to\infty$ limit.}

The dependence of the amplitude $m_{\ell}(t)$ on time $t$ and
coarse-graining size $\ell$ is consistent with what one expects for
a domain growth system. Firstly, if one fixes the coarse-graining size
$\ell$, the amplitude approaches $1$ as time $t$ is increased. So
the system looks ``more ordered'' at longer time scales.  On the other
hand, if the time $t$ is held fixed while the coarse-graining size
$\ell$ is increased, the amplitude of the magnetisation decreases
meaning that the system looks ``more disordered'' at larger length
scales.

\subsection{The distribution of coarse-grained two-time
correlations}

In the $O(N)$ model we can define local coarse-grained 
``correlations'' in the following way:
\begin{displaymath}
q_{V{\cal N}}(t,t')  \equiv
\frac{1}{{\cal N}} \sum_{\alpha=1}^{\cal N} 
\frac{1}{V_x} \sum_{\vec y\in V_x}
\phi_\alpha(\vec y,t) \phi_\alpha(\vec y,t')
\; .
\end{displaymath}
Strictly speaking this is not a correlation function but rather a
composite field. We shall use, however, both names in the following.
The first sum is an average over components of the vector $\vec \phi$
in its internal space. Clearly, when ${\cal N}=1$ we test the single
component correlation while when ${\cal N}=N$ we sum over all the
components of the $\vec \phi$ vector. In the following we shall
discuss these two limiting cases and the intermediate cases of finite
${\cal N}$.  The second sum is a coarse-graining in real space and it
runs over a neighbouring region of the point $\vec x$.  If $V_x=1$ we
have a strictly local quantity while for $V_x=V$ we recover the global
correlation.

\subsection{Local composite field}
\label{subsec:local}

Let us start by studying the strictly local composite field  
\begin{displaymath}
q_{\cal N} \equiv q_{V_x=1,{\cal N}}(t,t') =
\frac{1}{{\cal N}} \sum_{\alpha=1}^{\cal N}
\phi_\alpha(\vec x,t) \phi_\alpha(\vec x,t')
\;. 
\end{displaymath}
The {\sc pdf} of $q_{\cal N}$ is given by
\begin{eqnarray*}
&& 
p(q_{\cal N}) = \frac{1}{Z_0}\int {\cal D}\phi \;
\delta\left(
q_{\cal N}-\frac{1}{{\cal N}} \sum_{\alpha=1}^{\cal N}
\phi_\alpha(\vec x,t)\phi_\alpha(\vec x,t')
\right)
\nonumber\\
&& 
\;\;\;\;\;\;\;\;\;\;\;\;\;\;\;\;\;\;\;\;\;\;\;\;\;\;\;\;\;\;\;\;\;\;
\times \, 
e^{-\frac{1}{2\Delta^2}
\sum_\alpha\sum_k\;\phi^*_\alpha(\vec k,0)\;\phi_\alpha(-\vec k,0)}
\nonumber\\
&& 
= \int \frac{d\eta}{2\pi}\; 
e^{i\eta q_{\cal N}}\;
\frac{1}{Z_0}\int {\cal D}\phi \;
e^{-\frac{i\eta}{{\cal N}} \sum_\alpha\phi_\alpha(x,t)\; \phi_\alpha(x,t')}
\;
e^{-\frac{1}{2\Delta^2}
\sum_\alpha\sum_k\;\phi^*_\alpha(-\vec k,0)\;\phi_\alpha(-\vec k,0)}
\nonumber\\
&&
= \int \frac{d\eta}{2\pi}\; 
e^{i\eta q_{\cal N}}\;
\frac{Z_\eta}{Z_0}
\;,
\end{eqnarray*}
where
\begin{displaymath}
Z_\eta\equiv\int D\phi \;
e^{-\frac{1}{2\Delta^2}
\sum_\alpha\sum_{\vec k_1 \vec k_2}\;
\phi^*_\alpha(k_1,0)\;e^{-i\vec k_1  \vec x}
\;{\cal M}_{\eta}(\vec k_1, \vec k_2)
\;e^{i\vec k_2  \vec x}\;\phi_\alpha(\vec k_2,0)}
\;,
\end{displaymath}
with the symmetric matrix
\begin{equation}
{\cal M}_{\eta}(\vec k_1, \vec k_2)=
\delta_{\vec k_1 \vec k_2}+\frac{i\eta\Delta^2}{{\cal N}}
\left[r(k_1,t,0)\:r(k_2,t',0)+r(k_1,t',0)\:r(k_2,t,0)
\right]
\;.
\label{eq:M1}
\end{equation}
Notice that the second term in ${\cal M}_{\eta}$ contains the $r(k,t_1,t_2)$
terms for the time evolution of the $k$-component. Also notice that
$Z_0=Z_{\eta=0}$.

The calculation of the function $Z$ can be done
as follows. It is convenient to rescale the field $\vec\phi$,
\begin{displaymath}
\tilde{\vec\phi}(k)=\Delta^{-1}\;e^{i\vec k  \vec x}\;\vec \phi(\vec k,0)
\; , 
\end{displaymath}
in such a way that 
\begin{displaymath}
Z_\eta=\int {\cal D}\tilde{\vec\phi} \;
e^{-\frac{1}{2}
\sum_\alpha\sum_{k_1k_2}\;
{\tilde{\vec\phi}}^*_\alpha(\vec k_1,0)
\;{\cal M}_{\eta}(\vec k_1,\vec k_2)
\;\tilde{\vec \phi}_\alpha(\vec k_2,0)}
\;,
\end{displaymath}
up to a trivial (independent of $\eta$) multiplicative constant coming 
from the change of measure. It follows that
\begin{displaymath}
\frac{Z_\eta}{Z_0}=
\left(\frac{\det{\cal M}_\eta}{\det{{\cal M}_0}}\right)^{-{\cal N}/2}
\;,
\end{displaymath}
where we used that the $\alpha=1,\dots,{\cal N}$ components are independent.

In appendix \ref{sec:calc-M-1} the eigenmodes of the
matrix ${\cal M}_{\eta}(\vec k_1, \vec k_2)$ defined in eq.~(\ref{eq:M1})
are obtained; one finds two non-trivial eigenvalues
$\lambda_\pm=1+i\eta\;[C(t,t')\pm 1]$ and  $2L^{d}-2$ trivial
eigenvalues $\lambda=1$. Using these results we have
\begin{equation}
\frac{Z_\eta}{Z_0}=
\left\{
\left[1+\frac{i\eta}{{\cal N}}\;(C(t,t')+1)\right]
\left[1+\frac{i\eta}{{\cal N}}\;(C(t,t')-1)\right]
\right\}^{-\frac{\cal N}{2}}
\label{eq:last}
\end{equation}
where $C(t,t')$ is the global correlation function.
Thus, the {\sc pdf} $p(q_{\cal N})$ is solely parametrised by the
value of the  the global correlation function.

Lastly, let us note that it is straightforward to generalise the above
result to the case of a composite field associated with two different
points in space, $q_{\cal N}(\vec{x},\vec{y})={\cal N}^{-1}
\sum_{\alpha=1}^{\cal N} \phi_\alpha(\vec x,t) \phi_\alpha(\vec
y,t')$. One finds the same result as the one in 
eq.~(\ref{eq:last}) but with the
global correlation function being replaced by the global two-point
function $C(\vec{x},\vec{y};t,t')$.

\begin{figure}
\center{\includegraphics[scale=.7,]{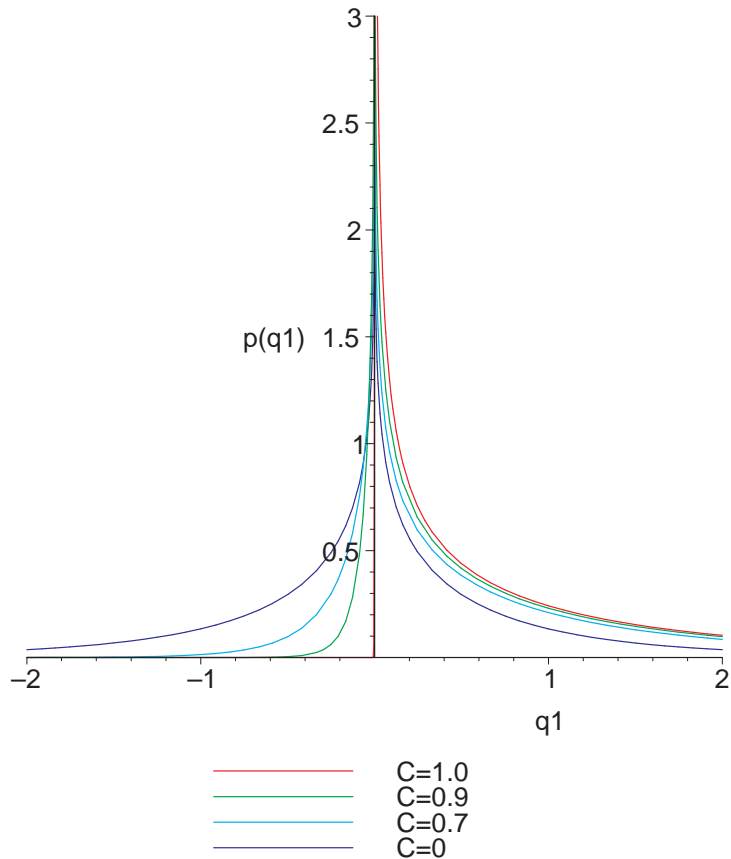}}
\caption{Probability distribution function of one-component 
local two-field composite operator, $\phi_1(\vec x,t)\phi_1(\vec x,t')$, 
 for several values of the pair of times 
$(t,t')$ such that the global correlation $C(t,t')$ takes the values 
given in the key.}
\label{fig1}
\end{figure}

\subsubsection{$N$-component averaged local composite field.}{$\;$}
\label{subsubsec-N-component}

When the average over all components of the $\vec \phi$ is 
considered, {\it i.e.} when ${\cal N}=N$, and the $N\to\infty$ limit 
is also taken, eq.~(\ref{eq:last}) implies
\begin{displaymath}
\frac{Z_\eta}{Z_0}
\to
e^{-i\eta\;C(t,t')}
\end{displaymath}
and 
\begin{displaymath}
p(q_N) = \int \frac{d\eta}{2\pi}\; e^{i\eta q_N}\;\frac{Z_\eta}{Z_0}
=
\int \frac{d\eta}{2\pi}\; e^{i\eta q_N}\;e^{-i\eta\;C(t,t')}
=
\delta[q_N-C(t,t')]
\;.
\end{displaymath}
As expected, the average over all internal components of the field 
erases all fluctuations and the local composite field 
is forced to take the value of the global correlation function
on each site. Note that this is a special feature
of the $N \to \infty$ limit.
It is clear that a further coarse-graining on real
space will have no effect on the form of the distribution. Thus, the 
scaling in time is trivially dictated by the global correlation
function in this case. 

We emphasise that $p(q_N)$ computed above is valid in $N \to \infty$ limit. For
some purposes $O(1/N)$ corrections \cite{Newman-Bray} of $p(q_N)$ can be
important. For example one may consider a spin-glass susceptibility-{\it
  like} quantity $\chi_{SG}=N (\langle q_{N} ^{2}\rangle - \langle q_{N}
\rangle ^{2})$ in an analogous way to the equilibrium one. In equilibrium, the
spin-glass susceptibility may be defined as follows. Consider two
replicas, say A and B, coupled by an interaction term $N\epsilon q$ in the
Hamiltonian where $q=(1/N)\sum_{\alpha} \phi^{A}_{\alpha}\phi^{B}_{\alpha}$
is the overlap between the two replicas.  The equilibrium spin-glass
susceptibility is then defined as $\chi^{eq}_{SG}=\partial \langle q
\rangle_{eq}/ \partial\epsilon=N (\langle q^{2} \rangle_{eq} -
\langle q \rangle_{eq}^{2})$.  Note that if $\langle q^{2}\rangle_{
  eq}-\langle q\rangle_{eq}^{2}=O(1/N)$, $\chi^{eq}_{SG}$ 
does not vanish
in the $N \to \infty$ limit. We may expect a similar non-trivial result
out-of-equilibrium.

\subsubsection{One component local composite field}{$\;$}

Let us now consider the opposite limit in which we 
take ${\cal N}=1$ and look at 
the distribution, $p(q_\alpha)$, of the $x$-dependent
composite field assembled from a {\it single} component of $\vec\phi$:
\begin{displaymath}
q_\alpha
\equiv
\phi_{\alpha}(\vec x,t)\; \phi_{\alpha}(\vec x,t')
\; .
\end{displaymath}
By setting ${\cal N}=1$ in eq.~(\ref{eq:last})
\begin{eqnarray*}
p(q_\alpha) 
&&
= \int \frac{d\eta}{2\pi}\; e^{i\eta q_\alpha}\;
\left[1+i\eta\;\;(C+1)\right]^{-1/2}
\left[1+i\eta\;\;(C-1)\right]^{-1/2}
\;.
\end{eqnarray*}
The distribution is non-Gaussian and it is a 
function of times only through the value of the 
global correlation function
$C=C(t,t')$.
In the above integral, the integrand has 
two branch points: one at $\eta=i/(1+C)$ and the other at $\eta=-i/(1-C)$.
Performing the integral we obtain
\begin{eqnarray*}
p(q_\alpha) &=&
e^{-\frac{|q_{\alpha}|}{1+{\rm sgn}(q_\alpha) C}}
\int_{0}^{\infty}\frac{dr}{\pi}
\frac{e^{-|q_{\alpha}|r}}{\sqrt{2r+(1-C^{2})r^{2}}}
=
\frac{e^{\frac{C q_{\alpha}}{1-C^{2}}}}{\pi \sqrt{1-C^{2}}}
K_{0}\left(\frac{|q_{\alpha}|}{1-C^{2}}\right)
\; ,
\end{eqnarray*}
with $K_0(x)$ the modified Bessel function which can be expressed as
$K_{0}(z)=\int_{1}^{\infty} dx \; e^{-zx}(x^{2}-1)^{-1/2}$. 
This function does not depend on the dimension of space explicitly, 
it only does through the form of $C$. It is sketched in Fig.~\ref{fig1}
for four values of the global correlation that are given in the key.

In the special limit $C \to 1$ we find
$p(q_\alpha)=e^{-q_\alpha/2}/\sqrt{2\pi q_\alpha}$
for $q_\alpha > 0$ and $p(q_\alpha)=0$ for $q_\alpha <
0$. In the extreme  limit of very separated times in which 
$C \to 0$, $p(q_\alpha)$ becomes a
symmetric function with respect to $q_\alpha=0$ which
is not, however, a delta function.

Note that this form is very similar to the result found by Fusco and
Zannetti for the equilibrium overlap distribution, $P(q)$, 
of the {\it mean-spherical} model
at zero temperature~\cite{Fusco}.

\subsubsection{Finite M-component averaged local composite field}{$\;$}
 
The distribution of {\it finite} $M$ component averaged local field,
$q_M \equiv M^{-1} \sum_{\alpha} \phi_\alpha(x,t) \phi_\alpha(x,t') $,
can be studied similarly by setting ${\cal N}=M$ in
eq.~(\ref{eq:last}).  The integral can be transformed into multiple
convolutions of the result for $M={\cal N}=1$. With the purpose of
presenting the result graphically we prefer to perform the integral
explicitly.  For simplicity we consider only even $M$, {\it i.e.}
$M=2n$ with integer $n=1,2,3,\ldots$.  The integrand has two simple
poles at $\eta=iM/(1+C)$ and $\eta=-iM/(1-C)$. We then obtain:
\begin{eqnarray*}
p(q_M) &=& \frac{n}{4^{n-1}}(1-C^{2})^{n-1}
e^{-\frac{2n|q_{M}|}{1+{\rm sgn}(q_M) C}} 
\nonumber\\
&&
\; \times \, 
\sum_{l=0}^{n-1} 
\left(
 \frac{4n|q_{M}|}{1-C^{2}} \right)^{n-1-l} 
\frac{(n-1+l)!}{(n-1-l)!l!(n-1)!}
\; ,
\end{eqnarray*}
whose mean is $C$ and the variance $\sigma$ is given by
\begin{eqnarray*}
\sigma^{2} &=& q^{2}_{+}+q^{2}_{-}-C^{2}
\nonumber\\
q^{2}_{\pm} &=&
\frac{n}{4^{n-1}} \left(\frac{1 \pm C}{2n}\right)^{3}\;\sum_{l=0}^{n-1} 
\frac{(n-1+l)!}{(n-1-l)!l!(n-1)!}
\left(\frac{2}{1\mp C}\right)^{n-1-l}\Gamma(n-l+2)
\end{eqnarray*}
where $\Gamma(x)$ is the {\it gamma} function.
Again, we see that the distribution function is parametrised solely
by the global correlation function $C=C(t,t')$.

\begin{figure}
\center{\includegraphics[scale=.7,]{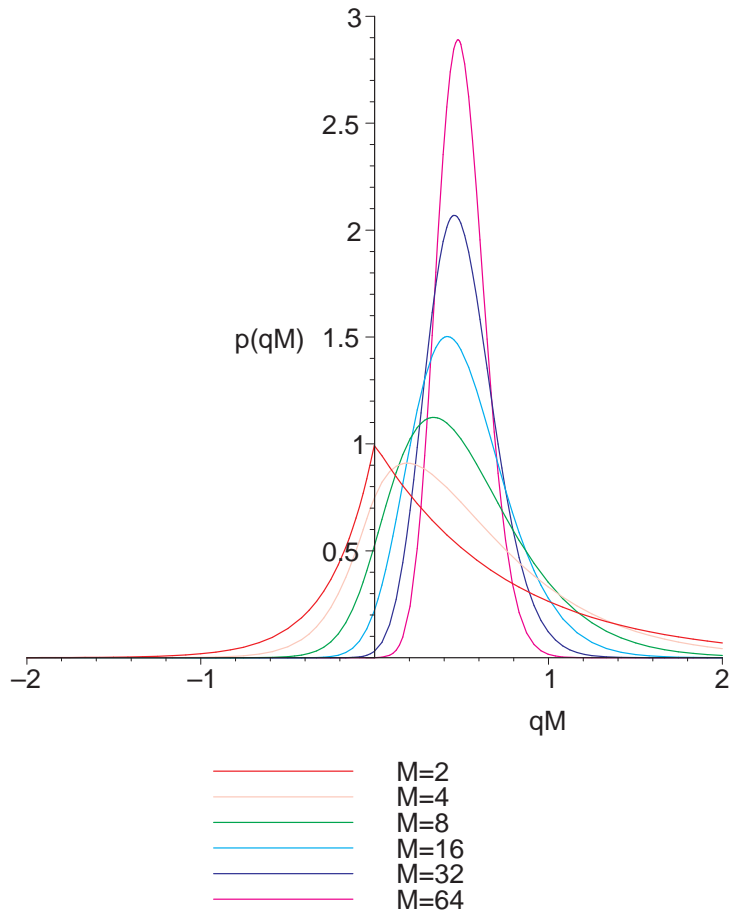}}
\caption{Probability distribution function of the $M$ component 
averaged local two-field composite operator, $q_M\equiv 
M^{-1} \sum_{\alpha=1}^M 
\phi_\alpha(\vec x, t) \phi_\alpha(\vec x, t')$,  
for several values of the number of components $M$
given in the key, at fixed global correlation $C(t,t')=0.5$.}
\label{fig2}
\end{figure}

Although the {\it mean} value of $p(q_M)$ is independent of 
$M$ and identical to the global correlation $C$, the functional 
form of this {\sc pdf} depends strongly on $M$. 
In Fig.~\ref{fig2} we show the functional form 
for six values of the number of components $M$
given in the key
and fixed global correlation $C$. It can be noticed that
for relatively small $M$, the position of the peak is different
from $C$. As $M$ increases the position of the peak
approaches $C$ and the width of the peak shrinks in such a way that
the {\sc pdf}  becomes the delta function $\delta(q-C)$ 
obtained in Sect.~\ref{subsubsec-N-component} 
in the $M=N \to \infty$ limit.

\begin{figure}
\center{\includegraphics[scale=.6,]{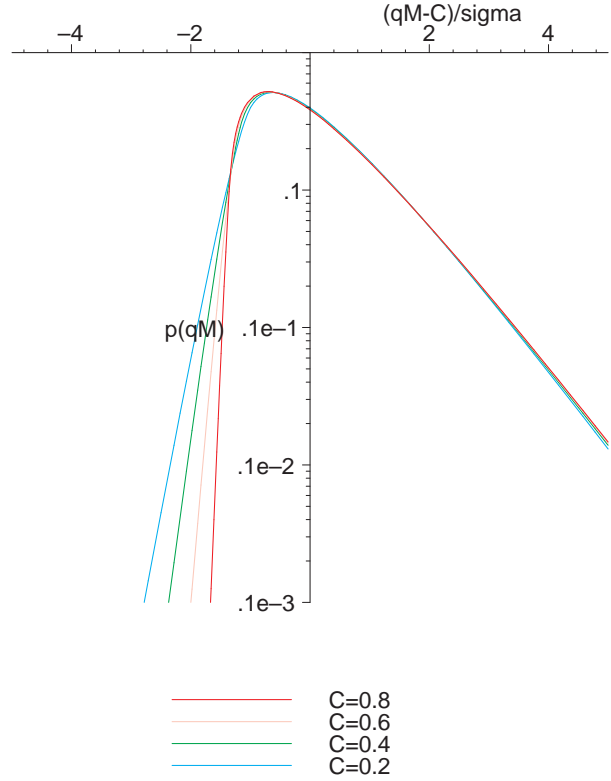}}
\caption{Probability distribution function of coarse-grained
M-component local two-field composite operator, $q_M=M^{-1}
\sum_{\alpha=1}^M \phi_\alpha(\vec x,t)\phi_\alpha(\vec x,t')$, with
$M=4$ and several values of the pair of times $(t,t')$ such that the
global correlation $C(t,t')$ takes the values given in the key. The
plot is double logarithmic scale and the $x$ axis has been put into
the normal form to compare the form of the {\sc pdf} for different
values of the global $C$.}
\label{fig3}
\end{figure}

\begin{figure}
\center{\includegraphics[scale=.6,]{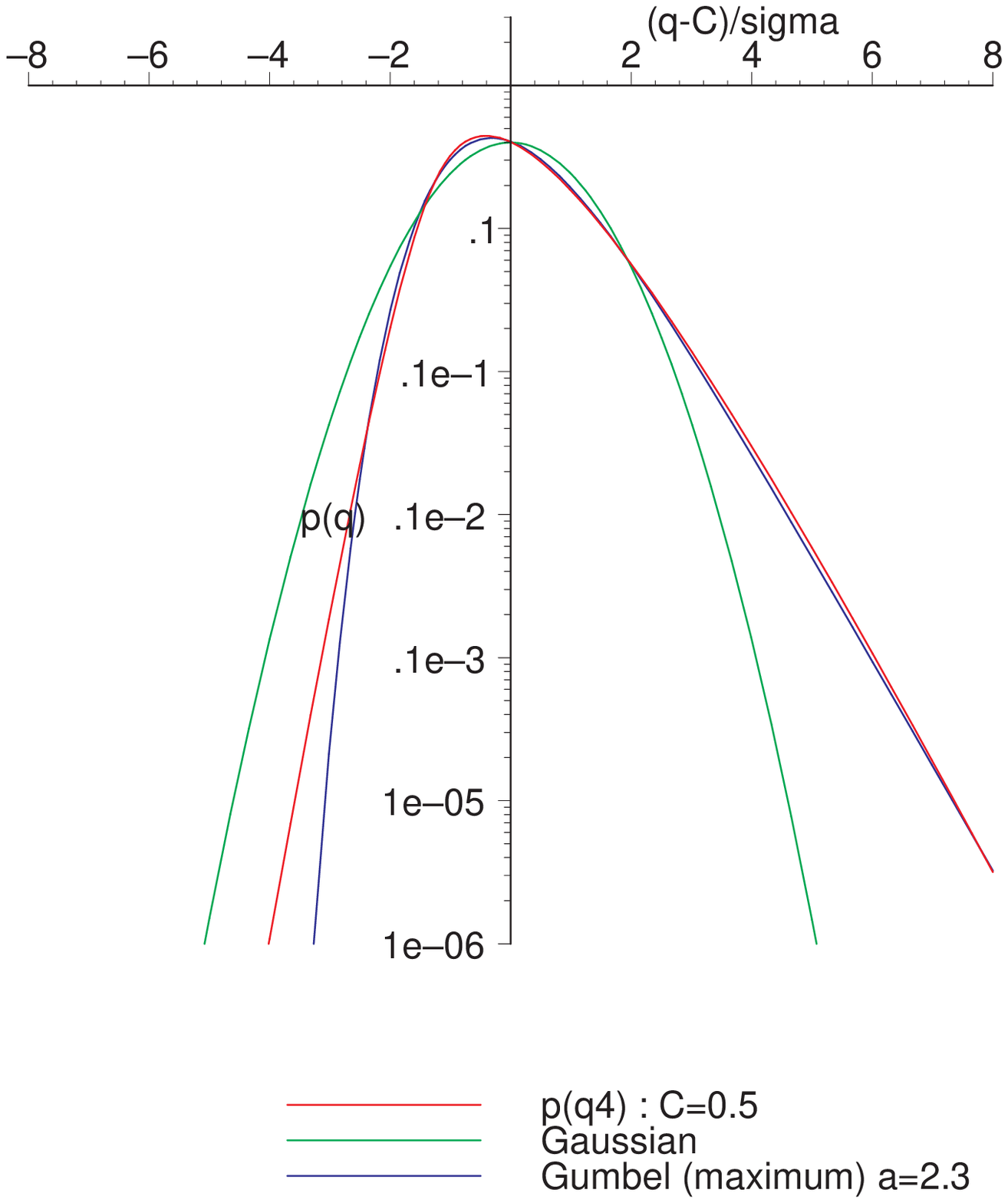}}
\caption{Log-log probability distribution function of coarse-grained 
M-component local ``correlations'', $q_M$ 
with $M=4$ and $C=0.5$ compared to a Gaussian and a Gumbel fit.}
\label{fig4}
\end{figure}

It is interesting to study the form of these {\sc pdf}s in more detail.
Figure~\ref{fig3} shows $p(q_M)$, with $q_M=M^{-1} \sum_{\alpha=1}^M
\phi_\alpha(\vec x,t)\phi_\alpha(\vec x,t')$, against $x=(q_M - \langle
\, q_M \, \rangle)/\sigma_{q_M}$. The number of components is $M=4$
and different curves correspond to several values of the global
correlation given in the key.  The plot is in double logarithmic scale. 
One sees that the curves are positively skewed with 
the right tail of the distribution being 
approximately independent of the value of $C$ while the left one is not.
In Fig.~\ref{fig4}  we compare the form of $p(q_M)$ to a Gaussian 
$e^{-x^{2}}/\sqrt{2\pi}$ and 
a Gumbel curve with positive parameter $a$ in such a way to make
it positively skewed. 
The normalised Gumbel distribution with mean zero
and variance $1$ is given by
\begin{eqnarray}
&& \Phi_{a}(x) =\frac{|\alpha|}{\Gamma(a)}
\; e^{a\log a} \; 
e^{a(\alpha(x-x_{0}))-e^{\alpha(x-x_{0})}}\;, 
\qquad \qquad \mbox{with} 
\nonumber\\
&& 
\alpha= \sqrt{\Psi'(a)} 
\;\;\; 
\mbox{and} 
\;\;\;
\alpha x_{0} =\log a- \Psi(a)
\; , 
\end{eqnarray}
where $\Gamma(x)$ is the {\it gamma} function and 
$\Psi(x)=\Gamma'(x)/\Gamma(x)$ is the {\it digamma} function.
For this intermediate value of $M$ the {\sc pdf} 
is clearly not Gaussian. The right tail is well fitted with the Gumbel 
form while the left tail is not.

\subsubsection{Effect of coarse-graining the ``correlation''}{$\;$}

With a similar analysis one shows that coarse-graining has no effect
if $V_x \ll [L(t')^d, L(t)^d]$ while for $V_x\gg [L(t')^d, L(t)^d]$ 
the probability distribution
function becomes a Gaussian as in the case in which we averaged over
all components of the field. Consider again the case ${\cal N}=1$, but
some general $V_x=\ell^d$:
\begin{displaymath}
q_{V_x} \equiv q_{V_x,{\cal N}=1}(\vec x; t,t') =
\frac{1}{V_x} 
\sum_{\vec y\in V_x}\phi_\alpha(\vec y,t)\; \phi_\alpha(\vec y,t')
\;. 
\end{displaymath}
The {\sc pdf} can be computed similarly as before but now
with 
\begin{eqnarray*}
{\cal M}_{\eta}(\vec k_1, \vec k_2)=
\delta_{\vec k_1 \vec k_2}+\frac{i\eta\Delta^2}{{V_x}}\sum_{\vec y\in V_x}
\Big[&&
r(k_1,t,0)\,
e^{i \vec k_1  \vec y}\:r(k_2,t',0)\,e^{-i \vec k_2  \vec y}
\\
&&\;\;\;\;\;\;
+r(k_1,t',0)\,
e^{i \vec k_1  \vec y}\:r(k_2,t,0)\,e^{-i \vec k_2  \vec y}
\Big]
\;.
\label{eq:M2}
\end{eqnarray*}

The eigenmodes of this matrix are studied in~\ref{sec:calc-M-2}.
Diagonalising is not easy for the general case but
the following two limiting cases can be considered.
\begin{itemize}
\item $\ell\ll L(t),L(t')$

The eigenvalues are the same as in the $V_{x}=1$ case:
one finds two non-trivial eigenvalues
$\lambda_\pm=1+i\eta\;[C(t,t')\pm 1]$ and $2V_x-2$ trivial
eigenvalues $\lambda=1$. Thus, the {\sc pdf} is the same as for 
$V_{x}=1$ (see Fig.~\ref{fig1}).

\item $\ell\gg L(t) \sim L(t')$

For simplicity we consider $L=L(t)\sim L(t')$.
Two non-trivial eigenvalues 
$\lambda_\pm\approx 1+i\eta(L/l)^d\;(C(t,t')\pm1)$ 
and $2V_x-2$ trivial eigenvalues $\lambda=1$ are obtained. 
Note that this is equivalent to the case $V_x=1$ 
if one substitutes $${\cal N}\sim(\ell/L)^d\;.$$ 
Thus, the {\sc pdf} is the same as those corresponding to composite fields
averaged over this number of field components
(see Fig.~\ref{fig2}-\ref{fig4}).

\end{itemize}

\subsection{The distribution of coarse-grained linear responses}

The distribution of local linear responses is surprisingly trivial
in quasi-quadratic systems such as the $O(N)$ model. 
Indeed, the linear response of each thermal run 
in Fourier space is given by 
\begin{displaymath}
\left.
\frac{\delta\phi_\alpha(\vec k,t)}{\delta h_\beta(-\vec k',t')} \right|_{h=0}
= 
\delta_{\alpha\beta} \; \delta^d(\vec k+\vec k') \, r(k;t,t')
\, \theta(t-t')
\; .
\end{displaymath}
This implies
\begin{eqnarray*}
\frac{\delta\phi_\alpha(\vec x,t)}{\delta h_\beta(\vec x,t')} 
&=& 
\delta_{\alpha\beta} \int \frac{d^d k}{(2\pi)^d} e^{i\vec k \vec x} 
\int \frac{d^d k'}{(2\pi)^d} e^{i\vec k' \vec x} 
\; 
\delta^d(\vec k+\vec k') \, r(k;t,t') \, \theta(t-t')
\nonumber\\
&=& 
\delta_{\alpha\beta} \int \frac{d^d k}{(2\pi)^d} \; r(k;t,t')
=
R_{\alpha\beta}(t,t')
\; ,
\end{eqnarray*}
{\it i.e.} a uniform result in space that is just equal to the 
global value. Again, this is independent of the spatial dimension 
$d$.

\subsection{The joint distribution of local correlations and responses}

Using the results above one concludes that the projection of the joint
{\sc pdf} on the $(C_x,\chi_x)$ plane at fixed pair of times $(t,t')$
such that they fall in the ageing regime (this is the extended {\sc
fdt} plot studied in \cite{Castillo-etal2}) is such that there are no
fluctuations in the vertical direction while there are in the
horizontal one.

\subsection{Summary}

In the strict $N\to\infty$ limit in which we have not taken into account
fluctuations of the constraint $N^{-1} \sum_{\alpha} \phi^2_\alpha(\vec
x,t)$, we found similarities and differences with the distributions of
coarse-grained local correlations and responses found in glassy systems. 
Let us discuss the points enumerated in the introduction to this Section 
in detail. 

{\it i.} In all cases the {\sc pdf}s of local composite fields in the
ageing regime depend on times only through the global correlation.
This property appears to be independent of there being time-reparametrisation 
invariance.

{\it ii.}  The form of the {\sc pdf}s of one-component composite
fields is definitely non-Gaussian but different from the one observed
in the $3d$ {\sc ea} model~\cite{Castillo-etal1,Castillo-etal2} and
kinetically constrained particle systems on the
lattice~\cite{Chamon-etal2}.  Before coarse-graining in real or
internal space the {\sc pdf} has a maximum at $q=0$ for all values of
$C$ (see Fig.~\ref{fig1}). This is simply due to the fact that in the
$O(N)$ model with $N\to \infty$ the configurations with many vanishing
components are very favourable.~\footnote{The configuration at each
point in real space is a vector in $N$ dimensions with fixed
length. For example, any such vector chosen from a flat distribution
typically has a few large components and many [$O(N)$] components with
vanishing value. This can be easily worked out when $N=2$, {\it i.e.}
for the XY model, for which the {\sc pdf} of the $x$ and $y$
components are Gaussians centred at zero.}  To understand the role of
the large $N$ limit one should compare the above results to, for
example, the same {\sc pdf}s in the XY problem.

Averageing over components or over real space washes out the 
weight on negative values ($q<0$) just as found in the spin models.
For finite value of ${\cal N}$ or for coarse-graining boxes that 
do not go beyond the domain length, 
the {\sc pdf}s remain, though, positively skewed for all values of 
$C$ even those corresponding to times that are close to each
other  (see Fig.~\ref{fig3}). 

We have also checked whether the distributions of $q_M$ can be approximated 
by a Gumbel-like form with negative parameter. We find that while the
tail on the right is quite well described with this functional form, 
the tail on the left is not (see Fig.~\ref{fig4}).

In the large coarse-graining volume, $\ell\gg [L(t),L(t')]$, 
or averaging over a diverging number 
of components, ${\cal N}=N\to\infty$, the 
{\sc pdf} becomes a delta function, $\delta(q-C)$.

{\it iii.}
There are no
fluctuations of the linear responses.
This is intimately related to the 
quasi quadratic nature of the model in
the limit $N\to \infty$. This result is clearly different 
from what found in glassy models, in which the local response 
functions do fluctuate form site to site though constrained to 
follow the global $\chi(C)$ curve. 
In the $O(N)$ model the projection of the joint {\sc pdf} of local 
correlations and responses also follows the global $\chi(C)$ curve but in a 
trivial manner, since the local responses take a single value.

It would be interesting to study whether these results are modified 
by $1/N$ corrections when the constraint is allowed to fluctuate
and yields an additional contribution to the linear 
response~\cite{Yoshino-etal}.

\section{Four point correlation function}
\label{sec:fourpoint}

A coarsening system is one in which the growing length 
is easily identified as the typical domain length. A scaling 
theory then predicts that all correlations should depend 
on distance and on times only 
through the value of the typical domain length. This is explicitly
realised by the $O(N)$ model and an example of such scaling 
law is given in eq.~(\ref{eq:scaling-corr}).

In spin-glasses and structural glasses the observation of such a
growing length has been elusive.  A growing correlation length in the
super-cooled liquid has been extracted from the analysis of the
connected correlation of fluctuating local composite operators in a
number of model
systems~\cite{Biroli-Bouchaud,Toninelli-etal}.  The
analysis of numerical simulations of several models as well as some
experiments indicate that this length takes very small values, of the
order of a few nanometres in the super-cooled liquid. A summary of
these results appeared recently in~\cite{Toninelli-etal}.

In an out of equilibrium system, such as the problem at hand, this
``four-point'' correlation function is naturally defined
as~\cite{Castillo-etal2,Mayer-etal}
\begin{eqnarray}
C_4(\vec x,\vec x'; t,t') &\equiv& 
[\langle \phi_{\alpha}(\vec x,t)\phi_{\alpha}(\vec x,t')
\phi_{\alpha}(\vec x',t)\phi_{\alpha}(\vec x',t')
\rangle]_{ic}   
\nonumber\\
&& \;\;\;\;\;\;
-
[\langle \phi_{\alpha}(\vec x,t)\phi_{\alpha}(\vec x,t') \rangle]_{ic} 
\; 
[\langle \phi_{\alpha}(\vec x',t)\phi_{\alpha}(\vec x',t') \rangle]_{ic} 
\; .
\label{eq:four-point}
\end{eqnarray}
Note that this quantity is nothing but the connected spatial correlation
function of the composite field $q_\alpha \equiv \phi_{\alpha}(\vec x,t)\;
\phi_{\alpha}(\vec x,t')$ (see Sect.~\ref{sec:local-corr}). Since noise and
initial condition averaged quantities are expected to be invariant under
translations of the space coordinates, this quantity should be equal to
\begin{eqnarray*}
C_4(\vec r; t,t') &\equiv& 
\frac{1}{V}\int d^d x \; 
[\langle \phi_{\alpha}(\vec x,t)\phi_{\alpha}(\vec x,t')
\phi_{\alpha}(\vec x',t)\phi_{\alpha}(\vec x',t')
\rangle]_{ic}   
\nonumber\\
&& -
\frac{1}{V^2} \int d^d x \; 
[\langle  \phi_{\alpha}(\vec x,t)\phi_{\alpha}(\vec x,t') \rangle]_{ic} 
\int d^d x' \; 
[\langle  \phi_{\alpha}(\vec x',t)\phi_{\alpha}(\vec x',t') \rangle]_{ic} 
\nonumber\\
&=&
\frac{1}{V} \int d^d x \; 
[\langle \phi_{\alpha}(\vec x,t)\phi_{\alpha}(\vec x,t')
\phi_{\alpha}(\vec x',t)\phi_{\alpha}(\vec x',t')
\rangle]_{ic}   
-
C^2_{\alpha\alpha}(t,t')
\end{eqnarray*}
with $\vec r\equiv \vec x- \vec x'$.
$C_4(\vec x,\vec x'; t,t')$
measures the probability that similar decorrelations
taking place between $t'$ and $t$ occur at a spatial distance $\vec r$ in 
the sample.

The volume integral of $A$ defines the quantity 
\begin{displaymath}
\chi_4(t,t') \equiv \int d^d r \; C_4(\vec r; t, t')
\; .
\end{displaymath}
that is loosely called a ``susceptibility'' advocating the use of a
fluctuation-dissipation theorem to relate the correlation of composite
operators to a linear response. When the operators and the
perturbations are composite ones depending on two (or more) times this
is however highly non-trivial. Some examples have been exhibited
in~\cite{Guilhem2}. In particular, $C_4$ is not equal to the response of
the composite observable $[\langle \phi_\alpha(\vec x, t)
\phi_\alpha(\vec x, t') \rangle]_{ic}$ to an infinitesimal field that
couples linearly to $\phi_\alpha(\vec x', t) \phi_\alpha(\vec x', t')$
(see~\ref{fdt}) as one would naively propose.  
In the low-temperature phase, where equilibrium
dynamics is lost, the relation between spontaneous and induced
fluctuations is still more complicated due to the fact that these are
not determined by the equilibrium measure.

With the aim of comparing to the results found in super-cooled liquids we
study the behaviour of $\chi_4$ during coarsening. The four point correlation
function eq. (\ref{eq:four-point}) is easily obtained using the solution to
the equation of motion, eq.~(\ref{eq:sol-k}).  Again, for simplicity we work
at $T=0$ and we find
\begin{eqnarray}
C_4(\vec r; t,t') &=& 
[ \phi_{\alpha}(\vec x,t)\phi_{\alpha}(\vec x',t) ]_{ic}
[ \phi_{\alpha}(\vec x,t')\phi_{\alpha}(\vec x',t') ]_{ic}
\nonumber \\
&& 
\;\;\;\;\;\;\;\;\;\;\;\;\;\;\;\;\;
+ [ \phi_{\alpha}(\vec x,t)\phi_{\alpha}(\vec x',t') ]_{ic} \; 
[ \phi_{\alpha}(\vec x',t)\phi_{\alpha}(\vec x,t') ]_{ic}
\nonumber \\
&=&
e^{-(r/L(t))^{2}}e^{-(r/L(t'))^{2}}
+C^{2}(t,t') \; e^{-2(r/L(t+t'))^{2}}
\label{eq:4point}
\end{eqnarray}
where $r=|\vec{x}-\vec{x}'|$, $C(t,t')$ is the global correlation
function and $L(t) \propto \sqrt{t}$ is the usual domain size. 
The first term
is a rather trivial contribution since it is just
the product of the (average) equal-time spatial correlation functions
at $t$ and $t'$. The last term depends on the domain length evaluated at 
the sum of the two times involved, $L(t+t')$.
Note that if ``reciprocity'' holds the last term
becomes $[ \phi_{\alpha}(\vec x,t)\phi_{\alpha}(\vec
x',t') ]_{ic}^{2}$.  
In the ageing regime the length scales $L(t)$ and $L(t')$ are of the 
same order. Moreover, since $L(t)\sim t^{1/2}$, 
$L(t+t')$ is also of the same order. Using $t'=\lambda t$ with $\lambda 
\in [0,1]$, $L(t') = \lambda^{1/2} L(t)$ and 
$L(t+t') \sim (1+\lambda)^{1/2} L(t)$.
Thus, for distances $r$ of the order of $L(t)$ all terms 
contribute. Note that $C_4(\vec r;t,t)$ does not vanish.

Using eq.~(\ref{eq:Cag-global}) for the global correlation in the ageing 
regime we note that $C_4(\vec r;t,t')$ can be put into the scaling form
\begin{displaymath}
C_4(\vec r;t,t') = 
f_{C_4}\left( \frac{L(t)}{L(t')}, \frac{r}{L(t')}\right)
= \tilde f_{C_4}\left( \frac{t}{t'}, \frac{r}{L(t')}\right)
\end{displaymath}
as expected from simple scaling arguments and found 
for the one-dimensional Ising chain~\cite{Mayer-etal}.

\begin{figure}
\center{\includegraphics[scale=.7]{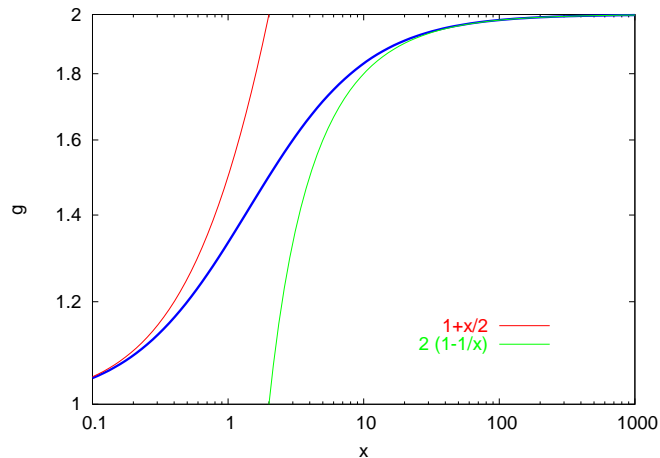}}
\caption{In blue the function $g(x)=2 [f_\chi(x)/(1+2^d)]^{\frac{2}{d}}$
in a log-log scale.
Also indicated with thin red and green curves are the behaviour close to 
$x\sim 0$ (short time differences) and asymptotically $x\to \infty$
(very long time-differences). 
}
\label{fig:scaling-chi4}
\end{figure}

From expression~(\ref{eq:4point}) we easily compute $\chi_4(t,t')$: 
\begin{eqnarray*}
\chi_4(t,t') \propto L^d(t') \;  f_{\chi_4}\left( \frac{\tau}{t'}\right) 
\;\;\;\;\;\;\;
\mbox{with} 
\;\;\;\;\;\;\;
f_{\chi_4}(x) = 2^{-\frac{d}{2}+2^\frac{d}{2}}
\left( \frac{1+x}{1+x/2} \right)^{\frac{d}{2}}
\; .
\end{eqnarray*}
This function has the form shown in Fig.~\ref{fig:scaling-chi4}.  It
does not have a maximum as a function of $\tau\equiv t-t'$ but it
monotonically increases towards a finite $t'$-dependent asymptote.
In this respect the behaviour is rather different from what has been
found in the supercooled-liquid phase of a number of glassy
systems~\cite{Biroli-Bouchaud,Toninelli-etal}
and in the coarsening foam studied in \cite{Mayer-etal}.

It is interesting to analyse the behaviour of the second term too. 
If one assumes, based on scaling arguments, that 
\begin{displaymath}
C_{ag}(t,t')
\sim \left(\frac{L(t')}{L(t)}\right)^{\overline\lambda}
\; ,
\end{displaymath}
{\it i.e.} that the very last decay is characterised by the
$\overline\lambda$ exponent~\cite{Huse-Jenssen,BK92}, the behaviour of
this term at very long time-differences depends on whether
$\overline\lambda$ is larger or equal than $d/2$, the lower bound
conjectured by Fisher and Huse~\cite{Fisher-Huse0}. When $\overline
\lambda=d/2$, as in the $O(N)$ model, this term is also finite and
contributes to the asymptotic value of $\chi_4$. For other
systems in which $\overline \lambda$ is larger than $d/2$ this terms
vanishes asymptotically (as implicitly assumed in \cite{Mayer-etal}).

Let us mention that the alternative definition of $C_4(t,t')$ proposed
in \cite{Mayer-etal} also has a finite asymptotic ($t'\to\infty$)
value in the $O(N)$ model. This is due to the fact that the last added
term is equal to the second term discussed in the previous paragraph
and does not vanish.

One could also define a connected spatio-temporal correlation function of the
original field $\phi_{\alpha}(\vec x,t)$. Due to the factorisation rules for
$N\to \infty$, these quantities vanish. However, there are $O(1/N)$
corrections which yield essential contributions to some related integral
susceptibilities \cite{Yoshino-etal}.

We can now compare to what 
has been observed in numerical simulations of the $3d$
Edwards-Anderson model~\cite{Castillo-etal2} using a slightly
different expression for the four-point correlation that differs from
(\ref{eq:four-point}) just in a normalisation. In \cite{Castillo-etal2}
$C_4$ was normalised to be one at $r=0$ for all times. For the $O(N)$
model this normalisation factor is $1+C^2(t,t')$. Thus, the space
integral of the $O(N)$ normalised four-point correlation also approaches a
finite limit when $\tau\to\infty$ and $t'$ is held fixed.

The normalised four point correlation in the $3d$ Edwards-Anderson
model was rather well described with the form $e^{-r/\xi(t,t')}$.
Even if the space and time dependence in the $O(N)$ model is more
complicated than a simple exponential decay, the qualitative behaviour
of $\xi(t,t')$ in \cite{Castillo-etal2} is similar to that of
$\chi_4(t,t')$ for the $O(N)$ model in that $\xi(t,t')$ {\it
increases} with both $t'$ and $t-t'$.

Finally, let us compare the four-point correlation function $C_4$ to 
the integrated response of the composite field, 
$\phi_\alpha(\vec x,t) \phi_\alpha(\vec x,t')$ to a composite perturbation
$h_\alpha(\vec x',t'',t''')$ \cite{Franz-Parisi}. 
In~\ref{fdt} we show 
\begin{eqnarray}
&& 
\left[ \, 
\frac{\delta \phi(\vec x, t) \phi(\vec x, t')}{\delta h(\vec x', t'',t''')}
\, \right]_{ic}
=
\nonumber\\
&& 
\;\;\;\;\;\;\;\;\;\;
R(\vec r;t,t'') C(\vec r;t',t''') \theta(t-r''')
+
R(\vec r;t,t''') C(\vec r;t',t'') \theta(t-t'') 
\nonumber\\
&& 
\;\;\;\;\;\;\;\;\;\;
+
R(\vec r;t',t'') C(\vec r;t,t''') \theta(t'-t''')
+
R(\vec r;t',t''') C(\vec r;t,t'') \theta(t'-t'')
\nonumber
\label{eq:composite-response0}
\end{eqnarray}
It is clear that this is not simply related to $C_4$.

In summary, we found that a model undergoing coarsening has a $\chi_4(t,t')$
that, as expected, depends on times only through $L(t)$ and $L(t')$, but 
does not decay to zero at long-time differences. We conclude that the 
existence of a maximum in $\chi_4(t,t')$ cannot be taken as evidence for a 
growing correlation length but other features of this quantity have to be 
analysed.

\section{Conclusions}
\label{sec:conclusions}

Neither the dynamic equations of the slow (coarsening) contributions
to the global correlation and response nor their effective action in
the large $N$ $O(N)$ model are invariant under generic time
reparametrisations. This symmetry is reduced to uniform time
re-scalings $t \to \zeta \, t$, with the advanced and retarded scaling
dimensions of the global correlation and response,
$\Delta_C^A=\Delta_C^R=0$ and $\Delta_R^A=0$, $\Delta_R^R=d/2$,
respectively, and similarly for the corresponding fluctuating fields.

The breakdown of time-reparametrisation invariance seems to be 
intimately related to the absence of a finite or well-defined 
effective temperature in $d>2$ and $d=2$, respectively. 
Indeed, the retarded scaling dimension $\Delta_R^R=d/2$ in
$d>2$ implies that the fluctuation-dissipation ratio vanishes
asymptotically in the low-temperature phase. Instead, $\Delta_R^R=1$
in $d=2$ implies that the fluctuation-dissipation ratio takes a
non-trivial $L(t')/L(t)$ dependent form but the evolution occurs in 
a single ageing scale in which the correlation itself varies as a 
function of this ratio. This is inconsistent with the natural
requirement of having a single value of the effective temperature per
correlation scale.

If we were to use the remaining time-rescaling
symmetry to characterise the fluctuations of local 
correlations of the $O(N)$ model when $N\to\infty$, 
as we did when we used time-reparametrisation 
invariance as a guideline to characterise fluctuations in glassy
systems~\cite{Chamon-etal,Castillo-etal1}, we should 
introduce the spatial dependence by rescaling time with a 
space-dependent parameter: $t \to \eta_x t$. However,
a simple multiplicative rescaling of time disappears from the 
correlations:
\begin{displaymath}
C_x(t,t') \sim f_C\left( \frac{\sqrt{\eta_x t'} }{\sqrt{\eta_x t}} \right) = 
f_C \left( \frac{\sqrt{t'}}{\sqrt{t}} \right)
=f_C \left( \frac{L(t')}{L(t)} \right)
\; .
\end{displaymath}
This means that no such fluctuations are generated.
Therefore spatio-temporal fluctuations in the $O(N)$ model have
a different origin.

We analysed several distribution functions with the aim of identifying
similarities and differences with the ones generated by
time-reparametrisation invariance. For simplicity we focused on the
zero temperature dynamics and we analysed the fluctuations induced by
random initial conditions. We concentrated in times such that the 
dynamics is in the coarsening -- ageing -- regime. 
Let us now summarise and discuss our findings.

Each $\phi_\alpha(\vec x,t)$, with $\alpha$ any component of the
$N$-dimensional vector $\vec \phi$, obeys a Gaussian {\sc pdf}. The local
one-component composite field, $\phi_\alpha(\vec x,t) \phi_\alpha(\vec
x,t')$, has a non-Gaussian distribution.  We derived the functional form of
this {\sc pdf} and we showed how it crosses over to a delta function under
coarsening over a sufficient large volume, of linear size larger than the
typical domain lengths, $\ell \gg L(t'), L(t)$. We also found that the
two-time observable made of a sum over a number, ${\cal N}$, of components of
the composite field has a similar behaviour to the distribution of the
one-component quantity coarse-grained over a volume of linear size $\ell^d \sim
L^d(t') \;{\cal N}$ when $L(t)$ and $L(t')$ are of the same order.

In all these cases, the {\sc pdf} of local composite fields {\it scales 
in time} just as the global correlation itself; that is to say, it is  
a function of the ratio between the two characteristic scales $L(t')$ and 
$L(t)$:
\begin{displaymath}
p(q_{V_x{\cal N}};t,t') = p(q_{V_x{\cal N}};C(t,t')) 
= p\left(q_{V_x{\cal N}};f_C\left(\frac{L(t')}{L(t)}\right)\right)
\; . 
\end{displaymath}
In~\cite{Chamon-etal}-\cite{Chamon-etal2} we argued that
uniform time-reparametrisation invariance and the simplest choices of
effective action for the local reparametrisations, $h_x(t)$, imply
this kind of scaling and, using numerical simulations, we found it
in the $3d$ Edwards-Anderson
model~\cite{Castillo-etal1,Castillo-etal2} and a kinetically
constrained lattice gas~\cite{Chamon-etal2}.  The solution of the $O(N)$
model when $N\to \infty$ shows that this property is not unique to
models with time-reparametrisation invariance.

The {\it form} of the {\sc pdf} of these local two-time quantities is
not the Gumbel-like form that we argued should describe the
fluctuations of local correlations of spin-like variables that are
associated to global time-reparametrisation. In particular, before any
coarse-graining -- or even after averaging over a small number of
components or a small coarse-graining box -- the {\sc pdf} has a peak
at very small values of the argument, $q_{V_x{\cal N}}\sim 0$. Under
further coarse-graining the peak moves towards positive values of
$q_{V_x{\cal N}}$ until reaching a Gaussian form centred on the
average -- global -- value $C$, that eventually becomes a delta
function.  Note that the form of the {\sc pdf}s does not depend on the
dimension of space explicitly -- it does only through $C$. 

We may then conjecture that the reason for finding a strong weight at small
values of $q_{V_x{\cal N}}$ is the continuous character of the order
parameter and its large dimensionality.  Indeed, a peak at small values of
the two-time composite field should be present in the {\sc pdf} of
$q_{V_x{\cal N}}$ with $V_x\ll \ell$ and ${\cal N}< N$ for all models with a
continuous order parameter and a spherical constraint. But this peak should
not be necessarily unique.  Indeed, preliminary numerical simulations of the
dynamics of the $2d$ XY model starting from a random initial condition and in
the low temperature phase show that the {\sc pdf} of, say, the horizontal
component of the local composite field has a second peak at one, when the two
times are not very far away and the global correlation takes a large value.
The fate of the two peaks, and thus of the full {\sc pdf}, under
coarse-graining needs to be analysed in more detail but it is not excluded
that it may then take a form and evolution similar to the one observed in the
$3d$ Edwards-Anderson and kinetically constrained lattice gas.  Note,
however, that the $2d$ XY model is critical in the full low temperature
phase; its non-equilibrium dynamics is then typical of a critical point with
a {\it multiplicative} separation of time-scales and an ageing regime that
eventually disappears in the long waiting-time limit~\cite{Mauro}. While in
the ageing regime, this model has the very appealing feature of having a
finite integrated response. It belongs to yet another class of models and it
is then a very interesting case to study in the context of our discussion.

The {\sc pdf} of local linear responses is deceptively trivial in the
quasi-quadratic large $N$ $O(N)$ model: these quantities do not
fluctuate at all and are just identical to the global value. We do not
expect this result to survive in such a trivial manner when including
$1/N$ corrections or for other coarsening problems that are not
(almost) quadratic. In particular, if full time-reparametrisation
invariance is broken there is no obvious
reason why the joint probability distribution of local responses and
correlations should follow the global $\chi(C)$ parametric curve
between integrated response and correlation. This is a problem that
deserves to be addressed analytically and/or numerically in other
coarsening models.

We computed the four-point correlation, $C_4$, that is usually used to
identify a growing correlation length in super-cooled liquids, now
during coarsening. Not surprisingly we found that it satisfies a
scaling relation in which times enter only through the typical domain
length, $L$. Contrary to what found in super-cooled liquids and
glasses, the integral over space of $C_4$ does not vanish at very long
time-differences, $t-t' \to \infty$ for any fixed $t'$. The
reason for this is the fact that in coarsening systems the
spatial correlation, $C(r,t)$, does not
vanish.  The same feature was signalled in \cite{Mayer-etal} in the
context of the ferromagnetic Ising chain. We also stressed the fact
that this quantity is not trivially related to a susceptibility
(see~\cite{Guilhem2,Bouchaud-Biroli} for a similar discussion).

It is interesting to compare the pure time transformations studied in this
paper to the common space-time invariance of domain growth~\cite{Bray}. The
exact solution of the $O(N)$ model is invariant, in the long times and large
scales limit, under simultaneously rescaling of time and space [see
eqs.~(\ref{eq:scaling-corr})], and the slow part of the dynamic action is
invariant under the related renormalisation group transformation [see
Sect.~\ref{sec:sym-action}]. However, it is not this space-time invariance
that is the relevant symmetry if one is interested in fluctuations within a
given domain. One should consider separations $r$ that are held fixed while
the long-time limit is taken. More precisely, one should consider fixed
ratios $t/t'$ while $L(t)\to \infty$, and thus $r/ L(t)\to 0$. It is in this
limit that reparametrisation invariance should be investigated, and there are
a number of issues that one must consider specifically in the case of the
$O(N)$ model. First, reparametrisation invariance cannot be an exact
symmetry of the solution to the $O(N)$ -- or any other similar dynamic
problem -- since a particular function $h(t)$ is bound to be chosen by the
evolution. It may only arise as an {\it approximate} invariance in the
asymptotic limit in which the non-invariant terms -- that act as ``pinning
fields'' and fix the time scaling $h(t)$ -- become less and less important.
This is indeed what happens in mean-field disordered models of the $p$ spin
type and, we argued~\cite{Chamon-etal}, in the $3d$ Edwards-Anderson
spin-glass. Second, we showed in this paper that reparametrisation invariance
{\it does not} develop in the $O(N)$ model, only a smaller
symmetry, simple time scale invariance, does. 
We arrived at these
results by studying the equations of motion for the global $C(t,t')$ and
$R(t,t')$ and the action for the slow flucting fields. 
Similar results would be obtained for the two space-point
correlation $C(r;t,t')$ and response $R(r;t,t')$, when $r$ is held fixed
while the long-time limit is taken.

Let us finally stress the main issue arising from this study, {\it i.e.} the
conjecture that an extreme violation of the fluctuation-dissipation theorem
is intimately related to the breakdown of time-reparametrisation invariance
at long times {\it in general}.  If this is correct, systems with a finite or
an asymptotically infinite 'effective temperature' belong to different
`universality' classes, as non-equilibrium fluctuations are
concerned~\cite{Racz}.  It would be interesting to put this conjecture to the
test in other solvable models. In particular, by comparing to similar
fluctuations in the XY model one should be able to identify the peculiar
features due to the $N\to \infty$ limit. The special $d=2$ case should be
particularly interesting. Another route is to analyse other coarsening
systems with a discrete order parameter: one then should be able to
disentangle the features that are due to $X\to 0$ from those that are due to
the continuous character of the field.

\vspace{1cm}

\noindent\underline{Acknowledgements} \\ We thank G. Biroli and
M. Picco for very useful discussions. L.F.C. is a member of the
Institut Universitaire de France. This research was supported in part
by NSF grants DMR-0305482, DMR-0403997, and INT-0128922 (C.C.), an
NSF-CNRS collaboration, the ACI-France ``Algorithmes d'optimisation et
syst\`emes desordonn\'es quantiques'', the STIPCO European
Community Network, and NSF Grant No. PHY99-07949  (L.F.C.).  
H. Yoshino acknowledges financial
support from the Japanese Society of Promotion of Science and CNRS.

\vspace{1cm}

\appendix

\section{The equation of motion for the global linear response}
\label{app:eq-of-motion}

In this appendix we show how to obtain the equations of motion for the
correlation and response, expressed in terms of $R(t,t')$ and $C(t,t')$
themselves, starting from the exact expressions 
for $R(t,t')$ and $C(t,t')$ obtained 
from the equations of motion for the field $\vec\phi(t)$.

The exact solution for the correlation and response follows from
the self-consistent solution of eqs.~(\ref{eq:sol-k}) and (\ref{eq:Y2}):
\begin{eqnarray}
C(t,t')&=& Y^{-1}(t)\;Y^{-1}(t')\;
\Big[
\Delta^2 \;\langle\langle e^{-\epsilon_k(t+t')}\rangle\rangle
\label{eq:Cexact}
\\
&&\;\;\;\;\;\;\;\;\;\;\;\;\;\;\;
+2T\int_0^{{\rm min}(t,t')} dt''\;\langle\langle
e^{-\epsilon_k(t+t'-2t'')}\rangle\rangle
\;Y^2(t'')
\Big]
\nonumber
\\
R(t,t')&=& Y^{-1}(t)\;Y^{}(t')\;
\langle\langle e^{-\epsilon_k(t-t')}\rangle\rangle\;\theta(t-t')
\;,
\label{eq:Rexact}
\end{eqnarray}
where $\epsilon_k$ is the dispersion and $\langle\langle f(k)
\rangle\rangle=\int\frac{d^dk}{(2\pi)^d}\; f(k)$.

For $t>t'$ we can write
\begin{eqnarray}
\frac{\partial C(t,t')}{\partial t} &=& -z(t)\;C(t,t')+
Y^{-1}(t)\;Y^{-1}(t')\;
\Big[
\Delta^2 \;\langle\langle -\epsilon_k\;e^{-\epsilon_k(t+t')}\rangle\rangle
\label{eq:diffCexact}
\\
&&\;\;\;\;\;\;\;\;\;\;\;\;\;\;\;
+2T\int_0^{{\rm min}(t,t')} dt''\;\langle\langle
-\epsilon_k\; e^{-\epsilon_k(t+t'-2t'')}\rangle\rangle
\;Y^2(t'')
\Big]
\nonumber
\\
\frac{\partial R(t,t') }{\partial t} &=& -z(t)\;R(t,t')+
Y^{-1}(t)\;Y^{}(t')\;
\langle\langle -\epsilon_k\;e^{-\epsilon_k(t-t')}\rangle\rangle\;\theta(t-t')
\;,
\label{eq:diffRexact}
\end{eqnarray}
where $z(t)={\partial\over \partial t}\;\ln Y(t)$.

In order to express eqs.~(\ref{eq:diffCexact}) and 
(\ref{eq:diffRexact}) in terms of
$R$'s and $C$'s, all one needs to do is express the right hand side of these
equations in terms of convolutions of $R$'s and $C$'s. The first step to do
so is to express the function
\begin{displaymath}
  \label{eq:D0}
D(t)=
\langle\langle -\epsilon_k\;e^{-\epsilon_k t}
\rangle\rangle \;\theta(t)
=
\int_0^\infty \!\!d\epsilon\;g(\epsilon)\;(-\epsilon)\;
e^{-\epsilon t}\;\theta(t)
\end{displaymath}
in terms of convolutions of the function
\begin{eqnarray*}
  \label{eq:G}
  G(t)=
  \langle\langle e^{-\epsilon_k t}\rangle\rangle\;\theta(t)
  &&=
  \int_0^\infty \!\!d\epsilon\;g(\epsilon)\;e^{-\epsilon t}\;\theta(t)
=
  \int_{-\infty}^{\infty}\frac{d\omega}{2\pi}\;e^{-i\omega t}\;{\tilde
  G}(\omega)
    \;,
\label{eq:G-tilde-G}
\end{eqnarray*}
where 
\begin{equation}
  \label{eq:tilde-G-def}
{\tilde
  G}(\omega)\equiv \int_0^\infty \!\!
d\epsilon\;\frac{g(\epsilon)}{\epsilon-i\omega}  
\label{eq:tilde-G}
\end{equation}
and $g(\epsilon)$ is the density of states with $\epsilon=\epsilon_k$.
In other words, we basically need to cast
\begin{equation}
  \label{eq:conv1}
D(t)=\sum_{n=1}^{\infty} A_{n-1} \;\underbrace{G*G*\cdots*G}_n \;(t)
\;\;.
\end{equation}
We start by writing
\begin{eqnarray*}
  \label{eq:n-G-conv}
\underbrace{G*G*\cdots*G}_n \;(t)
&&=
\int_{-\infty}^{\infty}\!\!\!d\tau_1
\int_{-\infty}^{\infty}\!\!\!d\tau_2
\dots
\int_{-\infty}^{\infty}\!\!\!d\tau_{n-1}
\;\;
G(t-\tau_1)\;G(\tau_1-\tau_2)
\\
&&
\:\:\:\:\:\:\:\:\:\:\:\:
\:\:\:\:\:\:\:\:\:\:\:\:
\:\:\:\:\:\:\:\:\:\:\:\:
s \;G(\tau_{n-2}-\tau_{n-1})\;G(\tau_{n-1})
\nonumber
\\
&&=
\int_{-\infty}^{\infty}\frac{d\omega}{2\pi}\;e^{-i\omega t}\left[{\tilde
  G}(\omega)\right]^n
\\
&&=
\int_0^\infty \!\!d\epsilon_1
s
\int_0^\infty \!\!d\epsilon_n
\int_{-\infty}^{\infty}\frac{d\omega}{2\pi}\;e^{-i\omega t}
\prod_{a=1}^n\;\frac{g(\epsilon_a)}{\epsilon_a-i\omega} 
\\
&&=
n
\int_0^\infty \!\!d\epsilon
\;g(\epsilon)\;e^{-\epsilon t}\;
\left[\int_0^\infty \!\!
d\epsilon'\;\frac{g(\epsilon')}{\epsilon'-\epsilon} \right]^{n-1}
\;\theta(t)
\;.
\end{eqnarray*}
Thus, in short, we have
\begin{displaymath}
  \label{eq:G-conv-h}
\underbrace{G*G*\cdots*G}_n \;(t)=
n
\int_0^\infty \!\!d\epsilon
\;g(\epsilon)\;e^{-\epsilon t}\;
\left[h(\epsilon)\right]^{n-1}
\;\theta(t)
\;,
\end{displaymath}
where the function $h(\epsilon)$ is defined as
\begin{equation}
  \label{eq:h-def}
h(\epsilon)=
\int_0^\infty \!\!d\epsilon'\;\frac{g(\epsilon')}{\epsilon'-\epsilon}
\;.
\end{equation}
Next, let us expand $\epsilon$ as a function of $h(\epsilon)$:
\begin{displaymath}
  \label{eq:expansion-e-h}
\epsilon=h^{-1}\circ h(\epsilon)=\sum_{n=0}^{\infty}a_n\; [h(\epsilon)]^n,
\quad {\rm with} \quad a_n=\frac{1}{n!}\;\frac{d^nh^{-1}}{dz^n}\Big|_{z=0}
\;.
\end{displaymath}
Therefore, we can write
\begin{eqnarray*}
  \label{eq:D-final}
D(t)
&&=
  \theta(t)\;\int_0^\infty \!\!d\epsilon\;g(\epsilon)\;e^{-\epsilon t}\;(-\epsilon)
\\
&&=
\theta(t)\;\int_0^\infty \!\!d\epsilon\;g(\epsilon)\;e^{-\epsilon t}\;
\left(-
\sum_{n=1}^{\infty}a_{n-1}\;
[h(\epsilon)]^{n-1}
\right)
\\
&&=
-\sum_{n=1}^{\infty} \frac{a_{n-1}}{n}\;
\underbrace{G*G*\cdots*G}_n \;(t)
\;,
\end{eqnarray*}
which is exactly eq.~(\ref{eq:conv1}), with $A_n=-a_n/(n+1)$.

Now that we have the expression for $D(t)$, let us show how one can
write, for example, an integral-differential equation for $R(t,t')$
[eq.~(\ref{eq:diffRexact})]. First, notice that from 
eq.~(\ref{eq:Rexact})
\begin{displaymath}
  \label{eq:G-from-R}
G(t-t')=\frac{Y^{}(t)}{Y^{}(t')}\;R(t,t')
\;.
\end{displaymath}
Hence,
\begin{eqnarray*}
\label{eq:n-G-conv-from-R}
\underbrace{G*G*\cdots*G}_n \;(t-t')
&&=
\int_{-\infty}^{\infty}\!\!\!d\tau_1
\int_{-\infty}^{\infty}\!\!\!d\tau_2
\dots
\int_{-\infty}^{\infty}\!\!\!d\tau_{n-1}
\;\;
\frac{Y(t)}{Y(\tau_1)}\;R(t,\tau_1)
\;
\\
&&
\:\:\:\:\:\:\:\:\:\:\:\:
\:\:\:\:\:\:\:\:\:\:\:\:
\:\:\:\:\:\:\:\:\:\:\:\:
s 
\;\frac{Y(\tau_{n-1})}{Y(t')}\;R(\tau_{n-1},t')
\nonumber
\\
&&=
\frac{Y(t)}{Y(t')}\;
\int_{-\infty}^{\infty}\!\!\!d\tau_1
\int_{-\infty}^{\infty}\!\!\!d\tau_2
\dots
\int_{-\infty}^{\infty}\!\!\!d\tau_{n-1}
\;\;
\;R(t,\tau_1)
\;
\\
&&
\:\:\:\:\:\:\:\:\:\:\:\:
\:\:\:\:\:\:\:\:\:\:\:\:
\:\:\:\:\:\:\:\:\:\:\:\:
s 
\;R(\tau_{n-1},t')
\\
&&=
\frac{Y(t)}{Y(t')}\;\underbrace{R*R*\cdots*R}_n \;(t,t')
\;,
\end{eqnarray*}
which allows us to write the last term in eq.~(\ref{eq:diffRexact})
as
\begin{eqnarray*}
  \label{eq:D-last-term-in-R}
\frac{Y(t')}{Y(t)}\;D(t-t')
&&=
\frac{Y(t')}{Y(t)}\;
\sum_{n=1}^\infty A_{n-1} \;\underbrace{G*G*\cdots*G}_n \;(t-t')
\\
&&=
\sum_{n=1}^\infty A_{n-1} \;\underbrace{R*R*\cdots*R}_n \;(t,t')
\;.
\end{eqnarray*}
Thus finally we have
\begin{equation}
\frac{\partial R(t,t')}{\partial t} = -z(t)\;R(t,t')+
\sum_{n=0}^{\infty} A_n \;\underbrace{R*R*\cdots*R}_{n+1} \;(t,t')
\;.
\label{eq:diffRexact-in-R}
\end{equation}

Lastly, let us note that the above equations can be easily extended to
the describe the evolution of the two-time two-point correlation
function $C(r;t,t')\equiv C(\vec x, \vec y;t,t')$ and response
function $R(r;t,t')\equiv R(\vec x,\vec y;t,t')$, with $r=|\vec x-\vec
y| > 0$.  One can easily verify that the generalisation can be done by
formally replacing the density of states $g(\epsilon)$ by
\begin{equation}
g(\epsilon;r) \equiv c g(\epsilon) \int_{-1}^{1} dy \; (1-y^{2})^{(d-3)/2}
\cos(\sqrt{\epsilon}ry) \;\;\;\;\;\;\;\;\;
\end{equation}
in $d \geq 3$. Here $c^{-1}=\int_{-1}^{1} dy \;
(1-y^{2})^{(d-3)/2}=2^{(d-2)}B((d-1)/2,(d-1)/2)$ is the normalization
constant.  In $d=1$ and $2$, one simply has to use
$g(\epsilon)\cos(\sqrt{\epsilon}r)$ and $g(\epsilon)\int_{0}^{\pi}
d\theta\cos(\sqrt{\epsilon}r \cos(\theta))/\pi$, respectively. The
closed set of equations of motion for $C(r;t,t')$ and $R(r;t,t')$ are series
expansions with coefficients $A_{n}(r)$ which now depend on the distance $r$
explicitly.

\section{The ageing limit of the equations of motion}
\label{app:eq-of-motion-aging}

To obtain the equations of motion for the response in the ageing limit, one
substitutes in eq.~(\ref{eq:series-with-A}) [or
eq.~(\ref{eq:diffRexact-in-R})]
\begin{equation*}
R(t,t')=R_{st}(t-t')+R_{ag}(t,t')  
\end{equation*}
and use that the stationary response decays to zero in time scales in which
the ageing component remains roughly constant. (See~\cite{Leto} for a detailed
explanation of this separation). For example, in the term
\begin{eqnarray*}
&& A_1 \int dt'' R(t,t'') R(t'',t') 
\nonumber\\
&& \sim A_1 \left[ 
\int_{t'}^t dt'' \; R_{ag}(t,t'') R_{ag}(t'',t') + 
\int_{t'}^{t'^+} dt'' \; R_{ag}(t,t'') R_{st}(t''-t') 
\right.
\nonumber\\
&& 
\;\;\;\;\;\;\;\;\;
\left.
+ 
\int_{t^-}^t dt'' \; R_{st}(t-t'') R_{ag}(t'',t') 
\right]
\nonumber\\
&&\sim A_1 \left[ 
\int_{t'}^t dt'' \; R_{ag}(t,t'') R_{ag}(t'',t') + 
2\,\chi_{st}\;R_{ag}(t,t') 
\right]
\;
\end{eqnarray*}
with
\begin{equation*}
\chi_{st}=\int_{t'}^{t'^+} dt'' \; R_{st}(t''-t') =
\int_{t^-}^t dt'' \; R_{st}(t-t'') 
\;.
\end{equation*}
Notice that if we start from a term with one time integral (the term with
coefficient $A_1$), then we collect in the ageing regime, in addition to the
term with one time integral, a term with no time integrals. Similarly,
starting from a term with $n$ integrals (the term with coefficient $A_n$), we
would generate in the ageing limit terms with $n,n-1,\dots,0$ integrals. We
can collect all these terms into a new series
\begin{equation*}
\sum_{n=0}^\infty {\tilde A}_n \int dt_n\int dt_{n-1} \dots 
\int dt_1 \; R_{ag}(t,t_1) R_{ag}(t_1,t_2) \dots R_{ag}(t_{n},t')
\;,
\end{equation*}
where the coefficients ${\tilde A}_n$ are related to the original $A_n$ by a
simple combinatorial argument, that goes as follows. Terms with $n$ time
integrals and $n+1$ $R_{ag}$'s are obtained starting with terms with $p\ge n$
integrals and $p+1$ $R$'s, where $p-n$ of the $R$'s are replaced by $R_{st}$
and the remaining $n+1$ $R$'s are replaced by $R_{ag}$'s. This allows us to
write
\begin{eqnarray*}
{\tilde A}_n
&&=\sum_{p=n}^\infty {A}_p \;
\left(
\begin{array}{ll}
p+1
\\
p-n
\end{array}
\right)
\; \chi_{st}^{p-n}
\\
&&=\frac{1}{(n+1)!}\sum_{p=n}^\infty {A}_p \;
(p+1)\,p\,(p-1)\dots (p-n+1) \; \chi_{st}^{p-n}
\\
&&=\frac{1}{(n+1)!} \left(\frac{d}{d\chi_{st}}\right)^n
\;\sum_{p=n}^\infty {A}_p \;
(p+1)\; \chi_{st}^{p}
\\
&&=\frac{1}{(n+1)!} \left(\frac{d}{d\chi_{st}}\right)^n
\;\sum_{p=0}^\infty {A}_p \;
(p+1)\; \chi_{st}^{p}
\end{eqnarray*}
Now, from \ref{app:eq-of-motion}, $a_p=-A_p\, (p+1)$ are the coefficients of
the series expansion of the function $\epsilon(h)$. Therefore we can simply
write
\begin{equation}
\label{eq:tilde-A-appendix}
{\tilde A}_n
=-\frac{1}{(n+1)!} \left(\frac{d}{d\chi_{st}}\right)^n \epsilon(\chi_{st})
\;.
\end{equation}

\section{The spherical spin-glass with Gaussian interactions}
\label{app:p=2}

The spherical spin-glass model with Gaussian distributed two-body
interactions has been studied in a series of papers~\cite{Ciuchi},
\cite{Horner}-\cite{Secu}. It was there shown that the 
asymptotic solution in the ageing regime scales as 
\begin{equation}
R_{ag}(t,t') \sim t^{-3/2} f_R\left(\lambda \right)
\; , 
\qquad
C_{ag}(t,t') \sim f_C\left(\lambda \right)
\; , 
\label{eq:scaling-p=2}
\end{equation}
and $0 \leq \lambda\equiv t'/t\leq 1$.
Here, we look at this problem from a
different angle, motivated by the generic discussion presented in
Sect.~\ref{sec:solution-dyn-eqs}. Let us analyse each term
in the equations for the global response and correlation by 
evaluating them
in the ageing regime using the scaling forms in~(\ref{eq:scaling-p=2}).
The equation for the global response reads
\begin{eqnarray*}
\frac{\partial R(t,t')}{\partial t} 
&=&
z(t) R(t,t') + \int_{t'}^t dt'' R(t,t'') R(t'',t') 
\; , 
\end{eqnarray*}
with the Lagrange multiplier $z(t)$ being fixed by the condition $C(t,t)=1$ 
that yields:
\begin{equation}
z(t) = T + 2 \int_0^t dt' \; C(t,t') R(t,t') 
\; .
\label{eq:z-eq-p=2}
\end{equation}
In the aging regime, the left-hand-side scales as 
\begin{equation}
- \left[\frac{3}{2} f_R(\lambda) + f'_R(\lambda) \right]  t^{-5/2} 
\; .
\end{equation}
The asymptotic scaling of the Lagrange multiplier is know from 
the exact solution to be 
\begin{equation}
z(t) \sim 2 + c t^{-1}
\end{equation}
with $c$ a numerical coefficient.
Let us derive this result from eq.~(\ref{eq:z-eq-p=2}) using the 
forms in (\ref{eq:scaling-p=2}). If, proceeding as usual, we separate 
 the integral in (\ref{eq:z-eq-p=2}) into a stationary 
and an aging part and we keep the leading contributions to each of these, we find
\begin{displaymath}
\lim_{t\to\infty} z(t) = z_\infty + \alpha \, t^{-1/2} \equiv
T + \frac{1}{T} (1-q_{ea}^2) +
t^{-1/2} \int_0^1 d\lambda' \; f_R (\lambda') f_C (\lambda')
\; . 
\end{displaymath}
If one uses the relation between $q_{ea}$ and $T$, 
the time independent term is consistent with $z_\infty=2$. However,
the approach to the asymptotic value is incorrect. The mistake we have 
done is that we neglected the correction to the constant value of the 
stationary contribution that cancels the leading aging one, and we 
neglected the correction to the leading aging contribution that 
yields the correct $t^{-1}$ decay.

The easiest and most general way of deriving the result above is 
to go back to the general representation of the solution for $C$ and 
$R$ and plug these into the integral term in (\ref{eq:z-eq-p=2}). 
After some algebra, and working at $T=0$ for simplicity, one finds
\begin{equation}
\int_0^t dt' \; C(t,t') R(t,t') 
=
Y^{-2}(t) \int d\epsilon \; e^{-2\epsilon t} g(\epsilon) h(\epsilon) 
\label{interm}
\end{equation}
where $g(\epsilon)$ is a generic density of states and 
$h(\epsilon)$ is the function defined in eq.~(\ref{eq:h-def}). 
Now, a density of states with a finite support in $[0,1]$  and 
power law decays on the two ends 
can be mimicked by the form
\begin{equation}
g(\epsilon) \propto \epsilon^\nu (1-\epsilon)^{1-\nu}
\end{equation}
that allows us to do the calculations explicitly. In particular, 
the semicircle case is mimicked by $\nu=1/2$.
Close to $\epsilon \sim 0$ the function $h(\epsilon)$ then reads
\begin{equation}
h(\epsilon) \sim \frac{\pi}{\sin\pi \nu} \; 
\left[ (1-\nu) - \epsilon -\cos(\pi\nu) \epsilon^\nu +\dots \right]
\; .
\end{equation}
Replacing in (\ref{interm}) and using the asymptotic form of $Y(t)$ one 
has
\begin{equation}
z(t) \sim a (1-\nu) - \alpha \cos\pi\nu \; t^{-\nu} - c \, t^{-1} + \dots
\end{equation}
Thus, for the special case $\nu=1/2$ the prefactor of the  
$t^{-\nu}$ term vanishes and one recovers the correct behaviour in
$t^{-1}$. A similar phenomenon occurs in the integral over the
two responses. The stationary contributions yields a term that is 
$O(t^{-3/2})$
and its cancellation with the constant asymptotic value of 
$z_\infty$ fixes the Edwards-Anderson order parameter
as a function of temperature:
\begin{equation}
T+\frac{1}{T} (1-q_{ea}^2) = \frac{2\, (1-q_{ea})}{T}
\label{eq:qea-p=2}
\end{equation}
that is equivalent to
$
T^2=(1-q_{ea})^2 \; \Rightarrow \; q_{ea} = 1-T
$ ($T\leq T_c=1$).
The next-to-leading order terms are
$O(t^{-2})$ but their prefactor vanishes. Finally, one
is left with a term that is $O(t^{-5/2})$, just as 
the time-derivative and the another term left from 
$z(t) R_{ag}(t,t')$. This non-trivial equation 
fixes the functions $f_C$ and $f_R$. 

The analysis of the equation for $C$ is similar. 
The leading terms are $O(1)$; their cancellation leads to an 
equation identical to (\ref{eq:qea-p=2}).
The next-to-leading order
terms are $O(t^{-1/2})$ but their overall prefactor vanishes. 
The time-derivative term is $O(t^{-1})$
and it combines with the remaining terms to yield a non-trivial equation.

Note that in the analysis above we used the correct asymptotic
behaviour of $R$ and $C$ in the ageing regime, that we know from 
the direct solution to the (linear set of) Langevin equations.
If $p\geq 3$ one cannot solve the dynamics exactly and one 
is forced to do an asymptotic analysis of the equations for 
$R$ and $C$ {\it assuming} a decay of the linear response and 
searching for a consistent solution. 
When $p\geq 3$ one proposes~\cite{Cuku} 
$R_{ag}(t,t') \sim t^{-1} f_R(\lambda)$ 
and $C_{ag}(t,t') \sim f_C(\lambda)$. In this case, 
the stationary and ageing contributions
to the Lagrange multiplier are both {\it finite}.
Moreover, all terms in the right-hand-side of the equations for $R$ and $C$
are of the same order, $O(t^{-1})$ and $O(1)$, respectively, while the time
derivatives are much smaller, $O(t^{-2})$ and $O(t^{-1})$, respectively.
Dropping the time-derivatives one finds a solution that is consistent
this the scaling assumption. In the $p=2$ one could have proposed a 
similar (wrong) scaling and look for its consequences.
It is interesting to
notice that if one naively pursues this calculation one finds $X=0$ as the
unique possible asymptotic solution [see eq.~(\ref{eq:Xdef}) for the
definition of $X$] which is consistent with the exact result, $X\sim
t^{-1/2}$, in the $t\to \infty$ limit.

\section{Diagonalising the matrix ${\cal M}_{\eta}(\vec k_1, \vec k_2)$}
\label{sec:calc-M}

In this appendix we study the eigenmodes the matrix 
${\cal M}_{\eta}(\vec k_1, \vec k_2)$ defined in 
eq.~(\ref{eq:M1}) for the case of $V_{x}=1$ (without coarse-graining) and
eq.~(\ref{eq:M2}) with finite coarse-graining volume $V_{x}$.

\subsection{Case $V_{x}=1$ (without coarse-graining)}
\label{sec:calc-M-1}

First we study the the matrix ${\cal M}_{\eta}(\vec k_1, \vec k_2)$ defined in 
eq.~(\ref{eq:M1}) for the case of $V_{x}=1$.
We show that two and only two eigenvalues of ${\cal M}_\eta$ depend on
$\eta$ and all the others are fixed to one.
For convenience, let ${\cal R}_{k}(t)
\equiv r(k,t,0)$ and ${\cal R}_{k}(t')=r(k,t',0)$ label
the $k$-indexed row of column vectors ${\mathbf R}(t)$ and ${\mathbf R}(t')$
(these vectors live in ${\cal D}=L^d$ dimensions). 
Note that the length of this vector is constant in time
$|{\mathbf R}(t)|^{2}=R^{2}$ 
(which ensures conservation of the
length of the $N$-component vector field in the $N \to \infty$ limit).
Let ${\mathbf v}^\lambda$
be, within the same notation, an eigenvector of ${\cal M}_\eta$ with
eigenvalue $\lambda$. Then
\begin{eqnarray*}
&& \lambda\;v^\lambda_{\vec k_1}
=
\sum_{\vec k_2}{\cal M}_\eta (\vec k_1,\vec k_2)\;v^\lambda_{\vec k_2}
\nonumber\\
&& 
=
\sum_{\vec k_2}
\delta_{\vec k_1 \vec k_2}\;v^\lambda_{\vec k_2}
+i\eta\;\frac{\Delta^2}{{\cal N}}
\left[
\;{\cal R}_{\vec k_1}(t)\;\sum_{\vec k_2} {\cal R}_{\vec k_2}(t')\;
v^\lambda_{\vec k_2}
+{\cal R}_{\vec k_1}(t')\;\sum_{\vec k_2} {\cal R}_{\vec k_2}(t)\;
v^\lambda_{\vec k_2}
\right]
\nonumber\\
&& 
=
v^\lambda_{\vec k_1}
+i\eta\;\frac{\Delta^2}{{\cal N}}
\;\left[({\mathbf R}(t'){\mathbf v}^\lambda)\; {\cal R}_{\vec k_1}(t)
+( {\mathbf R}(t){\mathbf v}^\lambda)\; {\cal R}_{\vec k_1}(t')
\right]
\;.
\end{eqnarray*}
This equation is equivalent to 
\begin{equation}
(\lambda-1)\;{\mathbf v}^\lambda=i\eta\;\frac{\Delta^2}{{\cal N}}\;
\left[({\mathbf R}(t'){\mathbf v}^\lambda)\;{\mathbf R}(t)
+({\mathbf R}(t){\mathbf v}^\lambda)\;{\mathbf R}(t')
\right]
\label{eq:eigen}
\end{equation}
and has ${\cal D}$ solutions. 

Only two eigenvalues of ${\cal M}_\eta$ are changed by
the presence of the $\eta$ term. One of them is
\begin{displaymath}
{\mathbf v}^{\lambda}\parallel{\mathbf R}(t)+{\mathbf R}(t')
\;\;\;
{\rm with}
\;\;\;
\lambda_{+}=1+\;\frac{i\eta}{{\cal N}}
\left(
\;{C(t,t')+1}
\right)
\; .
\end{displaymath}
the other is
\begin{displaymath}
{\mathbf v}^{\lambda}\parallel{\mathbf R}(t)-{\mathbf R}(t')
\;\;\;
{\rm with}
\;\;\;
\lambda_{-}=1+\;\frac{i\eta}{{\cal N}}
\left(
\;{C(t,t')-1}
\right)
\; .
\end{displaymath}
Here we used that expression for the global correlation at $T=0$,
\begin{displaymath}
C(t,t')=
\Delta^2
\;{\mathbf R}(t'){\mathbf R}(t)
=
\Delta^2
\;\sum_k r(k,t',0)\;r(k,t,0)
\;,
\end{displaymath}
and
\begin{displaymath}
\Delta^{2}R^{2}=1\;.
\end{displaymath}
The other ${\cal D}-2$ solutions are such that
\begin{displaymath}
{\mathbf v}^{\lambda}\perp 2 {\rm d} \; \mbox{plane spanned by
the above two eigenvectors with}
\;\; \lambda=1
\;.
\end{displaymath}


\subsection{Case of finite $V_{x}$}
\label{sec:calc-M-2}

Next we study the the matrix ${\cal M}_{\eta}(\vec k_1, \vec k_2)$ 
defined in eq.~(\ref{eq:M2}) for the case of finite coarse-graining
volume $V_{x}$.
Let ${\cal R}_{k}(t,\vec y) \equiv r(k,t,0)\,e^{i \vec k 
\vec y}$ label the $k$-indexed row of the column vector ${\mathbf
R}(t,\vec y)$. This allows us to write an eigenvalue equation for
${\cal M}_{\eta}$, similarly to what we have done above for the case $l=1$,
\begin{eqnarray}
(\lambda-1)\;{\mathbf v}^\lambda=i\eta\;\frac{\Delta^2}{V_x}\;
\sum_{\vec y\in V_x}
\Big[&&
({\mathbf R}(t',\vec y){\mathbf v}^\lambda)\;{\mathbf R}(t,\vec y)
\nonumber\\
&&\;\;\;\;\;\;
+({\mathbf R}(t,\vec y){\mathbf v}^\lambda)\;{\mathbf R}(t',\vec y)
\Big]
\label{eq:eigen-coarse}
\;,
\end{eqnarray}
where the inner (dot) product is here defined as ${\mathbf a} {\mathbf
b}=\sum_k a_k^*\,b_k$.

This eigenvalue equation has ${\cal D}-2V_x$ trivial solutions with
$\lambda=1$. The eigenvectors for such solutions satisfy ${\mathbf
R}(t',\vec y){\mathbf v}^\lambda=0$ and ${\mathbf R}(t,\vec
y){\mathbf v}^\lambda=0$, for $\vec y\in V_x$, and hence span the
orthogonal subspace to that spanned by the $2V_x$ vectors ${\mathbf
R}(t,\vec y)$ and ${\mathbf R}(t',\vec y)$ ($\vec y\in V_x$).

The remaining (non-trivial) eigenvectors can be written as
\begin{displaymath}
{\mathbf v}^\lambda=\sum_{\vec y\in V_x} \alpha^\lambda(\vec
y)\;{\mathbf R}(t,\vec y) +\beta^\lambda(\vec y)\;{\mathbf R}(t',\vec
y)
\end{displaymath}
for some $2V_x$ expansion coefficients $\alpha^\lambda(\vec y)$ and
$\beta^\lambda(\vec y)$ for $\vec y\in V_x$. Plugging this into
eq.~(\ref{eq:eigen-coarse}) leads to 
\begin{eqnarray}
(\lambda-1)\;\alpha^\lambda(\vec y)=i\eta\;\frac{\Delta^2}{V_x}\;
\sum_{\vec{y'}\in V_x}
\Big[&&
\alpha^\lambda(\vec{y'})\;({\mathbf R}(t',\vec y){\mathbf R}(t,\vec{y'}))
\nonumber\\
&&\;\;\;\;\;\;
+
\beta^\lambda(\vec{y'})\;({\mathbf R}(t',\vec y){\mathbf R}(t',\vec{y'}))
\Big]
\;,
\label{eq:alpha}
\end{eqnarray}
\begin{eqnarray}
(\lambda-1)\;\beta^\lambda(\vec y)=i\eta\;\frac{\Delta^2}{V_x}\;
\sum_{\vec{y'}\in V_x}
\Big[&&
\alpha^\lambda(\vec{y'})\;({\mathbf R}(t,\vec y){\mathbf R}(t,\vec{y'}))
\nonumber\\
&&\;\;\;\;\;\;
+
\beta^\lambda(\vec{y'})\;({\mathbf R}(t,\vec y){\mathbf R}(t',\vec{y'}))
\Big]
\;.
\label{eq:beta}
\end{eqnarray}
Using that 
\begin{equation*}
\Delta^2\;{\mathbf R}(t,\vec y){\mathbf R}(t',\vec{y'})=C(\vec y,\vec{y'};t,t')=
C(t,t')\;\exp\left[{\frac{|\vec y-\vec{y'}|^2}{L^2(t)+L^2(t')}}\right]
\;,
\end{equation*}
with the length scales $L(t)=2\sqrt{t}$ and $L(t')=2\sqrt{t'}$, and
substituting in eqs.~(\ref{eq:alpha}) and (\ref{eq:beta}), one obtains
\begin{eqnarray}
(\lambda-1)\;\alpha^\lambda(\vec y)=i\eta\;\frac{1}{V_x}\;
\sum_{\vec{y'}\in V_x}
&&\left\{
\alpha^\lambda(\vec{y'})\;
C(t,t')\;\exp\left[{\frac{|\vec y-\vec{y'}|^2}{L^2(t)+L^2(t')}}\right]
\right.
\nonumber\\
&&\left.
+
\beta^\lambda(\vec{y'})\;
\exp\left[{\frac{|\vec y-\vec{y'}|^2}{2L^2(t')}}\right]
\right\}
\;,
\label{eq:alpha2}
\end{eqnarray}
\begin{eqnarray}
(\lambda-1)\;\beta^\lambda(\vec y)=i\eta\;\frac{1}{V_x}\;
\sum_{\vec{y'}\in V_x}
&&\left\{
\alpha^\lambda(\vec{y'})\;
\exp\left[{\frac{|\vec y-\vec{y'}|^2}{2L^2(t)}}\right]
\right.
\nonumber\\
&&\left.
+
\beta^\lambda(\vec{y'})\;
C(t,t')\;\exp\left[{\frac{|\vec y-\vec{y'}|^2}{L^2(t)+L^2(t')}}\right]
\right\}
\;.
\label{eq:beta2}
\end{eqnarray}

These equations are difficult to solve for generic ratios of the length
scales $L(t)$ and $L(t')$ to the coarse-graining box size $\ell$. 
In the following we consider some limiting cases.

\subsubsection{Case $\ell\ll L(t), \; L(t')$}$\;$

One simple
situation is given by $\ell\ll L(t),L(t')$, in which case $|\vec y-\vec{y'}|\ll
L(t),L(t')$ and eqs.~(\ref{eq:alpha2}) and (\ref{eq:beta2}) simplify to
\begin{equation}
(\lambda-1)\;\alpha^\lambda(\vec y)=i\eta\;\frac{1}{V_x}\;
\sum_{\vec{y'}\in V_x}
\left[
C(t,t')\;\alpha^\lambda(\vec{y'})+\beta^\lambda(\vec{y'})\;
\right]
\;,
\label{eq:alpha3}
\end{equation}
\begin{equation}
(\lambda-1)\;\beta^\lambda(\vec y)=i\eta\;\frac{1}{V_x}\;
\sum_{\vec{y'}\in V_x}
\left[
\alpha^\lambda(\vec{y'})+C(t,t')\;\beta^\lambda(\vec{y'})\;
\right]
\;.
\label{eq:beta3}
\end{equation}
The eigenvalues can now be found if one adds and subtracts 
eqs.~(\ref{eq:alpha3}) and (\ref{eq:beta3}) and then 
sums both sides over $\vec y$, obtaining
\begin{eqnarray}
&& (\lambda-1)\;
\left[\sum_{\vec y\in V_x}\alpha^\lambda(\vec y)\pm \sum_{\vec y\in
    V_x}\beta^\lambda(\vec y)\right]
\nonumber\\
&&
\;\;\;\;\;\;\;\;\;\;\;\;\;\;\;\;\;\;\;\;\;
=i\eta\;[C(t,t')\pm1]\;
\left[\sum_{\vec y\in V_x}\alpha^\lambda(\vec y)\pm \sum_{\vec y\in
    V_x}\beta^\lambda(\vec y)\right]
\;,
\label{eq:alpha-beta}
\end{eqnarray}
which has two non-trivial solutions
$$\lambda_\pm=1+i\eta\;[C(t,t')\pm 1]
\;,$$ and $2V_x-2$ trivial
solutions such that $\lambda=1$ and $\sum_{\vec y\in
V_x}\alpha^\lambda(\vec y)=\sum_{\vec y\in V_x}\beta^\lambda(\vec
y)=0$. Thus, in the case $\ell\ll L(t),L(t')$ we recover the same
eigenvalues, and hence the same distribution as in the case
$V_x=1$. This result was to be expected since coarse-graining of
completely correlated regions should not affect the distribution
obtained for a single site.

\subsubsection{Case $\ell\gg L(t),L(t')$}$\;$

One can seek approximate solutions of eqs.~(\ref{eq:alpha2}) and
(\ref{eq:beta2}) in this limit if one assumes that the
$\alpha^\lambda(\vec y)$ and $\beta^\lambda(\vec y)$ are slowly
varying functions of $\vec y$, in which case one must solve the
approximate equations
\begin{equation}
(\lambda-1)\;\alpha^\lambda(\vec y)\approx i\eta\;\frac{L^d}{V_x}\;
\left[
C(t,t')\;\alpha^\lambda(\vec y)+\beta^\lambda(\vec y)\;
\right]
\;,
\label{eq:alpha4}
\end{equation}
\begin{equation}
(\lambda-1)\;\beta^\lambda(\vec y)\approx i\eta\;\frac{L^d}{V_x}\;
\left[
\alpha^\lambda(\vec y)+C(t,t')\;\beta^\lambda(\vec y)\;
\right]
\;,
\label{eq:beta4}
\end{equation}
where for simplicity we considered $L=L(t)\sim L(t')$. These equations
admit non-trivial solutions
$$\lambda_\pm\approx
1+i\eta\;\left(\frac{L}{\ell}\right)^d\;(C(t,t')\pm1)
\;.
$$ 
Naively,
there are as many of these solutions as the number of $\vec y$ points in
$V_x$, for each of $\lambda_\pm$. However, the assumption that the
$\alpha^\lambda(\vec y)$ and $\beta^\lambda(\vec y)$ are slowly varying correlates
them, and thus one cannot expect that the non-trivial solutions span
the whole of the $2V_x$ dimensional space. The number of independent
non-trivial solutions should be only order
$V_x/L^d=(\ell/L)^d$ for each of $\lambda_\pm$.

 
\section{The response of composite operators}
\label{fdt}

In this Appendix we compute the response of the composite operator
$\phi(\vec x,t) \phi(\vec x,t') $ to a perturbation that couples to
the same composite operator evaluated at a different spatial point and
the same times~\cite{Franz-Parisi}. In the Langevin equation such a
perturbation is represented by an additional deterministic
time-dependent force:
\begin{displaymath}
F(\vec x,t) = \int_0^t dt'' [h(\vec x; t'',t)+
h(\vec x; t,t'')]
\phi(\vec x,t'') 
\; .
\end{displaymath}
In the following we work at zero temperature. The perturbed field $\phi_h$ is
\begin{displaymath}
\phi_h(\vec k,t) = r(k;t,0)  \phi(\vec k,0)
+ \int_0^t dt'' \;  r(k;t,t'') F(\vec k,t'')
\; .
\end{displaymath}
The variation of the force $F$ 
with respect to the perturbation $h$ is
\begin{displaymath}
\frac{\delta F(\vec k, t)}{\delta h(\vec k', t_1,t_2)}
=
\phi(\vec k-\vec k',t_2) \delta(t-t_1) \theta(t-t_2) +
\phi(\vec k-\vec k',t_1) \delta(t-t_2) \theta(t-t_1) 
\end{displaymath}
The response we are interested in is given by 
\begin{eqnarray*}
\left[ \, 
\frac{\delta \phi(\vec x, t) \phi(\vec x, t')}{\delta h(\vec x', t'',t''')} 
\, \right]_{ic}
=
\left[ \, \phi(\vec x,t)
\frac{\delta \phi(\vec x, t')}{\delta h(\vec x', t'',t''')}
\, \right]_{ic}
+
\left[ \, 
\frac{\delta \phi(\vec x, t) }{\delta h(\vec x', t'',t''')}
\phi(\vec x,t')
\, \right]_{ic}
\; .
\end{eqnarray*}
After some rather straightforward calculations one finds
\begin{eqnarray}
\left[ \, 
\frac{\delta \phi(\vec x, t) \phi(\vec x, t')}{\delta h(\vec x', t'',t''')}
\, \right]_{ic}
&=& 
\; R(\vec x-\vec x';t,t'') C(\vec x-\vec x';t',t''') \theta(t-r''')
\nonumber\\
&& +
R(\vec x-\vec x';t,t''') C(\vec x-\vec x';t',t'') \theta(t-t'') 
\nonumber\\
&& +
R(\vec x-\vec x';t',t'') C(\vec x-\vec x';t,t''') \theta(t'-t''')
\nonumber\\
&& +
R(\vec x-\vec x';t',t''') C(\vec x-\vec x';t,t'') \theta(t'-t'')
\nonumber
\label{eq:composite-response}
\end{eqnarray}
where $R$ and $C$ are the usual two-point, two-time linear 
response and correlation.
One can readily verify that this expression is not 
simple related to time-variations of the four-point correlation
$C_4$ contrary to what one might have naively expected.
Note that this expression has the expected $t=t'$ and $t''=t'''$ limit.

\end{document}